\RequirePackage{fix-cm}
\documentclass[smallextended]{svjour3}     
\smartqed  
\usepackage{graphicx}
\usepackage[utf8]{inputenc}
\usepackage[english]{babel}
\usepackage{float}
\usepackage{booktabs}
\usepackage{adjustbox}
\usepackage{amsmath,amssymb,amsfonts}
\usepackage{subcaption}
\usepackage{graphicx}
\usepackage{textcomp}
\usepackage{multirow}
\usepackage{url} 
\usepackage{verbatim}
\usepackage{xcolor}
\usepackage{threeparttable}
\usepackage{framed}
\usepackage[numbers]{natbib}
\usepackage{balance}
\usepackage{rotating}
\usepackage{amsmath}
\usepackage[hidelinks]{hyperref}
\usepackage{enumitem}
\usepackage{listings}
\usepackage{algpseudocode}
\usepackage{float}
\usepackage{makecell}
\usepackage{textcomp}
\usepackage{hhline}
\usepackage{lscape}
\usepackage{rotating}
\usepackage{textcomp}
\usepackage{tablefootnote}
\usepackage{tabularx}
\usepackage{acronym}
\usepackage{natbib}
\usepackage{hyperref}
\setcitestyle{numbers,square,comma}
\mathchardef\mhyphen="2D 
\algnewcommand{\LineComment}[1]{\Statex \(\triangleright\) \#1}

\def\BibTeX{{\rm B\kern-.05em{\sc i\kern-.025em b}\kern-.08em
    T\kern-.1667em\lower.7ex\hbox{E}\kern-.125emX}}

\colorlet{shadecolor}{gray!10}
\colorlet{framecolor}{black}
\newenvironment{frshaded}{%
	\MakeFramed {\FrameRestore}}%
{\endMakeFramed}

\paperwidth=\dimexpr
1in + \oddsidemargin
+ \textwidth
+ 1in + \oddsidemargin
\relax
\paperheight=\dimexpr
1in + \topmargin
+ \headheight + \headsep
+ \textheight
+ 1in + \topmargin
\relax
\usepackage[pass]{geometry}\relax

\begin{document}

\title{Towards Understanding the Challenges of Bug Localization in Deep Learning Systems}

\titlerunning{Towards Understanding the Challenges of Bug Localization in Deep Learning Systems}        

\author{Sigma Jahan        \and Mehil B. Shah \and
        Mohammad Masudur Rahman  
}

\institute{Sigma Jahan \at
              Dalhousie University, Canada \\
              \email{sigma.jahan@dal.ca}  
            \and
          Mehil B. Shah \at
              Dalhousie University, Canada \\
              \email{shahmehil@dal.ca} 
           \and
           Mohammad Masudur Rahman \at
              Dalhousie University, Canada \\
              \email{masud.rahman@dal.ca} 
}

\date{the date of receipt and acceptance should be inserted later}

\maketitle

\begin{abstract}

Software bugs cost the global economy billions of dollars annually and claim $\sim$50\% of the programming time from software developers. Locating these bugs is crucial for their resolution but challenging. It is even more challenging in deep-learning systems due to their black-box nature. Bugs in these systems are also hidden not only in the code but also in the models and training data, which might make traditional debugging methods less effective. In this article, we conduct a large-scale empirical study to better understand the challenges of localizing bugs in deep-learning systems. First, we determine the bug localization performance of five existing techniques using 2,365 bugs from deep-learning systems and 2,913 from traditional software. We found these techniques significantly underperform in localizing deep-learning system bugs. Second, we evaluate how different bug types in deep learning systems impact bug localization. We found that the effectiveness of localization techniques varies with bug type due to their unique challenges. For example, tensor bugs were more accessible to locate due to their structural nature, while all techniques struggled with GPU bugs due to their external dependencies. Third, we investigate the impact of bugs' extrinsic nature on localization in deep-learning systems. We found that deep learning bugs are often extrinsic and thus connected to artifacts other than source code (e.g., GPU, training data), contributing to the poor performance of existing localization methods.


\end{abstract}

\section{Introduction} \label{sec:loc_introduction}

Software bugs are human-made errors in the code that prevent it from working correctly \cite{arcuri2008automation}. They are often prevalent in modern software systems and could range from hundreds to thousands in a single system \cite{karampatsis2020often}. Due to the bugs in software systems, the global economy loses billions of dollars every year \cite{r50, cisq2022report}. Developers also spend about 50\% of their programming time dealing with software bugs and failures \cite{r50}. To correct any bug, the developers first need to identify the location of a bug within a software system, which is known as \textit{bug localization} \cite{zhou2012should}. According to a recent survey, 49.20\% of 327 software practitioners from several major technology companies (e.g., Google, Meta, Amazon, and Microsoft) consider the localization of bugs as one of the most challenging tasks during software development and maintenance \cite{survey}. 

While localizing bugs in traditional software systems (a.k.a, non-deep learning systems) remains a challenge, it could even be more challenging in deep learning systems. Unlike bugs in non-deep learning systems, deep learning-related bugs could be hidden in the source code, training data, trained models, or even deployment scripts \cite{chen2020comprehensive, amershi2019software, gonzalez2020state}. Besides, the use of various deep learning frameworks (e.g., PyTorch, Caffe, and TensorFlow) could make these bugs even more complex \cite{foutse}. 

Given the prevalence and costs of software bugs, any automated support to localize the bugs can greatly benefit software practitioners. Over the years, many approaches have been designed to localize bugs in traditional software systems using information retrieval \cite{buglocator, bluir, blizzard, amalgam}, dynamic program analysis \cite{moreno2014use, perez2014dynamic}, and deep learning \cite{DNNLOC, deeplocalize, deeplocator}. However, due to the significant differences between traditional and deep learning bugs, these existing solutions might not be adequate for localizing bugs in deep learning systems. 

To date, there exist only a few techniques for detecting bugs in deep learning systems. Most of them concentrate on specific types of bugs (e.g., model bugs, training bugs) without considering the broader spectrum of deep learning systems. \citet{deeplocalize} propose a dynamic approach to localize different types of model bugs in the Deep Neural Network (DNN). They identify the faulty layers containing numerical bugs by customizing the Keras' callback function and analyzing the dynamic behaviors of a model. However, their solution focuses on only model bugs from deep learning systems, strongly coupled with the Keras library, and achieves a low accuracy, which presents significant challenges for widespread adoption by the industry. In another study, \citet{wardat2022deepdiagnosis} propose a heuristic-based approach to diagnose two main categories of bugs -- model bugs and training bugs. They also recommend actionable fixes of the bugs based on the diagnosis. Since their approach depends on a set of hard-coded rules, it might be limited in terms of scalability and context-awareness. In a recent work, \citet{cao2022deepfd} introduces a technique that leverages the dynamic properties of a model and an ensemble of three machine learning classifiers (e.g., KNN, Decision Tree, Random Forest) to localize five types of training bugs (e.g., loss, gradient) from deep learning systems. Their technique might also not be able to address a broader array of bug types from deep learning systems, highlighting the issues of scalability. On the other hand, \citet{irblrose} use basic Information Retrieval (IR) algorithms, such as rVSM and BM25, to localize bugs in deep-learning systems. They report poor performance but do not perform any comprehensive analysis to understand the poor performance of IR-based techniques. 

Interestingly, at least 30 techniques adopt IR algorithms to locate bugs in traditional software systems due to their computational efficiency and lightweight nature \cite{buglocator, bluir, amalgam, blizzard, BLIA, irblrose}. They were also reported to perform comparably to the complex models (e.g., LDA) \cite{rahman2018improving}. Unlike deep learning-based techniques, IR-based techniques rely on the textual similarity between bug reports and source code as a proxy of suspiciousness, which is simple and explainable. However, IR-based techniques suffer from vocabulary mismatch issues \cite{r51} and can only capture linear relationships between two items. On the contrary, deep learning-based techniques can capture the non-linear relationships between two items \cite{obulesu2018machine, almeida2002predictive, bitvai2015non}. Thus, they have the potential to capture more nuances in the relevance between enriched information from the source code and bug reports. However, they also suffer from poor outlier handling, class imbalance problems, and a lack of monitoring \cite{deng2018deep}. Thus, the potential of existing solutions for localizing bugs in deep learning applications is neither well understood nor well investigated to date. Our work in this article fills in this important gap in the literature.

In this article, we conduct a large-scale empirical study to better understand the challenges of locating bugs in deep learning systems. First, we collect a total of 2,365 bugs from deep-learning systems and 2,913 bugs from traditional software systems (a.k.a non-deep-learning systems), and \textit{empirically} show how existing techniques (e.g., BugLocator \cite{buglocator}, BLUiR \cite{bluir}, BLIA \cite{BLIA}, DNNLOC \cite{DNNLOC}), bjXnet \cite{han2023bjxnet}) perform in locating bugs from deep learning systems. Our work utilizes these traditional techniques as baselines, adapting their core principles to the specific nuances of deep learning bugs. Second, we categorize our collected bugs based on an existing bug taxonomy \cite{humbatova2020taxonomy} and found that certain bugs from deep learning systems (e.g., GPU bugs) are more difficult than others to locate due to their multifaceted heterogeneous dependency issues. Finally, we found that deep learning bugs are connected to artifacts other than source code (e.g., GPU, training data, external dependencies) and are prone to be extrinsic in nature, which might explain the poor performance of existing techniques for these bugs. We thus answer three important research questions in our study as follows.

\begin{enumerate}
\itemsep3pt
\item [(a)]\textbf{RQ$\mathbf{_1}$: How effective are the existing approaches in localizing bugs from deep learning systems?}
\\
We evaluated the performance of five existing approaches (BugLocator \cite{buglocator}, BLUiR \cite{bluir}, BLIA \cite{BLIA}, DNNLOC \cite{DNNLOC}, and bjXnet \cite{han2023bjxnet}) using two datasets – Denchmark \cite{denchmark} and BuGL \cite{BuGL}. First, we found that their performance measures are poorer (e.g., 31.59\% less MAP for BugLocator, 33.25\% for BLUiR, 34.14\% for BLIA, 31.43\% for DNNLOC, 27.87\% for bjXnet ) in localizing bugs from deep learning systems than that of non-deep learning systems. Our statistical tests (t-test \cite{ttest}, Cohen's D \cite{cohend}) also report that their performance is significantly lower. Second, we found that localizing bugs from the deep learning frameworks is more challenging than libraries or tools due to the frameworks' inherent complexity. Although our findings reinforce the existing understanding and belief about the challenges of the bugs in deep learning systems \cite{islam2019comprehensive, irblrose, deeplocalize}, we also substantiate them with solid empirical evidence and demonstrate the performance gap of existing solutions in localizing the two categories of bugs.

\item [(b)]\textbf{RQ$\mathbf{_2}$: How do different types of bugs in deep learning systems impact bug localization?
}
\\
We use an existing taxonomy \cite{humbatova2020taxonomy} of bugs to classify the bugs in deep learning systems and evaluate the performance of five existing techniques for each type of bug. First, we found that 64.80\% of the bugs from deep learning systems are related to deep learning (e.g., model, training), whereas the remaining ones are not. Second, we found that bjXnet and DNNLOC demonstrated better results in locating model and tensor bugs, possibly due to their ability to capture comprehensive contextual information specific to these bugs using deep neural networks. We also found that BLUiR performs comparably to DNNLOC for training bugs, which might be attributed to its structured information retrieval. However, all five baseline techniques experienced significant difficulty localizing GPU bugs. Thus, our analysis offers valuable insights regarding the nature of different types of bugs in deep learning systems and highlights the specific strengths and weaknesses of existing techniques, which could be useful to advance debugging support for deep learning systems. 

\item [(c)]\textbf{RQ$\mathbf{_3}$: What are the implications of extrinsic bugs in deep learning systems for bug localization?
}
\\
Bugs triggered by external entities (e.g., third-party libraries, GPU) are called \emph{extrinsic bugs} \cite{extrinsic&intrinsic}. Given the frequent use of deep learning libraries and their external dependencies, the bugs in deep learning systems could be extrinsic \cite{foutse}. Since the existing techniques mostly focus on intrinsic bugs (i.e., triggered by bug-introducing change), we investigate how they deal with extrinsic bugs from deep learning systems. First, we found deep learning systems have 40.00\% extrinsic bugs, which is almost four times higher than that of non-deep learning systems. Second, we found that the localization performance of existing techniques degrades significantly for extrinsic bugs (e.g., 15.20\% less MAP for DNNLOC) (Table \ref{tab:performance-extrinsic-intrinsic-dl}). We also found that deep neural network-based solutions (e.g., DNNLOC) are not particularly helpful for locating extrinsic bugs because they are designed to detect code patterns, not issues from external sources or environments. Finally, we found a significant correlation between the bugs in deep learning systems and the extrinsic factors using appropriate statistical analysis (Chi-Square test), which delivers valuable insight for designing effective solutions to find bugs in deep learning systems.
\end{enumerate}

\section{Background}\label{chapter:Background}

In this section, we introduce the necessary terminologies and concepts to follow the remainder of the article. We introduce extrinsic bugs, intrinsic bugs, the taxonomy of deep-learning bugs, and the categories of projects in deep-learning systems. 

\setlength{\arrayrulewidth}{0.1mm}
\setlength{\tabcolsep}{11pt}
\renewcommand{\arraystretch}{1.0}
\begin{table}[h]
\centering
\caption{Example of deep learning-related and non deep-learning-related extrinsic bugs \cite{fairseq_issue_1860, ignite_issue_1426}}
\label{tab:bug_extrinsic}
\begin{tabular}{|l|}
\hline
\multicolumn{1}{|c|}{\textbf{\begin{tabular}[c]{@{}l@{}}Deep Learning-related Extrinsic Bug \cite{fairseq_issue_1860}\\ \end{tabular}}} \\ \hline
\begin{tabular}[c]{@{}l@{}}\textbf{Title:} \\ wmt19 model cannot run on GPU except \#0.\\ \textbf{Description:}\\ I am running the tutorial. I successfully loaded the model from\\ the hub and tried to run it on the second GPU (id=1). \\, Which raised an exception that data and models are stored \\ on different GPUs. With GPU (id=0) works fine.\\ Code sample:\\ import torch\\ en2de = torch.hub.load('pytorch/fairseq', \\ 'transformer.wmt19.en-de'\\ checkpoint\_file='model1.pt:model2.pt:model3.pt:model4.pt'\\ tokenizer='moses',bpe='fastbpe').to(torch.device('cuda:1'))\\ result = en2de.translate({[}'hello'{]})\\ Environment:\\ fairseq Version==0.9.0, PyTorch Version ==1.4.0, \\ OS: Ubuntu 18.04,vPython version: 3.6, \\ CUDA/cuDNN version: 10.2, \\ GPU models and configuration: RTX 2080 x 2\end{tabular}\\ \hline
\multicolumn{1}{|c|}{\textbf{\begin{tabular}[c]{@{}l@{}}Non Deep Learning-related Extrinsic Bug \cite{ignite_issue_1426}\\ \end{tabular}}}\\ \hline
\begin{tabular}[c]{@{}l@{}}\textbf{Title:}\\ GitHub CI on Windows is broken.\\ \textbf{Description:}\\ Normally, we should skip distributed tests on Windows with \\ SKIP\_DISTRIB\_TESTS=1 \\ CI\_PYTHON\_VERSION="3.7" \\ sh tests/run\_cpu\_tests.sh, \\ but a distributed test was executed:\\ tests/ignite/contrib/engines/test\_common.py\\ ::test\_distrib\_cpu ERROR {[}2\%{]}\\ Related to beta support of distributed on Windows in Pytorch 1.7\end{tabular} \\ \hline
\end{tabular}
\end{table}

\subsection{Extrinsic bug} A bug caused by factors external to a software system, such as changes to the operating environment, requirements, or third-party libraries, is known as \textit{extrinsic bug}. Rodriguez-Perez et al. suggests three heuristics based on bug reports to identify extrinsic bugs as follows \cite{extrinsic&intrinsic}. 

\textit{(a) Environment:} An extrinsic bug is caused by a modification to the environment in which the software system operates. The environment could be an operating system, a physical machine, or even a cloud infrastructure.

\textit{(b) Requirement:} An extrinsic bug is triggered by a change outside the project's version control system. During software development, if a user requirement changes after implementation, the development team might implement the new requirement without discarding the old feature. The old, unexpected feature will then be considered an extrinsic bug. 

\textit{(c) Third-party library:} The bug found in the project's third-party library is considered an extrinsic bug. For example, if a software project uses a third-party library for processing images for a mobile application, and the app crashes when processing certain image formats due to a bug in that third-party library, that bug will then be considered an extrinsic bug.

\subsection{Intrinsic bug}
The external factors do not cause an intrinsic bug; rather, it is caused by a bug-introducing change in the version control system \cite{extrinsic&intrinsic}. For example, if a messaging application fails to deliver messages due to a logical error in a recent code change, that would be an intrinsic bug.

\subsection{Taxonomy of bugs in deep learning systems} Software bugs in deep learning systems can be divided into two categories -- DL bug and NDL bug \cite{humbatova2020taxonomy}.

\textbf{Deep Learning (DL) bug} refers to a software error that is connected to the deep learning module embedded in the software system, causing inaccurate or unexpected output. According to the existing literature \cite{humbatova2020taxonomy}, DL bugs can be divided into five main categories: Model, Training, Tensor \& Input, API, and GPU. 

\begin{itemize}
 \item \textbf{Model bug} is connected to the structure and properties of a deep learning model (Table \ref{tab:model} \cite{texar_pytorch_issue_313}). An example of a model bug is an incorrect model initialization caused by an input image size mismatch, resulting in inaccurate output in a computer vision application. 

\item \textbf{Training bug} occurs during the training phase of a deep learning application (Table \ref{tab:training} \cite{fastai_issue_3048}). For instance, during the training of a deep learning model for object detection, if the loss function is incorrectly defined, the model will learn to detect objects with very poor accuracy, leading to incorrect output from the system.

\item \textbf{Tensor \& Input bug} (a.k.a tensor bug) occurs due to wrong tensor input or tensor calculation issues (Table \ref{tab:tensor} \cite{mxnet_issue_13760}). For instance, if the tensor input shape is declared incorrectly, it will lead to output errors.

\item \textbf{API bug} occurs due to incorrect use of an API in the deep learning software system (Table \ref{tab:API} \cite{mxnet_issue_13862}). For example, an API bug might occur if a developer mistakenly calls the wrong API function from the deep learning framework (e.g., Tensorflow), causing inaccurate results in the output. 

\item \textbf{GPU bug} is connected to the Graphics Processing Unit (GPU) used in the system (Table \ref{tab:GPU} \cite{autokeras_issue_1238}). For example, if the model's memory requirements exceed the available GPU memory or the GPU is not compatible with the DL framework, then they could lead to errors during model training.

\end{itemize}

\textbf{Non-Deep Learning (NDL) bug} refers to a software error that is not related to the deep learning module but still leads to unexpected behaviors in deep learning applications. An example of NDL bugs could be a logical error in the source code that leads to a deadlock, making the program being stuck in an infinite loop.

As shown in Table \ref{tab:bug_extrinsic}, Bug 1860 \cite{fairseq_issue_1860} is a deep learning-related extrinsic bug triggered by the change in the environment. When the WMT19 model runs on multiple GPUs, the execution fails since the same GPU cannot store both the model and data. It is clearly related to the deep learning module. On the other hand, this bug is not related to the Fairseq library (a.k.a., deep learning application); rather, it is related to external factors (e.g., GPU), which indicates its extrinsic nature.

In Table \ref{tab:bug_extrinsic}, Bug 1426 \cite{ignite_issue_1426} is another extrinsic bug connected to the Windows OS environment. The bug triggers when the tests from the CI pipeline are distributed over multiple Windows machines. It is clearly not related to deep learning (a.k.a., Non-DL bug), but the triggering factors are outside of the version control system, which indicates an extrinsic nature.

\subsection{Categories of projects in deep learning systems} Deep learning-based projects can be categorized into Frameworks, Tools, Libraries, Applications, Engines, Platforms, and Compilers \cite{denchmark}. The core elements—Frameworks, Tools, and Libraries—comprise the majority of relevant bug reports, accounting for 96.6\% of the dataset \cite{denchmark}.

\setlength{\arrayrulewidth}{0.1mm}
\setlength{\tabcolsep}{11pt}
\renewcommand{\arraystretch}{1.0}
\begin{table}[h]
\centering
\caption{Example of a bug report from deep learning framework \cite{mxnet_issue_10224}}
\label{tab:framework_example}
\begin{tabular}{c}
\hline
\multicolumn{1}{|c|}{\textbf{Framework Bug (Bug ID: 10224)}} \\ \hline
\textbf{Title} \\ \hline
\multicolumn{1}{|l|}{Language model example cannot be run} \\ \hline
\textbf{Description} \\ \hline
\multicolumn{1}{|l|}{\begin{tabular}[c]{@{}l@{}}The language model example cannot be run without manually \\
    creating a data folder. There are also inconsistencies between the \\ documentation 
    and the code. Optional argument -- data DATA \\ location of the data corpus does not appear in the code train.py\\
    ...\\ Detailed BR: https://github.com/apache/mxnet/issues/10224 \\
    \end{tabular}} \\ \hline
\end{tabular}
\end{table}

A \textbf{deep learning framework} is a software platform that provides the environment for designing, training, and deploying deep learning models \cite{nguyen2019machine}. Examples include TensorFlow\footnote{\url{https://www.tensorflow.org/}}, PyTorch\footnote{\url{https://pytorch.org/}}, and Apache MXNet\footnote{\url{https://mxnet.apache.org/versions/1.9.1/}}. These frameworks come with pre-defined modules and functions and offer a structured way to implement deep learning architectures using high-level programming interfaces \cite{khan2018deep}. The example bug in Table \ref{tab:framework_example} \cite{mxnet_issue_10224} is characterized by the inability to run a language model without manually creating a data folder. It represents a framework bug because it directly impacts the core functionalities of the framework, specifically the design and training of models. Frameworks are expected to provide seamless, user-friendly environments for developing deep learning models, and issues that hinder the ease of use, such as documentation inaccuracies and additional manual setup steps, indicate problems at the framework level. 

\setlength{\arrayrulewidth}{0.1mm}
\setlength{\tabcolsep}{11pt}
\renewcommand{\arraystretch}{1.0}
\begin{table}[h]
\centering
\caption{Example of a bug from deep learning library \cite{texar_pytorch_issue_313}}
\label{tab:library_example}
\begin{tabular}{c}
\hline
\multicolumn{1}{|c|}{\textbf{Library Bug (Bug ID: 313)}} \\ \hline
\textbf{Title} \\ \hline
\multicolumn{1}{|l|}{A bug in GPT2Tokenizer} \\ \hline
\textbf{Description} \\ \hline
\multicolumn{1}{|l|}{\begin{tabular}[c]{@{}l@{}}GPT2Tokenizer fails to recover a sentence \\ \textbackslash{}"BART is a seq2seq model.\textbackslash{}"\\ with encoded ids of it. \\ The output sentence is \textbackslash{}"BART is a seqseq model.\textbackslash{}".\\ It should be related to numbers' processing. \\ ... \\
URL: https://github.com/tanyuqian/texar-pytorch\\ /blob/master/examples/bart/gpt2\_tokenizer\_bug.py \\
Detailed BR:https://github.com/asyml/texar-pytorch/issues/313
\end{tabular}} \\ \hline
\end{tabular}
\end{table}

A \textbf{deep learning library} is a collection of functions that facilitate specific tasks within deep learning. Libraries like Keras\footnote{\url{https://keras.io/}} and cuDNN\footnote{\url{https://developer.nvidia.com/blog/tag/cudnn/}} can either be integrated into frameworks or can operate independently \cite{wang2019various}. From Table \ref{tab:library_example} \cite{texar_pytorch_issue_313}, the bug in the `GPT2Tokenizer' within the \texttt{texar-pytorch} project is classified as a library bug due to its specific component focus and the nature of functionality. It uses the TokenizerBase class from the texar library, designed to work with the PyTorch framework. The tokenizer's failure to accurately process text data leads to the bug, which indicates a library issue. 

\setlength{\arrayrulewidth}{0.1mm}
\setlength{\tabcolsep}{11pt}
\renewcommand{\arraystretch}{1.0}
\begin{table}[h]
\centering
\caption{Example of a bug from deep learning tool \cite{tensorboard_issue_5596}}
\label{tab:tool_example}
\begin{tabular}{c}
\hline
\multicolumn{1}{|c|}{\textbf{Tool Bug (Bug ID: 5596)}} \\ \hline
\textbf{Title} \\ \hline
\multicolumn{1}{|l|}{Tensorboard can not load all Hyperparameters keys} \\ \hline
\textbf{Description} \\ \hline
\multicolumn{1}{|l|}{\begin{tabular}[c]{@{}l@{}}Encountered an issue in TensorBoard where it couldn't load \\ all hyperparameter keys when I used writer.add-hparams with \\ different hparam-dict parameters in multiple experiments. This \\ problem made it difficult to track and compare different \\ hyperparameter settings across these experiments, affecting the \\ overall functionality of TensorBoard. I provided code snippets and \\ responses to showcase the inconsistency in the display of \\ hyperparameters in TensorBoard's interface. This issue \\ is significant as it hinders the effective use of TensorBoard \\ for experiment tracking and analysis. \\ ...\\ Detailed BR: https://github.com/tensorflow/tensorboard/issues/5596\\ \end{tabular}} \\ \hline
\end{tabular}
\end{table}

Finally, \textbf{deep learning tools} refer to utilities that assist with the tasks related to deep learning, such as visualization or model optimization. An example would be TensorBoard\footnote{\url{https://www.tensorflow.org/tensorboard}}, which is often used for TensorFlow visualization \cite{erickson2017toolkits}. Each of the frameworks, libraries, and tools plays unique but complementary roles in the context of deep learning. From Table \ref{tab:tool_example} \cite{tensorboard_issue_5596}, the bug in TensorBoard qualifies as a tool bug since TensorBoard is a visualization toolkit in the TensorFlow ecosystem. The bug refers to the inability of TensorBoard to load all hyperparameter keys when \texttt{writer.add-hparams} is used with varying \texttt{hparam-dict} parameters, directly impacting its core functionality as a tool. This feature is essential for monitoring and contrasting various experimental settings of TensorBoard.

\section{Study Methodology}\label{sec:loc_study_methodology}

\begin{figure*}[htbp]
  \centering
  \includegraphics[width=1.0\textwidth]{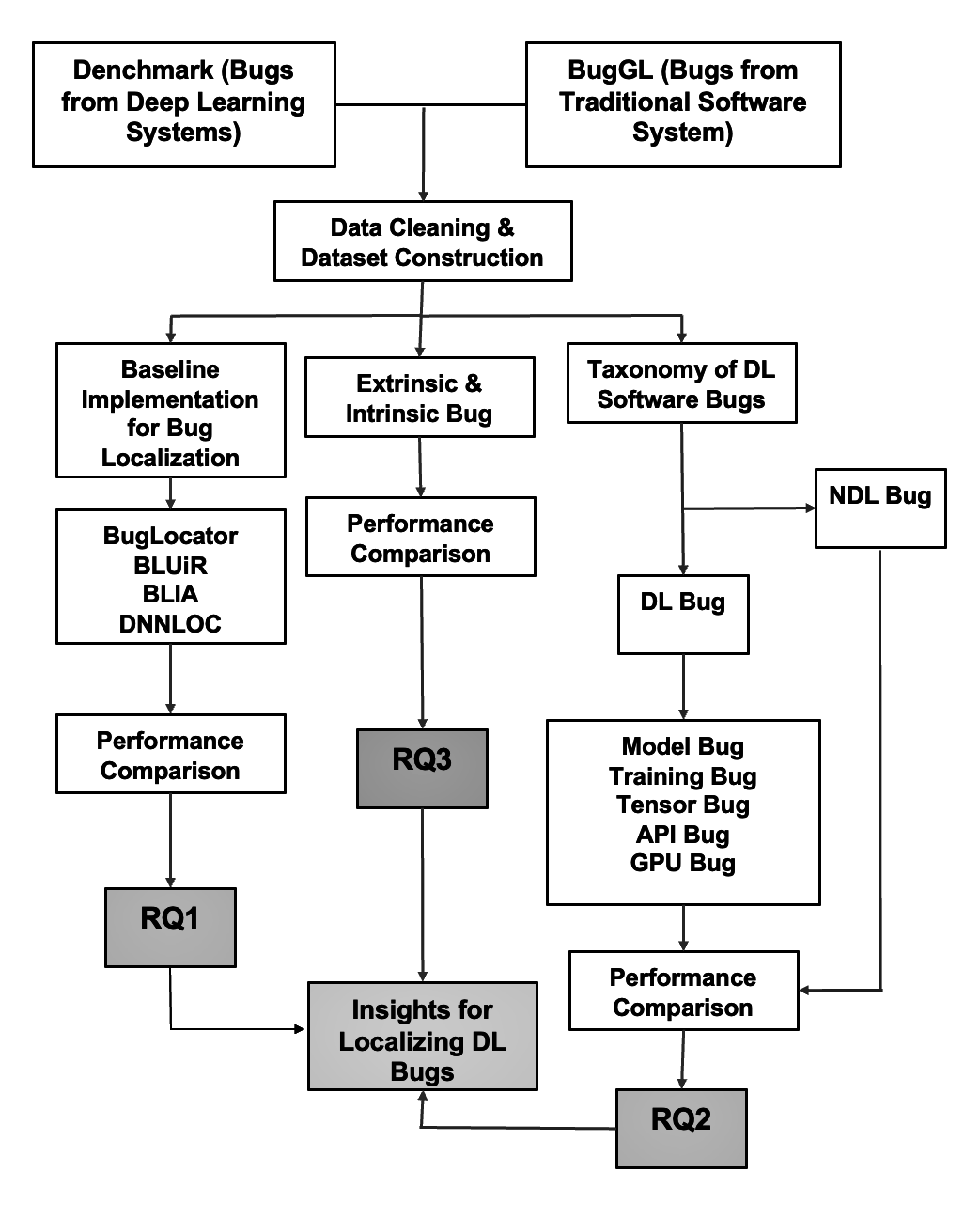}
  \caption{Schematic diagram of our empirical study}
  \label{fig:localiztion_schematic_diagram}
\end{figure*}

Fig.~\ref{fig:localiztion_schematic_diagram} shows the schematic diagram of our conducted study. 
First, we collect bug reports from two benchmark datasets for two different software systems: deep learning systems \cite{denchmark} and traditional software systems \cite{BuGL}. Then, we contrast the performance of five existing techniques~\citep{buglocator, bluir, BLIA, DNNLOC} in locating bugs between deep learning systems \cite{denchmark} and traditional software systems \cite{BuGL}. Second, we perform an in-depth analysis to understand the challenges of localizing different deep-learning bugs. Finally, we investigate the influence of extrinsic factors on deep learning bugs and their impact on bug localization. This section discusses the major steps of our study design, which are as follows:

\subsection{Construction of dataset}

\textbf{Dataset Collection.} In our study, we use two benchmark datasets: BuGL and Denchmark. BuGL contains traditional software bugs, while Denchmark focuses on bugs in deep learning systems. BuGL \cite{BuGL} provides a wide range of traditional software bugs across multiple programming languages (e.g., C++, Java, Python) and includes a total of 2,913 bug reports from 12 Python projects. In contrast, Denchmark \cite{denchmark}, published in the MSR 2021 Dataset Track, focuses on bugs in deep learning systems written in ten different programming languages, including JavaScript, Python, Java, Go, C++, Ruby, TypeScript, PHP, C\#, and C. It contains 2,365 bug reports from 136 deep learning-based projects written in Python. While the Denchmark dataset may not cover all possible systems, it is representative of the current state of deep learning system debugging, encompassing a wide range of 136 DL projects (e.g., computer vision, natural language processing, image processing, and visualization).

To ensure a fair comparison between deep learning and traditional software bugs, we only included Python-based projects and bugs from both datasets. Additionally, multiple studies \cite{BuGL, irblrose} have used Denchmark and BuGL for bug localization experiments, making them suitable benchmarks for our experiments as well.

\begin{table}[ht]
\centering
\caption{Original \cite{denchmark} and Experimental Distribution of Projects and Bugs}
\label{tab:denchmark_org}
\begin{tabular}{|l|c|c|}
\hline
\multicolumn{3}{|c|}{\textbf{Original Distribution}} \\ \hline
\textbf{Category} & \textbf{Projects} & \textbf{Bugs} \\ \hline
Framework         & 25                & 836           \\ \hline
Platform          & 8                 & 150           \\ \hline
Engine            & 4                 & 47            \\ \hline
Compiler          & 2                 & 33            \\ \hline
Tool           & 31                & 510           \\ \hline
Library           & 44                & 666           \\ \hline
Application       & 12                & 124           \\ \hline
\textbf{Total}    & \textbf{136}      & \textbf{2365} \\ \hline
\multicolumn{3}{|c|}{\textbf{Experimental Dataset (DL Systems)}} \\ \hline
\textbf{Category} & \textbf{Projects} & \textbf{Bugs} \\ \hline
Framework         & 17                & \multicolumn{1}{c|}{746} \\ \hline
Tool           & 28                & \multicolumn{1}{c|}{455} \\ \hline
Library           & 41                & \multicolumn{1}{c|}{594} \\ \hline
\textbf{Total}    & \textbf{86}      & \textbf{1795}\\ \hline
\end{tabular}
\end{table}

We adopted a set of filtration steps to construct our experimental dataset, which consists of 86 projects from DL systems (shown in Table \ref{tab:denchmark_org}). 

\begin{itemize}
    \item \textit{Original Denchmark Dataset:} The original Denchmark dataset contains 136 projects and 2,365 bug reports. These projects are divided into several categories by the original authors \cite{denchmark}: Frameworks, Tools, Libraries, Applications, Engines, Platforms, and Compilers.
    \item \textit{Filter 1:} We focused on the core elements of deep learning – Frameworks, Tools, and Libraries, as they represent the majority of relevant bug reports in the dataset (96.6\%). In the original dataset, there were 25 projects from Framework, 31 projects from Tool, and 44 projects, which gave us a total of 100 projects. We excluded projects with fewer than ten reported bugs to ensure a robust dataset for meaningful comparisons of bug localization techniques. After applying this filter, we had 18 framework projects, 29 tool projects, and 42 library projects, which led to a total of 87 projects. 
    \item \textit{Filter 2:} We noted that some projects had multiple categories (e.g., a tool that also functions as a library). After carefully reviewing the official documentation and analyzing the role of these projects, we removed overlapping projects to ensure that each project was uniquely classified as either a framework, tool, or library. 
\end{itemize} 
    After applying these filtering steps, the final experimental dataset consists of 86 projects and 1,795 bug reports - Frameworks: 17 projects, 746 bugs, Tools: 28 projects, 455 bugs, Libraries: 41 projects, 594 bugs.

\textbf{Data cleaning and pre-processing.} After collecting the data from two benchmark datasets, we cleaned and preprocessed them using a set of steps.\\
\indent
\textit{Corpus creation.} We first download the latest version of the code repositories, ensuring that we have the most up-to-date code for analysis. Next, to accurately link bug reports to their corresponding buggy code, we used the heuristic of \citet{irblrose}, focusing on commit messages and bug reports. This involved analyzing commit messages for keywords indicative of bug fixes (e.g., `fix', `bug', `error', `resolve') and connecting them to corresponding bug reports. To identify the correct bug-fix commit, we cross-referenced these commit messages with bug IDs from the reports, ensuring a precise match. To extract the buggy part from the bug-fix commit, we employed PyDriller, suggested by \citet{irblrose}. We also capture each project’s most recently released version as of the bug report’s date and collect the appropriate version of the buggy code, especially in cases where the bug reports did not have clear buggy version information. \\
\indent
\textit{Query construction.} In IR-based bug localization, bug reports are treated as queries that can be executed with a search engine to detect the relevant source documents from the corpus. We construct a repository of bug reports by parsing the original datasets (Denchmark \& BuGL) and extracting important information such as bug IDs, descriptions, and timestamps. We construct the queries by extracting tokens from the title and description of bug reports, removing stop words, stemming each word, and splitting the tokens.\\
\indent
\textit{Meta data extraction.} We also capture the historical context of bugs by extracting their commit history information from the repositories, including commit messages, authors, timestamps, and code changes history. This information is extracted to replicate the existing technique BLIA \cite{BLIA}, which provides valuable insights into the evolution of the codebase and the bugs over time.\\
\indent
\textit{Ground truth construction.} Both benchmark datasets provide the ground truth that contains the correct locations of bugs in the code against the bug reports. To evaluate the performance of the bug localization approaches, we collect ground truth files from both of the original datasets.\\
\indent
To ensure a fair performance comparison of the bug localization techniques between DLSW and NDLSW, we have selected an equal amount of data from both datasets using probability sampling (1795 bug reports from each dataset), which have 95\% confidence interval and 5\% error margin \cite{hernandez2006effect}. We use the principle of randomization for selecting the subsets \cite{acharya2013sampling} to avoid any bias. We also manually analyzed the projects to avoid any overlap between the two datasets. We spent $\approx$5 hours on the manual analysis.

\begin{table}[t!]
    \centering
    \caption{Study dataset for bug localization \cite{denchmark, BuGL}}
    \label{tab:dataset stats}
    \begin{adjustbox}{max width=\linewidth}
        \begin{threeparttable}
            \begin{tabular}{|c|c|c|c|c|c|c|c|c|}
                \hline
                \multirow{2}{*}{\textbf{Data}} & \multirow{2}{*}{\textbf{\#P}} & \multirow{2}{*}{\textbf{\#BR}} & \multicolumn{2}{c|}{\textbf{\#SF}} & \multicolumn{2}{c|}{\textbf{\#BF}} & \multicolumn{2}{c|}{\textbf{\#Versions}} \\ \cline{4-9} 
                 &  &  & \textbf{Mean} & \textbf{Max} & \textbf{Mean} & \textbf{Max} & \textbf{Mean} & \textbf{Max} \\ 
                \hline \hline
                Dench. & 136 & 1795 & 441.60 & 3,559 & 2.60 & 227 & {6.7} & {80} \\ 
                \hline
                BuGL & 12 & 1795 & 420.50 & 3,306 & 2.35 & 198 & {6.2} & {65} \\ 
                \hline
            \end{tabular}
               \centering
                \small
                 \textbf{P}: Projects, \textbf{BR}: Bug Reports, \textbf{SF}: Source Files, \textbf{BF}: Buggy Files
        \end{threeparttable}
    \end{adjustbox}
\end{table}

\subsection{Replicating of existing techniques for experiments}

To answer our first research question, we needed to replicate existing techniques that localize bugs in deep learning systems. We thus select suitable representatives from the existing literature on bug localization. In particular, we choose baseline methods from the frequently used methodologies -- Information Retrieval (IR) and Deep Learning (DL).

Over the past decade, Information Retrieval (IR)-based bug localization techniques have gained significant attention, and some of these techniques have been proven effective in localizing bugs from traditional software systems \cite{buglocator, bluir, amalgam, BRTracer, BLIA}. Our extensive literature review shows that more than 30 peer-reviewed research papers have utilized IR for bug localization and highlighted its benefits \cite{irblrose, lee2018bench4bl}. These techniques rely on textual similarities between bug reports and source code to identify relevant files or methods. Our motivation was to explore whether the success of these well-established IR-based methods could extend to DL systems. 

While spectrum-based bug localization techniques are widely studied and capable of method-level or line-level localization \cite{jones2005empirical, eniser2019deepfault}, they might face significant challenges when applied to DL systems. Specturm-based techniques analyze program traces from passing and failing tests, but DL systems often present complexities such as non-deterministic behavior, silent bugs (bugs that do not cause crashes or generate stack traces) \cite{islam2019comprehensive, tambon2024silent}, and external dependencies (e.g., GPUs), which make obtaining such traces difficult or computationally prohibitive. In contrast, IR-based techniques do not require execution traces and instead rely on textual similarity between bug reports and code, which can be particularly effective in DL projects where bugs often stem from code, configuration, or data script issues \cite{wardat2021deeplocalize}. Although IR techniques localize bugs at a coarser file level, this granularity serves as a practical initial step for large-scale DL projects, narrowing down the search space in extensive codebases and complex systems with multiple components \cite{cao2022understanding}. Given these advantages and the computational efficiency of IR-based approaches, we focused our study on evaluating their applicability in DL systems.

We selected three IR-based techniques as our baselines for bug localization: BugLocator \cite{buglocator}, BLUiR \cite{bluir}, and BLIA \cite{BLIA}. These techniques represent critical advancements in the field. BugLocator is a seminal work that first introduced the application of Information Retrieval (IR) through the Vector Space Model (VSM) for bug localization. BugLocator paved the way for subsequent research in IR-based localization by using both source code and historical bug reports. On the other hand, BLUiR is built upon BugLocator by introducing structured information retrieval, which integrates structural elements from both source code and bug reports, marking a significant improvement over earlier methods reliant solely on textual similarity. Thirdly, BLIA extended BLUiR's approach by incorporating additional meta-components such as stack traces, code change histories, and version control information. This combination has shown superior performance over earlier techniques like Locus \cite{wen2016locus}, AmaLgam \cite{amalgam}, and BRTracer \cite{BRTracer}, making BLIA a comprehensive and effective baseline for our study. These three techniques not only represent key advancements in IR-based bug localization but are also computationally efficient, making them feasible for large-scale DL projects where obtaining execution traces is often impractical.

In addition to IR-based techniques, we selected DNNLOC \cite{DNNLOC}, a hybrid approach that combines traditional IR methods with Deep Neural Networks (DNNs). Unlike pure IR-based methods, DNNLOC leverages the ability of deep learning models to capture non-linear and complex relationships between bug reports and source code. This hybrid approach brings the computational efficiency of IR-based methods while enhancing localization through deep learning's ability to uncover hidden patterns in large datasets. Moreover, we included bjXnet \cite{han2023bjxnet}, a state-of-the-art model that applies an attention mechanism with a code property graph to localize buggy files. This ensures that our study covers a wide range of methodologies, from traditional IR-based approaches to more advanced deep learning methods with attention mechanisms.

We selected these five techniques to ensure a comprehensive evaluation of bug localization methodologies. They represent a spectrum of approaches, ranging from early IR-based methods to cutting-edge deep learning models, each contributing novel ideas to the field. We chose these baselines based on their relevance to our study goals, computational feasibility, representativeness of the existing literature, and the interpretability of their results. Even DNNLOC, despite incorporating a deep learning component, maintains computational efficiency due to its hybrid nature. The explainability of IR-based approaches offers another significant advantage, as developers can easily interpret the results based on textual similarities—a crucial factor when debugging complex DL systems. The details of each baseline are as follows.
\\

\indent
\textbf{BugLocator \cite{buglocator}} uses rVSM (revised Vector Space Model) that takes the document length into consideration to optimize the classic VSM model for bug localization to detect relevant source code documents against a bug report. It also calculates the SimiScore, which is a measure of similarity between a newly reported bug and previously fixed bugs based on their bug reports. SimiScore is combined with rVSM to calculate the final relevancy score. The relevant source code files are then ranked based on their combined scores, and the top-K documents are marked as buggy. Many subsequent IR-based techniques \cite{amalgam, blizzard, BLIA} adopted this method due to its simplicity and explainability. Hence, we chose this method as our first baseline.
\\

\indent
\textbf{BLUiR \cite{bluir}} uses AST parsing to extract four items: class, method, variable, and comment -- from each source code document. It also captures two fields from each of the bug reports (summary \& description). Then, a total of eight separate similarities are calculated between these two sets using the BM25 algorithm \cite{bluir}. Then, these document scores are summed as the suspiciousness score to rank buggy files against a given bug report. BLUiR technique is the first one that leverages structured elements from both source code and bug reports to localize bugs using IR. We thus choose this as another baseline technique for our study.
\\

\indent
\textbf{BLIA \cite{BLIA}} integrates several items such as textual similarity between bug reports and source documents \cite{buglocator}, code structures \cite{bluir}, version control history \cite{amalgam}, stack trace analysis  \cite{BRTracer}, and code change analysis in the IR-based bug localization. While bug reports and source code are useful, code change history can also assist in bug localization by providing the changes likely to induce a bug. BLIA has outperformed several previous techniques: BugLocator \cite{buglocator}, BLUiR \cite{bluir}, Amalgam \cite{amalgam}, BRTracer \cite{BRTracer}, which makes it suitable as the third baseline technique for our study.

IR-based localization can be adapted to different granularity levels (e.g., method, file).  We chose file-level granularity since each of the selected baselines frequently used this granularity. 
\\

\indent
\textbf{DNNLOC \cite{DNNLOC}} uses a hybrid method that incorporates both rVSM \cite{buglocator} from IR and Deep Neural Networks (DNNs) from DL. DNNs establish associations between specific terms in bug reports and the corresponding code tokens and terms in the source files. While DNNs alone do not achieve high accuracy due to dimensionality reduction in the projection process as they lose information, the integration with rVSM enhances their capability to correlate bug reports with relevant buggy files. Buggy source documents may not share textual similarities with bug reports, where IR-based techniques struggle. Thus, to address the challenges of IR-based techniques (e.g., lexical mismatch problem \cite{r51}) and to include non-linearity in relevance estimation, we chose DNNLOC as the final baseline for our study. 
\\

\indent
\textbf{bjXnet \cite{han2023bjxnet}} employs a multimodal representation learning framework that integrates the Code Property Graph (CPG) \cite{yamaguchi2014modeling} and attention mechanism \cite{ashish2017attention} to improve bug localization. CPG provides a rich, structured representation of the source code by incorporating different aspects such as abstract syntax trees (ASTs), control flow graphs (CFGs), and data flow graphs (DFGs). This comprehensive representation captures both the syntactic and semantic properties of the code, facilitating a deeper understanding that is crucial for identifying buggy parts. The attention mechanism in bjXnet allows the model to focus on the most critical components within the CPG, effectively highlighting important patterns and dependencies that are likely associated with bugs. By weighting specific nodes or edges more heavily, bjXnet can localize bugs more accurately than methods relying solely on textual similarity. bjXnet has outperformed several earlier techniques - such as BugLocator \cite{buglocator}, DeepLocator \cite{deeplocator}, DeepLoc \cite{xiao2019improving}, and DreamLoc \cite{qi2021dreamloc}. Given its cutting-edge methodology and empirical success, we selected bjXnet for our empirical study.

Since the original authors' replication packages were unavailable, we used the publicly available versions to replicate BugLocator \cite{buglocator}, BLUiR \cite{bluir}, and DNNLOC \cite{DNNLOC, DNNLOCreplicationpackage} that have been validated in peer-reviewed papers \cite{lee2018bench4bl}. We also carefully adapted the BLIA \cite{BLIA} and bjXnet \cite{han2023bjxnet} from their original replication packages to our datasets. Our replication package is provided for further experimental details \cite{Jahan2023replicationpackage}. 

We validated our replication of the baseline techniques by conducting experiments on a common subject system – SWT. All baseline techniques \cite{buglocator, bluir, BLIA, DNNLOC, han2023bjxnet} utilized the SWT dataset to assess performance in their original work, which makes it a viable choice for our validation. 

\begin{table}[h!]
\centering
\begin{tabular}{|l|c|c|c|}
\hline
\textbf{Model} & \textbf{MAP (Original)} & \textbf{MAP (Replication)} & \textbf{Difference (\%)} \\
\hline
BugLocator & 0.45 & 0.44 & 2.22 \\
BLUiR & 0.58 & 0.57 & 1.72 \\
BLIA & 0.65 & 0.63 & 3.08 \\
DNNLoc & 0.37 & 0.36 & 2.70 \\
bjXnet & 0.55 & 0.53 & 3.64 \\
\hline
\end{tabular}
\caption{Comparison of Original and Replicated Techniques in Terms of MAP Performance on the SWT Dataset}
\label{tab:baseline_comparison}
\end{table}

From Table \ref{tab:baseline_comparison}, we notice that the MAP differences range from 1.72\% to 3.64\% and fall within the margin of error, confirming the reliability of our replications. We also conducted a paired t-test, which resulted in a p-value of \textit{0.12}, showing the differences are statistically insignificant and validating the accuracy of our replication process.

Our study focuses on localizing bugs in deep learning systems, which are predominantly Python-based. To maintain consistency and eliminate confounding factors in our comparative analysis, we also examine traditional software systems written in Python. Although the baseline techniques \cite{buglocator, bluir, BLIA, DNNLOC, han2023bjxnet} were initially developed for Java projects, recent empirical evidence suggests that their core principles are largely language-agnostic \cite{Rezaalipour2024}. However, none of these techniques rely on language-specific features for bug localization. Our primary objective is to evaluate the effectiveness of existing bug localization techniques in deep learning systems. The results we obtained from traditional Python projects (NDLSW) are comparable to those reported for Java projects when these techniques were first introduced (Table \ref{tab:result_IR_loc}). This comparability suggests that the programming language may not significantly influence the validity or applicability of our findings. By focusing on Python, we address the predominant language in deep learning while ensuring our results remain relevant to a broader context of software engineering.

\subsection{Performance Evaluation}
We use three performance metrics for our study --- Top-K accuracy (Top@K), Mean Average Precision (MAP), and Mean Reciprocal Rank (MRR). These metrics have been frequently used by the relevant literature \cite{buglocator, bluir, amalgam, BLIA, DNNLOC, BRTracer}. 

\subsubsection{Top@K}
Top-K accuracy (Top@K) measures the percentage of bug reports for each of which at least one of the buggy files was present in the top-k retrieved files. We have used K= 1, 5, 10 for this study. 

\subsubsection{Mean Average Precision}
Precision@K measures the precision of each buggy source document's occurrence within a ranked list. Average Precision@K (AP) computes the average precision for all buggy documents within the ranked list against a search query (a.k.a. bug report). Mean Average Precision (MAP) is the average AP@K value across all queries in a system. \\
\begin{equation}
    \text{AP =} \frac{1}{D} \sum_{k=1}^{D} P_k \times \text{buggy}(k)
\end{equation}

\begin{equation}
    \text{MAP =} \frac{1}{|Q|} \sum_{q\in Q} \text{AP}(q)
\end{equation}

Here, $\text{AP}$ represents the Average Precision, and $D$ refers to the number of total results for a query. $k$ represents the position in the ranked list, $P_k$ denotes the precision calculated at the $k$-th position and $\text{buggy} (k)$ determines whether the $k$-th result in the ranked list is buggy or not.

\subsubsection{Mean Reciprocal Rank}
Mean reciprocal rank (MRR) calculates the average of the reciprocal ranks for a set of queries.
\begin{equation}
    \text{{MRR}(Q) =} \frac{1}{|Q|} \sum_{q\in Q} \frac{1}{\text{firstRank}(q)}
\end{equation}
where $\text{MRR}(Q)$ represents the Mean Reciprocal Rank for a set of queries $Q$, $|Q|$ represents the total number of queries in the set $Q$, $q\in Q$ represents each query in the set $Q$, $\text{firstRank}(q)$ represents the rank of the first correctly retrieved buggy document for the query $q$.

\section{Study Finding}\label{sec:loc_result}
\subsection{\textit{Answering RQ$\mathbf{_1}$: How effective are the existing approaches in localizing bugs from deep learning systems?}}

\subsubsection{\textbf{Performance comparison between DLSW and NDLSW}} Table \ref{tab:result_IR_loc} compares the performance of our baseline techniques in bug localization between deep learning systems and non-deep learning systems (a.k.a traditional software systems). We used three different evaluation metrics -- Top@k, MRR, and MAP, for our comparative analysis.

Our results reveal notable performance differences between the two types of systems across all five methods. Specifically, for BugLocator, the differences in MAP and MRR are \textit{31.59\%} and \textit{29.46\%}, respectively. BLUiR exhibits differences of \textit{33.25\%} in MAP and \textit{30.24\%} in MRR. BLIA shows even larger difference, with MAP and MRR differences of \textit{34.14\%} and \textit{30.77\%}, respectively. DNNLOC, while improving overall performance, still presents significant gaps with differences of \textit{31.43\%} in MAP and \textit{26.25\%} in MRR. Notably, bjXnet, despite being a state-of-the-art technique that leverages advanced features like Code Property Graphs (CPG) and attention mechanisms, also demonstrates a substantial performance gap. The differences in MAP and MRR for bjXnet are \textit{27.87\%} and \textit{24.24\%}, respectively, which is consistent with the performance gaps observed in other techniques included in our study. 

\begin{figure}[t!]
 \centering
 \includegraphics[width= 3.8in]{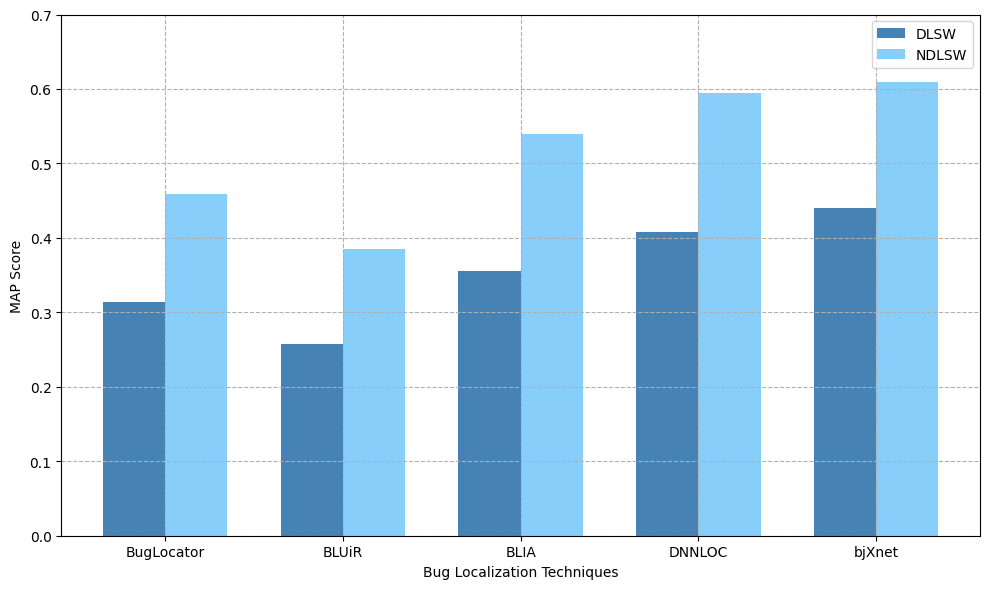}
    \caption{Performance comparison of existing approaches (BugLocator, BLUiR, BLIA, DNNLOC, bjXnet) between Deep Learning Systems (DLSW) and Non-Deep Learning Systems) (NDLSW)} 
 \label{fig:DLSW_NDLSW}
\end{figure}

We calculated these performance differences using the \textit{PerformanceDiff} metric from \citet{wattanakriengkrai2020predicting}. Fig. \ref{fig:DLSW_NDLSW} visualizes the MAP measures, and the differences are clearly visible. Overall, the results show that all five approaches perform lower when localizing bugs in deep learning systems, and the trend is consistent across all three metrics.

We also perform statistical tests to determine the significance of the performance gap between the two types of systems -- deep learning systems and non-deep learning systems (Table \ref{tab:significance_test_loc}). We took the Reciprocal Rank (RR) and Average Precision (AP) results of all the samples for each of the five approaches. Then, we performed \emph{Shapiro-Wilk normality test} \cite{royston1992approximating}, which reported normal distribution for those metrics. Then, we used appropriate significance and effect size tests to compare the result values from the two types of systems. For the normal distribution, we used t-test as the parametric test \cite{ttest}. In all significance tests, the p-values were less than the threshold (0.05) for each of the five approaches. Thus, the null hypothesis can be rejected for all comparisons. In other words, the performances of all baseline techniques significantly differ between the two types of systems. 

\begin{table*}[t!]
\centering
\caption{Performance of existing approaches (BugLocator, BLUiR, BLIA, DNNLOC, bjXnet) in bug localization}
\label{tab:result_IR_loc}
\resizebox{4.5in}{!}{
    \begin{threeparttable}
\begin{tabular}{cccccc}
\hline
\multicolumn{1}{|c|}{\textbf{Method}} & \multicolumn{1}{c|}{\textbf{Top@1}} & \multicolumn{1}{c|}{\textbf{Top@5}} & \multicolumn{1}{c|}{\textbf{Top@10}} & \multicolumn{1}{c|}{\textbf{MRR}} & \multicolumn{1}{c|}{\textbf{MAP}} \\ \hline 
\multicolumn{6}{c}{\textbf{DLSW}} \\ \hline
\multicolumn{1}{|c|}{BugLocator} & \multicolumn{1}{c|}{0.344} & \multicolumn{1}{c|}{0.547} & \multicolumn{1}{c|}{0.615} & \multicolumn{1}{c|}{0.371} & \multicolumn{1}{c|}{0.314} \\ \hline
\multicolumn{1}{|c|}{BLUiR} & \multicolumn{1}{c|}{0.201} & \multicolumn{1}{c|}{0.472} & \multicolumn{1}{c|}{0.585} & \multicolumn{1}{c|}{0.316} & \multicolumn{1}{c|}{0.257} \\ \hline
\multicolumn{1}{|c|}{BLIA} & \multicolumn{1}{c|}{0.411} & \multicolumn{1}{c|}{0.609} & \multicolumn{1}{c|}{0.719} & \multicolumn{1}{c|}{0.423} & \multicolumn{1}{c|}{0.355} \\ \hline
\multicolumn{1}{|c|}{DNNLOC} & \multicolumn{1}{c|}{0.468} & \multicolumn{1}{c|}{0.682} & \multicolumn{1}{c|}{0.786} & \multicolumn{1}{c|}{0.455} & \multicolumn{1}{c|}{0.408} \\ \hline
\multicolumn{1}{|c|}{bjXnet} & \multicolumn{1}{c|}{0.500} & \multicolumn{1}{c|}{0.700} & \multicolumn{1}{c|}{0.800} & \multicolumn{1}{c|}{0.500} & \multicolumn{1}{c|}{0.440} \\ \hline
\multicolumn{6}{c}{\textbf{NDLSW}} \\ \hline
\multicolumn{1}{|c|}{BugLocator} & \multicolumn{1}{c|}{0.419} & \multicolumn{1}{c|}{0.671} & \multicolumn{1}{c|}{0.794} & \multicolumn{1}{c|}{0.526} & \multicolumn{1}{c|}{0.459} \\ \hline
\multicolumn{1}{|c|}{BLUiR} & \multicolumn{1}{c|}{0.311} & \multicolumn{1}{c|}{0.575} & \multicolumn{1}{c|}{0.686} & \multicolumn{1}{c|}{0.453} & \multicolumn{1}{c|}{0.385} \\ \hline
\multicolumn{1}{|c|}{BLIA} & \multicolumn{1}{c|}{0.512} & \multicolumn{1}{c|}{0.716} & \multicolumn{1}{c|}{0.820} & \multicolumn{1}{c|}{0.611} & \multicolumn{1}{c|}{0.539} \\ \hline
\multicolumn{1}{|c|}{DNNLOC} & \multicolumn{1}{c|}{0.617} & \multicolumn{1}{c|}{0.786} & \multicolumn{1}{c|}{0.855} & \multicolumn{1}{c|}{0.617} & \multicolumn{1}{c|}{0.595} \\ \hline
\multicolumn{1}{|c|}{bjXnet} & \multicolumn{1}{c|}{0.650} & \multicolumn{1}{c|}{0.800} & \multicolumn{1}{c|}{0.880} & \multicolumn{1}{c|}{0.660} & \multicolumn{1}{c|}{0.610} \\ \hline
\end{tabular}
\centering
\small
\textbf{DLSW}= Deep Learning Systems, \textbf{NDLSW}=Non-Deep Learning Systems
\end{threeparttable}}
\end{table*}

\begin{table}[t!]
    \centering
    \caption{Statistical tests for the performance gap of existing approaches between deep learning systems (DLSW) and non-deep learning systems (NDLSW)} 
    \label{tab:significance_test_loc}
    \begin{adjustbox}{max width=\linewidth}
        \begin{threeparttable}
            \begin{tabular}{|c|c|c|c|}
                \hline
                \textbf{Method} & \textbf{Metric} & \textbf{Sig. (p-val)} & \textbf{Effect Size} \\ 
                \hline \hline
                \multirow{2}{*}{BugLocator} & RR & 0.00008721** & Medium (0.3834) \\ \cline{2-4} 
                 & AP & 0.00004211** & Medium (0.4023) \\ 
                \hline
                \multirow{2}{*}{BLUiR} & RR & 0.00395* & Medium (0.2709) \\ \cline{2-4} 
                 & AP & 0.00339* & Medium (0.3154) \\ 
                \hline
                \multirow{2}{*}{BLIA} & RR & 0.00000734*** & Large (0.5012) \\ \cline{2-4} 
                 & AP & 0.00000819*** & Large (0.4896) \\ 
                \hline
                \multirow{2}{*}{DNNLOC} & RR & 0.00000345*** & Large (0.5237) \\ \cline{2-4}
                 & AP & 0.00000389*** & Large (0.5110) \\
                \hline
                \multirow{2}{*}{bjXnet} & RR & 0.00000250*** & Large (0.5400) \\ \cline{2-4}
                 & AP & 0.00000275*** & Large (0.5280) \\
                \hline
            \end{tabular}
            \centering
            \small
            \textbf{Sig.}: Significance, \textbf{p-val}: p-value, \textbf{RR}: Reciprocal Rank, \textbf{AP}: Average Precision, \textbf{***}: Large, \textbf{**}: Medium, \textbf{*}: Small
        \end{threeparttable}
    \end{adjustbox}
\end{table}

While the significance of a result indicates how probable it is that it is due to chance, the effect size indicates the extent of the difference \cite{effectsize}. Hence, we performed the Cohen's D effect size test \cite{cohend}, and our analysis found a \textit{medium} to \textit{large} effect size for all cases (Table \ref{tab:significance_test_loc}). Thus, our results from effect size tests reinforce the above finding from significance tests. Even the state-of-the-art technique bjXnet shows the gap is statistically significant with a large effect size. This indicates that the disparity is not only statistically meaningful but also practically significant, underscoring the challenges of localizing bugs in DL systems. 

To better understand the challenges of bug localization in deep learning systems, we performed a deeper analysis and selected bjXnet, a state-of-the-art technique. Although it has outperformed other baselines, its performance remains inadequate for deep learning systems. Our analysis identified several key factors contributing to bjXnet's reduced effectiveness in deep learning systems. By examining a random subset of bugs (100 bugs) from the Denchmark dataset, we found that DL systems heavily rely on high-level abstractions provided by frameworks like TensorFlow and PyTorch. 

\begin{figure}[ht]
\centering
\includegraphics[width=0.9\textwidth]{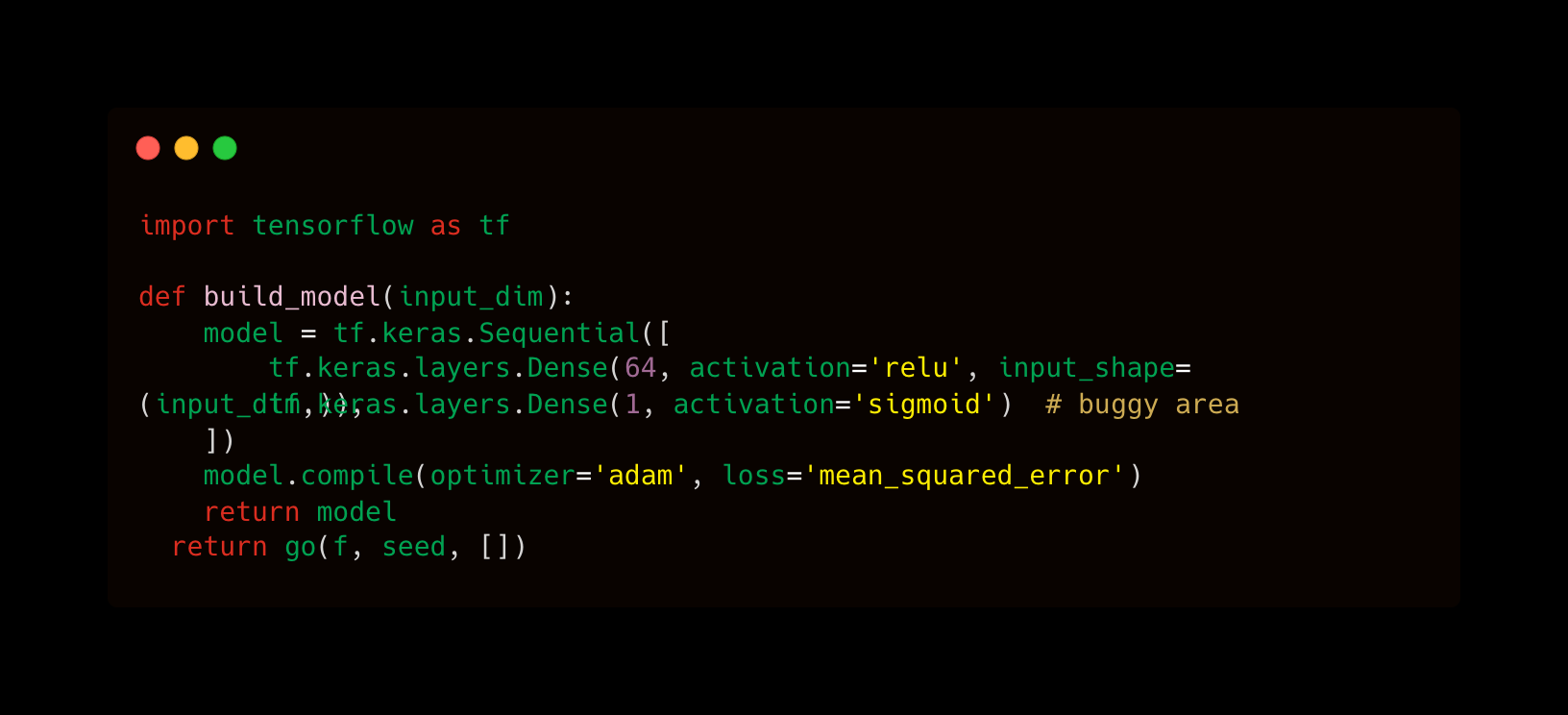}
\caption{A code snippet from our DLSW dataset showing error in activation function}
\label{fig:bjxnet}
\end{figure}

These abstractions include predefined components such as neural network layers (e.g., \texttt{Dense}, \texttt{Conv2D}), activation functions (e.g., \texttt{relu}, \texttt{sigmoid}), and optimizers (e.g., \texttt{Adam}, \texttt{SGD}). They encapsulate complex mathematical operations—like matrix multiplications and gradient computations—into single library calls. This encapsulation means that essential computational logic is hidden within these library functions rather than being explicitly defined in the source code, which might lead to bjXnet's low performance because the technique relies on analyzing source code to trace errors. In 87\% of the cases we examined, core functionalities such as model training, gradient calculations, and loss computations were implemented using these high-level library calls. For example, Fig. \ref{fig:bjxnet} shows a bug that involves the incorrect selection of activation functions in a dense layer—using \texttt{'sigmoid'} in regression tasks.

Moreover, DL systems build computation graphs where each layer and operation represents a node. In frameworks like TensorFlow, the computation graph is typically static, defined before runtime. However, in newer versions (e.g., TensorFlow\footnote{https://www.tensorflow.org/api} 2+ with eager execution) and other frameworks like PyTorch, computation graphs can evolve dynamically. This dynamic behavior contrasts with traditional systems, making it difficult for static analysis tools like bjXnet to trace errors. These tools struggle to handle the dynamic flow of data and operations, especially when key logic is abstracted in high-level library functions. Other IR-based baselines like BLUiR also have this limitation where only static code analysis is inadequate for DL systems. 

In short, even though our findings above match natural intuition, we performed extensive experiments using five different baselines, which resulted in strong empirical evidence. Thus, not only do our findings reinforce the existing understanding and belief about the challenges of the bugs in deep learning systems, but also they substantiate them with solid empirical evidence and demonstrate the performance gap of existing solutions in localizing the two categories of bugs.
\\

\textbf{Potential Confounding Factors:} To ensure fairness in comparison and highlight potential confounding factors, such as code scale, code complexity, and fault representativeness, which could influence the effectiveness of bug localization techniques, we conducted additional analyses as follows. \\
\indent

\textbf{1. Code Scale:} We analyzed the Lines of Code (LOC) from both datasets to investigate whether code scale biases our results. We found that the average LOC per file is higher for NDLSW (325.32) compared to DLSW (280), and the median LOC per file is also higher for NDLSW (244) than for DLSW (210). Such differences in LOC might pose a confounding factor in our experiment. To mitigate this, we repeated a part of our experiment using source documents with similar LOC.

We defined a consistent LOC threshold using the median values from both datasets (NDLSW and DLSW). The median was chosen as a robust central measure to handle the skewed distribution, a standard practice in similar studies \cite{howell2012statistical}. We averaged the medians (244 and 210 LOC, respectively) to get a common median of 227 LOC, then applied a 25\% margin (upper bound: 284 LOC, lower bound: 170 LOC) to define our LOC range. This thresholding method follows established guidelines for managing skewed data distributions \cite{howell2012statistical}. We selected 439 files from NDLSW and 618 files from DLSW within the defined LOC threshold. Then, we randomly sampled 350 files from each dataset (95\% confidence interval, 5\% margin of error) to measure bug localization performance while controlling for code scale.

\begin{table}[t!]
    \centering
    \caption{Performance of bug localization approaches on the subsample of 350 files}
    \label{tab:performance_subset}
    \begin{adjustbox}{max width=\linewidth}
        \begin{threeparttable}
            \begin{tabular}{|l|c|c|c|c|c|}
                \hline
                \textbf{Method} & \textbf{Top@1} & \textbf{Top@5} & \textbf{Top@10} & \textbf{MRR} & \textbf{MAP} \\
                \hline \hline
                \multicolumn{6}{|c|}{\textbf{DLSW Subset}} \\
                BugLocator & 0.325 & 0.523 & 0.600 & 0.355 & 0.300 \\
                BLUiR & 0.180 & 0.450 & 0.550 & 0.290 & 0.240 \\
                BLIA & 0.390 & 0.590 & 0.695 & 0.410 & 0.340 \\
                DNNLOC & 0.450 & 0.670 & 0.770 & 0.440 & 0.390 \\
                bjXnet & 0.470 & 0.690 & 0.780 & 0.460 & 0.410 \\
                \hline
                \multicolumn{6}{|c|}{\textbf{NDLSW Subset}} \\
                BugLocator & 0.400 & 0.650 & 0.770 & 0.500 & 0.450 \\
                BLUiR & 0.285 & 0.550 & 0.670 & 0.420 & 0.350 \\
                BLIA & 0.470 & 0.700 & 0.800 & 0.570 & 0.510 \\
                DNNLOC & 0.580 & 0.750 & 0.830 & 0.580 & 0.560 \\
                bjXnet & 0.600 & 0.770 & 0.850 & 0.600 & 0.580 \\
                \hline
            \end{tabular}
        \end{threeparttable}
    \end{adjustbox}
\end{table}

\begin{table}[t!]
    \centering
    \caption{Statistical tests for the performance gap of existing approaches between DLSW subset and NDLSW subset}
    \label{tab:statistical_tests_scale}
    \begin{adjustbox}{max width=\linewidth}
        \begin{threeparttable}
            \begin{tabular}{|l|c|c|c|}
                \hline
                \textbf{Method} & \textbf{Metric} & \textbf{Sig. (p-val)} & \textbf{Effect Size} \\
                \hline \hline
                BugLocator & RR & 0.00009123** & Medium (0.3652) \\
                BugLocator & AP & 0.00005467** & Medium (0.3824) \\
                \hline
                BLUiR & RR & 0.00412* & Small (0.2551) \\
                BLUiR & AP & 0.00374* & Small (0.2992) \\
                \hline
                BLIA & RR & 0.00000912*** & Large (0.4850) \\
                BLIA & AP & 0.00000899*** & Large (0.4723) \\
                \hline
                DNNLOC & RR & 0.00000381*** & Large (0.5101) \\
                DNNLOC & AP & 0.00000419*** & Large (0.4956) \\
                \hline
                bjXnet & RR & 0.00000280*** & Large (0.5250) \\
                bjXnet & AP & 0.00000290*** & Large (0.5120) \\
                \hline
            \end{tabular}
            \centering
            \small
            \textbf{Sig.}: Significance, \textbf{p-val}: p-value, \textbf{RR}: Reciprocal Rank, \textbf{AP}: Average Precision, \textbf{***}: Large, \textbf{**}: Medium, \textbf{*}: Small
        \end{threeparttable}
    \end{adjustbox}
\end{table}

We evaluated the performance of our bug localization techniques on these selected subsets. The results are presented in Table \ref{tab:performance_subset}. The performance trends in the deep learning subset are similar to those observed in the full deep learning dataset. DNNLOC and bjXnet perform best across all metrics, followed by BLIA, BugLocator, and BLUiR. Although the overall performance is slightly lower due to the smaller sample size, our overall findings remain consistent. In the non-deep learning subset, all methods achieve higher scores compared to the deep learning subset, with DNNLOC and bjXnet performing significantly better. We also conducted statistical tests (refer to Table \ref{tab:statistical_tests_scale}) to determine if the performance differences between DLSW and NDLSW are significant when the code scale confounding factor is mitigated. The p-values indicate that all methods' performance differences between DLSW and NDLSW are statistically significant, with medium to large effect sizes.

To summarize, we controlled for differences in LOC between the two datasets. After re-running our experiments using comparable file sizes, the performance trends remained consistent, and the results showed that the differences between DLSW and NDLSW are still statistically significant. This indicates that our findings are not influenced by code scale variations. \\
\indent

\begin{table}[t!]
    \centering
    \caption{Code Complexity Metrics for deep learning systems (DLSW) and non-deep learning systems (NDLSW)}
    \label{tab:code_complexity_metrics}
    \begin{adjustbox}{max width=\linewidth}
        \begin{threeparttable}
            \begin{tabular}{|l|c|c|c|c|c|}
                \hline
                \textbf{Dataset} & \textbf{Median LOC} & \textbf{AVG LOC} & \textbf{AVG CC} & \textbf{AVG Halstead} & \textbf{AVG MI} \\
                \hline \hline
                BuGL (NDLSW) & 244 & 325.32 & 8.72 & 1456.23 & 75.30 \\
                Denchmark (DLSW) & 210 & 280 & 9.5 & 1482.30 & 68.44 \\
                \hline
            \end{tabular}
            \centering
            \small
            \textbf{LOC}: Lines of Code, \textbf{CC}: Cyclomatic Complexity, \textbf{MI}: Maintainability Index
        \end{threeparttable}
    \end{adjustbox}
\end{table}

\textbf{2. Code Complexity:} We also considered code complexity as a potential confounding factor. To compare the code complexity between DLSW (Denchmark) and NDLSW (BuGL), we used established software metrics: Cyclomatic Complexity (CC) \cite{ebert2016cyclomatic}, Halstead Complexity \cite{hariprasad2017halstead}, and the Maintainability Index (MI) \cite{najm2014measuring}. 

Cyclomatic Complexity measures the number of linearly independent paths through a program's source code, providing insight into control flow complexity. Halstead Complexity assesses cognitive complexity based on the number of operators and operands, reflecting the effort required to understand the code. The Maintainability Index evaluates code maintainability by considering factors like LOC, CC, and Halstead metrics. 

Our findings, summarized in Table \ref{tab:code_complexity_metrics}, reveal that the average Cyclomatic Complexity per file is slightly higher for DLSW (9.5) compared to NDLSW (8.72). Similarly, the average Halstead Complexity is comparable between the two datasets (DLSW: 1482.30, NDLSW: 1456.23), and the Maintainability Index is slightly higher for NDLSW (75.30) than DLSW (68.44). We conducted statistical tests (e.g., independent two-sample t-tests) based on the sample size and the distribution of the data for each of the complexity metrics. 

The t-test results are as follows: Cyclomatic Complexity: t(698) = 1.45, p = 0.15; Halstead Complexity: t(698) = 0.78, p = 0.44; Maintainability Index: t(698) = -1.32, p = 0.19. These p-values are all above 0.05, indicating that the differences in code complexity metrics between DLSW and NDLSW are not statistically significant. 

Our analysis shows that both BuGL and Denchmark exhibit similar levels of code complexity. While their structures differ—BuGL being more modular and Denchmark relying on computational tasks—the statistical analysis of Cyclomatic Complexity, Halstead Complexity, and Maintainability Index indicates that both datasets are comparable in terms of code complexity.\\
\indent

\begin{table}[t!]
    \centering
    \caption{API Bug Localization Performance (Subset of 100 Bugs) for deep learning systems (DLSW) and non-deep learning systems (NDLSW)}
    \label{tab:api_bug_localization}
    \begin{adjustbox}{max width=\linewidth}
        \begin{threeparttable}
            \begin{tabular}{|l|c|c|}
                \hline
                \textbf{Method} & \textbf{MRR} & \textbf{MAP} \\
                \hline \hline
                \multicolumn{3}{|c|}{\textbf{DLSW Subset}} \\
                \hline
                BugLocator & 0.365 & 0.430 \\
                BLUiR & 0.410 & 0.270 \\
                BLIA & 0.430 & 0.400 \\
                DNNLOC & 0.510 & 0.480 \\
                bjXnet & 0.530 & 0.500 \\
                \hline
                \multicolumn{3}{|c|}{\textbf{NDLSW Subset}} \\
                \hline
                BugLocator & 0.470 & 0.520 \\
                BLUiR & 0.460 & 0.350 \\
                BLIA & 0.520 & 0.490 \\
                DNNLOC & 0.590 & 0.580 \\
                bjXnet & 0.610 & 0.600 \\
                \hline
            \end{tabular}
            \centering
            \small
            \textbf{MRR}: Mean Reciprocal Rank, \textbf{MAP}: Mean Average Precision
        \end{threeparttable}
    \end{adjustbox}
\end{table}

\begin{figure}[ht]
\centering
\includegraphics[width=0.9\textwidth]{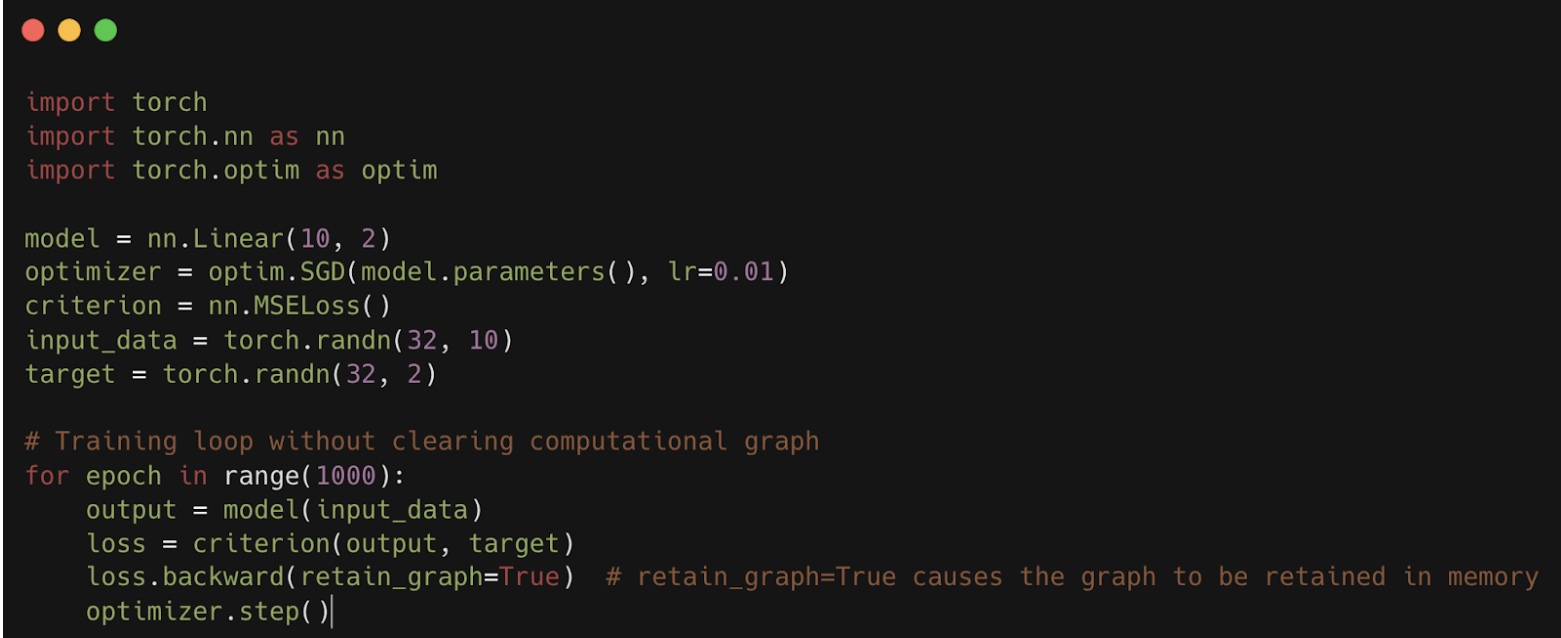}
\caption{A code snippet from the DLSW dataset showing a PyTorch API misuse leading to a memory leak}
\label{fig:api_rq1}
\end{figure}

\textbf{3. Fault Representation:} We also considered fault representativeness in our extended analysis. To address this concern, we focused on API bugs, which are common in both DLSW and NDLSW. API bugs, such as incorrect interactions with external libraries or frameworks, are prevalent in both deep learning systems (e.g., TensorFlow, PyTorch) and traditional software systems (e.g., Matplotlib, Requests). We identified 195 API bugs in DLSW and 102 in NDLSW through keyword searches and manual analysis by two authors. We selected 100 API bugs from each dataset and evaluated the bug localization performance of our techniques. Despite mitigating the confounding factor by focusing on API bugs, bug localization remains more challenging in DL systems. From Fig. \ref{fig:api_rq1}, in PyTorch, missing `optimizer.zero\_grad()` causes a memory leak, leading to out-of-memory errors after prolonged training, making the bug harder to locate using the baselines.

Our comprehensive analysis addresses potential confounding factors such as code scale, code complexity, and fault representation. Even after controlling for these factors, our findings show that bug localization remains significantly more challenging in deep learning systems compared to traditional software systems. This suggests that, despite the presence of similar bug types (e.g., API misuse), bugs in deep learning systems often manifest in more complex and indirect ways. As a result, traditional methods may be insufficient for effectively localizing bugs in deep learning systems, highlighting the need for new approaches tailored to their unique characteristics. \\

\subsubsection{\textbf{Comparison among the categories of deep learning systems}} 

The original Denchmark dataset comprises 136 projects and 2,365 bug reports, categorized into Frameworks, Tools, Libraries, Applications, Engines, Platforms, and Compilers \cite{denchmark}. Our focus centered on the core components of deep learning systems—Frameworks, Tools, and Libraries—as these are essential and represent the majority of relevant bug reports. We aimed to examine whether there are differences in bug localization performance across these system categories, and we classified the dataset categories as follows.\\
\indent
\textit{Filter 1:} From the original dataset, we selected 25 Frameworks, 31 Tools, and 44 Libraries, totaling 100 projects. To ensure each project provided sufficient data for robust analysis, we excluded projects with fewer than ten reported bugs. This criterion was crucial for enabling meaningful comparisons of bug localization techniques across projects with substantial debugging histories. After applying this filter, we retained 18 Frameworks, 29 Tools, and 42 Libraries, resulting in 89 projects for further analysis. \\
\indent 
\textit{Filter 2:} We identified overlaps where projects belonged to multiple categories (e.g., a tool functioning as a library). To resolve this, we meticulously reviewed official documentation and the functional roles of these projects, assigning each to a unique category. This refinement led to a final dataset of 86 projects and 1,795 bug reports, distributed as follows: 17 Frameworks (746 bugs), 28 Tools (455 bugs), and 41 Libraries (594 bugs).

To better understand the challenges in localizing deep learning bugs, we analyzed the performance of our baseline techniques across different categories of bugs in DL systems. We employed stratified random sampling to select a balanced number of bug reports from each category for performance comparison, ensuring fair representation and reducing selection bias. We repeated the process three times with different samples to ensure robustness. The results were averaged across these evaluations, providing reliable insights into the efficacy of the bug localization methods.

\begin{table}[h]
\centering
\caption{Comparative analysis of bug localization techniques across various deep learning project classes}
\label{tab:comparison}
\begin{threeparttable}
\begin{tabular}{|c|c|c|c|}
\hline
\textbf{Project Class} & \textbf{Method} & \textbf{MRR} & \textbf{MAP} \\ \hline
\multirow{5}{*}{Framework} & BugLocator & 0.404 & 0.334 \\ \cline{2-4}
                           & BLUIR      & 0.205 & 0.155 \\ \cline{2-4}
                           & BLIA       & 0.466 & 0.392 \\ \cline{2-4}
                           & DNNLOC     & 0.493 & 0.412 \\ \cline{2-4}
                           & bjXnet     & 0.560 & 0.510 \\ \hline
\multirow{5}{*}{Library}   & BugLocator & 0.559 & 0.453 \\ \cline{2-4}
                           & BLUIR      & 0.632 & 0.539 \\ \cline{2-4}
                           & BLIA       & 0.583 & 0.486 \\ \cline{2-4}
                           & DNNLOC     & 0.656 & 0.597 \\ \cline{2-4}
                           & bjXnet     & 0.685 & 0.640 \\ \hline
\multirow{5}{*}{Tool}      & BugLocator & 0.499 & 0.407 \\ \cline{2-4}
                           & BLUIR      & 0.469 & 0.387 \\ \cline{2-4}
                           & BLIA       & 0.546 & 0.449 \\ \cline{2-4}
                           & DNNLOC     & 0.592 & 0.514 \\ \cline{2-4}
                           & bjXnet     & 0.605 & 0.550 \\ \hline
\end{tabular}
\end{threeparttable}
\end{table}

\begin{figure}[t!]
 \centering
 \includegraphics[width= 3.5in]{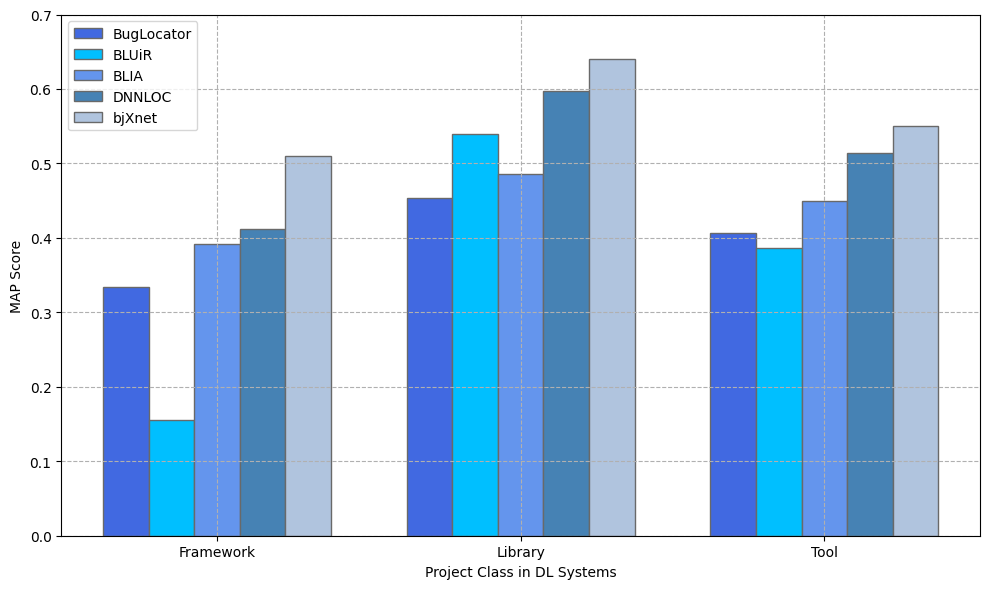}
    \caption{Performance comparison of existing approaches (BugLocator, BLUiR, BLIA, DNNLOC, bjXnet) for various deep learning project classes} 
 \label{fig:class}
\end{figure}

According to our experimental results in Table \ref{tab:comparison}, existing techniques perform higher in localizing bugs from deep learning libraries than that of frameworks and tools. This could be attributed to the modular design of libraries, aimed at handling specific tasks independently \cite{wang2019various}. In the example bug report (Table \ref{tab:library_BL_example} \cite{pytorch_issue_87085}), the autograd module of the Torch.Library\footnote{\url{https://pytorch.org/docs/stable/library.html}} plays a crucial role in calculating gradient across all tensor operations, whereas autograd.gradcheck\footnote{\url{https://pytorch.org/docs/stable/generated/torch.autograd.gradcheck.html}} is a utility function for verifying the accuracy of these computed gradients. Here, a specific error concerning gradient computation in gradcheck was reported during sparse matrix multiplication. Such problems can be localized more efficiently when occurring within a clearly defined module like autograd.gradcheck. 

Both DNNLOC and bjXnet retrieved the buggy file at Top@1 by leveraging this modularity. 
DNNLOC's success in accurately locating bugs can be linked to its ability to capture complex patterns using deep learning, which works effectively with the modular design of library bugs. On the other hand, bjXnet uses attention mechanisms to focus on critical functions or variables, like autograd.gradcheck, which might lead to more precise localization and demonstrate the strength of attention-based techniques in isolating and targeting key components in modular architectures.

\setlength{\arrayrulewidth}{0.1mm}
\setlength{\tabcolsep}{11pt}
\renewcommand{\arraystretch}{1.0}
\begin{table}[h]
\centering
\caption{Example of a bug from deep learning library for bug localization \cite{pytorch_issue_87085}}
\label{tab:library_BL_example}
\begin{tabular}{c}
\hline
\multicolumn{1}{|c|}{\textbf{Library Bug (Bug ID: 87085)}} \\ \hline
\textbf{Title} \\ \hline
\multicolumn{1}{|l|}{gradcheck failure with sparse matrix multiplication} \\ \hline
\textbf{Description} \\ \hline
\multicolumn{1}{|l|}{\begin{tabular}[c]{@{}l@{}}Sparse @ dense matrix multiplication fails to pass gradcheck.\\ A manual inspection of the gradient seems to indicate that \\ this is a bug with gradcheck rather than the matrix multiplication itself. \\ Steps to reproduce: ... \\
Detailed BR: https://github.com/pytorch/pytorch/issues/87085\\ 
\end{tabular}} \\ \hline
\end{tabular}
\end{table}

In contrast, we notice from Fig.\ref{fig:class} that the performance of bug localization techniques dropped for framework bugs, especially when using the BLUiR method. Frameworks consist of a broad architecture that guides the flow of control, involving multiple layers and components that interact in complex ways \cite{nguyen2019machine, foutse}. The example bug in Table \ref{tab:library_BL_example} \cite{tensorflow_issue_61297} deals with CTC Loss in a Keras model with LSTM on a TensorFlow Processing Unit (TPU), highlighting a framework-level issue. For this example bug, DNNLOC retrieved the correct buggy file at the top 38 position, while BugLocator retrieved it at the top 65 position, indicating limitations in relying only on textual similarities between framework code and bug reports. BLUiR, emerging as the least effective, ranked the correct file at the top 93 positions, highlighting its insufficiency in localizing framework-level bugs with only structural code analysis. bjXnet retrieved the correct buggy file at the top 28 position. All baseline techniques struggle with framework bugs due to the complex, multi-layered architecture of frameworks, which lacks the clear modular boundaries found in libraries. Even DNNLOC and bjXnet face significant challenges because the complex dependencies and interactions across multiple components in frameworks make it harder to trace and isolate the exact location of faults.

\setlength{\arrayrulewidth}{0.1mm}
\setlength{\tabcolsep}{11pt}
\renewcommand{\arraystretch}{1.0}
\begin{table}[h]
\centering
\caption{Example of a bug from deep learning framework for bug localization \cite{tensorflow_issue_61297}}
\label{tab:framework_BL_example}
\begin{tabular}{c}
\hline
\multicolumn{1}{|c|}{\textbf{Framework Bug (Bug ID: 61297)}} \\ \hline
\textbf{Title} \\ \hline
\multicolumn{1}{|l|}{CTC Loss errors on TPU} \\ \hline
\textbf{Description} \\ \hline
\multicolumn{1}{|l|}{\begin{tabular}[c]{@{}l@{}}The Keras Model with LSTM+ CTC loss runs normally on GPU,\\
but VM TPU prints grappler errors. The errors \\ usually don't stop the execution, but the loss is nan. \\ It sometimes also crashes with core dump, but it's not consistent. \\ I created a public Kaggle notebook with the code producing the issue: \\ https://www.kaggle.com/code/shaironen/ctc-example/notebook\\
The grappler errors are:
\\....\\
In addition, I read it's recommended to use \\model.compile(jit-compile=True) while on GPU to \\pre-diagnose TPU issues. It gives similar errors and terminates.\\
(with jit-compile=False, it runs normally on GPU only).\\
According to tf.nn.ctc-loss documentation, it should work on tpu.
\\ 
Detailed BR: https://github.com/tensorflow/tensorflow/issues/61297\\
\end{tabular}} \\ \hline
\end{tabular}
\end{table}

Lastly, the performance in bug localization slightly improved for tools, but it was not as good as with libraries. This improvement could be attributed to the fact that tools (e.g., TensorBoard) are more independent and have fewer dependencies than the entire framework (e.g., TensorFlow). While tools like TensorBoard focus on specialized tasks such as visualization, they often integrate with complex frameworks like TensorFlow. The example bug in Table \ref{tab:tool_BL_example} refers to an incorrect rendering of floating point metrics in the TensorBoard HParams dashboard. It can be classified as a tool bug, as TensorBoard is a visualization utility within the TensorFlow ecosystem. In this instance, bjXnet retrieves the correct buggy file at the top 25 position. DNNLOC identified the correct buggy file at the top 27 position, BugLocator located the correct file at the 40th position, and BLUiR identified the correct file at the top 52 position. These results suggest that while tools are more independent, their connection to complex frameworks still creates challenges. bjXnet and DNNLOC perform slightly better by understanding the structure more deeply, but all techniques find it difficult to handle the interaction between tools and frameworks.

\setlength{\arrayrulewidth}{0.1mm}
\setlength{\tabcolsep}{11pt}
\renewcommand{\arraystretch}{1.0}
\begin{table}[h]
\centering
\caption{Example of a bug from deep learning tool for bug localization \cite{tensorboard_issue_5948}}
\label{tab:tool_BL_example}
\begin{tabular}{c}
\hline
\multicolumn{1}{|c|}{\textbf{Tool Bug (Bug ID: 5948)}} \\ \hline
\textbf{Title} \\ \hline
\multicolumn{1}{|l|}{TB.dev HParams dashboard shows floating point metrics incorrectly} \\ \hline
\textbf{Description} \\ \hline
\multicolumn{1}{|l|}{\begin{tabular}[c]{@{}l@{}}TensorBoard renders incorrectly locally when uploading \\ an experiment to tensorboard.dev some of the floating point values \\ are mangled, and zero values are shown as +/-Infinity
\\....\\ 
Detailed BR: https://github.com/tensorflow/tensorboard/issues/5948\\
\end{tabular}} \\ \hline
\end{tabular}
\end{table}

In short, our findings indicate that the modular design of libraries \cite{erickson2017toolkits} facilitates better bug localization, whereas the complexities inherent in frameworks and tools hurt the bug localization performance of the existing techniques.

\subsubsection{\textbf{Preliminary study on spectrum-based bug localization}}
\label{sec:sbfl-study}

Specturm-based bug localization techniques have been widely used for decades in traditional software systems \cite{jones2005empirical}. However, they might face significant challenges when applied to deep learning systems. Spectrum-based bug localization techniques analyze the execution traces of passing and failing tests to measure the frequency with which each program element is executed in both cases \cite{jones2005empirical, akbar2020large}. They rank elements by suspiciousness to prioritize fault-prone locations. For example, Tarantula \cite{jones2005empirical} calculates the suspiciousness score based on the ratio of failed to total executions of an element, highlighting those more frequently associated with failures. However, capturing execution traces against passing and failed tests from deep learning systems could be highly challenging. Besides, not all bugs in deep learning systems cause crashes or generate a stack trace \cite{islam2019comprehensive}. Instead, some lead to incorrect behavior, known as silent bugs \cite{tambon2024silent}. To better understand the challenges of spectrum-based bug localization techniques, we conducted a case study to assess their applicability in deep learning systems. \\

\begin{figure}[htbp]
    \centering
    \includegraphics[width=0.60\textwidth]{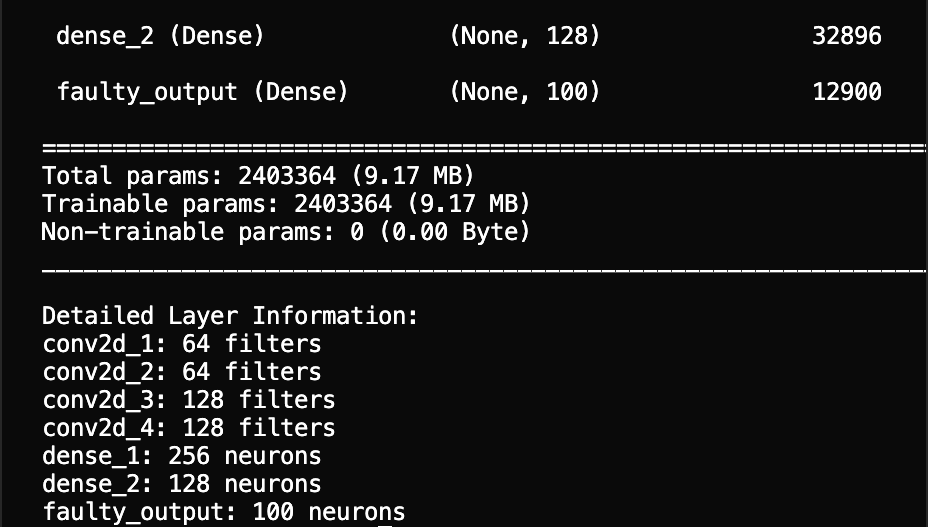}
    \caption{Model architecture of the selected deep learning model}
    \label{fig:sbfl-layer}
    \vspace{-0.5em}
\end{figure}

\textit{Bug selection:} We randomly selected a training bug from a classification model for our case study (see Fig.\ref{fig:sbfl-layer}). Here, the final layer of the model is defined as \texttt{activation=None} (buggy) instead of \texttt{activation='softmax'} (correct). The bug causes the final layer to output unnormalized logits. Although the cross-entropy loss is still computed, these raw outputs hinder the expected decrease in loss over multiple epochs, thereby hampering model training.\\
\indent
\textit{Ground truth for bug localization:} The final layer’s activation function is incorrect, which makes the layer buggy (a.k.a., ground truth). We aim to determine whether spectrum-based localization techniques can correctly locate this buggy layer from the model. \\
\indent
\textit{Choice of spectrum-based technique:} To perform the preliminary study, we chose a seminal work on spectrum-based bug localization -- Tarantula \cite{jones2005empirical}. Tarantula uses the execution profiles of test cases to compute suspiciousness scores for program components (e.g., statements). Several contemporary studies \cite{Eniser2019} have extended Tarantula’s spectra-based approach by incorporating machine learning and deep learning techniques. However, we wanted to directly assess Tarantula’s effectiveness and limitations in locating faults in deep learning systems without introducing any confounding factors from hybrid methods. Hence, we selected Tarantula for our case study.\\
\indent
\begin{figure}[htbp]
    \centering
    \includegraphics[width=0.99\textwidth]{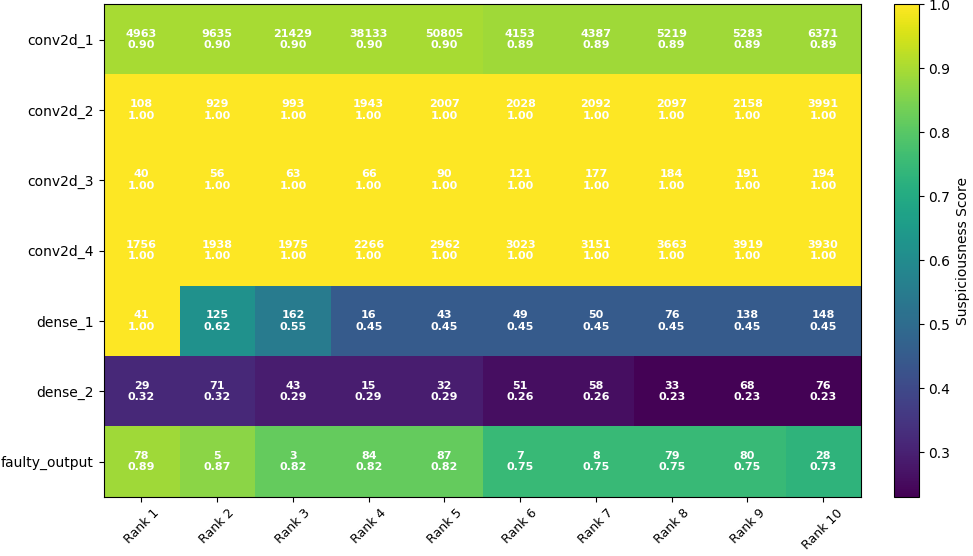}
    \vspace{-0.7em}
    \caption{Top 10 Neuron-Level Suspiousness Scores by Layers}
    \label{fig:sbfl-neuron-scores}
\end{figure}
\begin{table}[htbp]
    \centering
    \caption{Pass/Fail Results from 20 Training Runs}
        \vspace{-0.7em}
    \label{tab:passfail-results}
    \begin{tabular}{c c | c c | c c}
        \hline
        \textbf{Run} & \textbf{Result} & \textbf{Run} & \textbf{Result} & \textbf{Run} & \textbf{Result} \\
        \hline
        1  & Failed  & 8  & Passed  & 15  & Passed  \\
        2  & Failed  & 9  & Passed  & 16  & Failed  \\
        3  & Passed  & 10 & Passed  & 17  & Failed  \\
        4  & Passed  & 11 & Passed  & 18  & Passed  \\
        5  & Failed  & 12 & Failed  & 19  & Failed  \\
        6  & Failed  & 13 & Passed  & 20  & Failed  \\
        7  & Failed  & 14 & Failed  &      &        \\
        \hline
    \end{tabular}
                \vspace{-2em}
\end{table}
\textit{Test cases:} The training of deep learning models are inherently stochastic due to factors such as random weight initialization and data shuffling \cite{Zhuang2022}. Hence, designing test cases for deep learning models is challenging \cite{Ma2018}. However, to simulate the traditional approach for spectrum-based bug localization, we needed to create both failing and passing test cases. Since the final layer of our target model is defined without a softmax activation, its output remains unnormalized. This unnormalized output is processed by the model’s loss function (e.g., categorical cross-entropy), but the loss is calculated using raw logits, which is undesirable. However, in some training runs, favorable random initialization and data ordering may cause the loss to decrease, even when the bug (i.e., incorrect activation function) is present. In other cases, the loss might only decrease. Building on prior research \cite{Ma2018, Tian2017, Zhuang2022}, we conducted 20 separate training runs, each with 10 epochs and different random seeds, to capture this variability. We define our passing and failing test cases as follows:\\
\noindent
\textbf{- Failing test cases}: Training runs where the loss stagnates across \(N\) epochs (here, \(N = 3\), with a total of 10 epochs). \\
\textbf{- Passing test cases}: Training runs where the loss decreases despite the model containing the fault. \\
\indent
\textit{Coverage Metric:} Our study captures test coverage at two levels: \emph{layer level} and \emph{neuron level}. First, we record whether each layer in the model is executed during each training run. This binary coverage metric aligns with traditional spectrum-based techniques, where suspiciousness scores are calculated based on how components are covered by the passing and failing test cases. Similarly, we calculate the suspiciousness score for a given layer \( l \) as follows:\\
\indent
\textbf{- Failed(\( l \))}: Number of failing runs that execute layer \( l \). \\
\indent
\textbf{- Passed(\( l \))}: Number of passing runs that execute layer \( l \).

Tarantula integrates these statistics into its standard formula to compute a suspiciousness score, ideally flagging layers with a higher ratio of failing-to-passing executions as more suspicious. 
Second, we record the activation of individual neurons within each layer to assess the contribution of neurons to the overall model behavior. Here, a neuron is considered covered if it is activated (i.e., its output is greater than zero) during a training run. We calculate a Tarantula score for each neuron based on its activation frequency in failing versus passing test cases. Our approach is conceptually similar to the granular coverage criteria introduced by \citet{Ma2018}. \\

\begin{table}[ht]
    \setlength{\abovecaptionskip}{0em} 
    \setlength{\belowcaptionskip}{0em} 
    \setlength{\extrarowheight}{0pt} 
    \centering
                \vspace{-0.6cm}
    \caption{Suspiciousness Scores Per Layer}
    \label{tab:suspicious-scores}
    \begin{tabular}{l c}
        \hline
        \textbf{Layer Name} & \textbf{Suspiciousness Score} \\
        \hline
        \texttt{conv2d}           & 0.5000 \\
        \texttt{conv2d\_1}        & 0.5000 \\
        \texttt{max\_pooling2d}   & 0.5000 \\
        \texttt{conv2d\_2}        & 0.5000 \\
        \texttt{conv2d\_3}        & 0.5000 \\
        \texttt{max\_pooling2d\_1}& 0.5000 \\
        \texttt{flatten}          & 0.5000 \\
        \texttt{dense}            & 0.5000 \\
        \texttt{dense\_1}         & 0.5000 \\
        \texttt{dense\_2 (faulty)}& 0.5000 \\
        \hline
    \end{tabular}
\end{table}

\textit{Results and Discussion:} Table \ref{tab:passfail-results} shows that out of 20 training runs, 9 runs failed (i.e., loss did not decrease), while 11 runs passed (i.e., loss decreased). Our findings (see Table \ref{tab:suspicious-scores}) show that Tarantula assigns the same suspiciousness score to all layers in the model. Since every layer is activated during the forward and backward passes, there is no variability in their passing/failing test coverage. We also incorporated neuron-level coverage for more granular analysis and visualized the results in the color-coded heat map (see Fig.\ref{fig:sbfl-neuron-scores}). In this plot, each row represents a layer, and each column shows the top ten neurons ranked by descending suspiciousness (from 0.0 to 1.0).  We chose the top ten neurons for a concise visualization because our conv2d layers contain 64–128 filters, while the dense layers have up to 256 neurons each (see Fig.\ref{fig:sbfl-layer}), making it impractical to display every neuron in a single figure. 

From Fig.\ref{fig:sbfl-neuron-scores}, we see that specific layers, such as \textit{conv2d\_2}, \textit{conv2d\_3}, and \textit{conv2d\_4}, have suspicious neurons, suggesting their strong association with failing runs. In contrast, \textit{conv2d\_1} and \textit{faulty\_output }(i.e., the last layer where the bug exists\textit{)} exhibit high but varying levels of suspiciousness. It should be noted that the final\_output layer contains the bug. Although neuron-level coverage reveals these differences, suggesting that not all neurons in a layer behave the same, the overall patterns indicate uniformly high or near-universal activations. Since traditional spectrum-based bug localization techniques rely on such scores to pinpoint the buggy location, the reported trends make bug localization difficult.

Our preliminary study thus confirms the limitation of spectrum-based bug localization techniques, such as Tarantula, in the context of deep learning systems. The results emphasize that while neuron-level coverage offers more detailed insights than layer-level coverage alone, both coverage metrics struggle to isolate the bug in the model.

\subsubsection{\textbf{Preliminary study on dynamic techniques for bug localization}}

Deep learning systems differ from traditional software systems due to their reliance on high-dimensional data, hardware accelerators, and complex model architectures \cite{islam2019comprehensive}. Bugs in deep learning systems can arise from diverse sources, including model configurations, training procedures, and runtime dependencies, making their localization particularly challenging. While static analysis techniques have been widely applied to software bug localization, their effectiveness in deep learning systems is often limited, as many issues only manifest during execution \cite{deeplocalize}. To address this gap, dynamic fault localization techniques leverage runtime analysis to detect faults by monitoring execution traces and identifying abnormal behaviors \cite{deeplocalize}. Thus, in this study, we systematically evaluate three state-of-the-art dynamic techniques -- DeepLocalize \cite{deeplocalize}, DeepDiagnosis \cite{wardat2022deepdiagnosis}, and DeepFD \cite{cao2022deepfd} -- to assess their effectiveness in localizing bugs in deep learning systems. While these techniques leverage runtime analysis to detect and locate bugs, their practical applicability is constrained by several factors, as follows. 

\textit{Computational and Practical Challenges:} Dynamic-based approaches like DeepLocalize \cite{deeplocalize}, DeepDiagnosis \cite{wardat2022deepdiagnosis}, and DeepFD \cite{cao2022deepfd} rely on runtime data collection, which is computationally expensive and often impractical for large-scale deep learning programs. Moreover, many bug reports in the Denchmark \cite{denchmark} dataset are linked to large, industry-scale deep learning programs that often miss critical information (e.g., dataset details required to run the model) or include numerous external dependencies \cite{extrinsic&intrinsic}. These factors make dynamic analysis both time-consuming and, in many cases, impractical.

\textit{Nature of the Techniques:} Dynamic approaches differ fundamentally from the static, query-based techniques that we employ. These tools (e.g., DeepLocalize \cite{deeplocalize}, DeepDiagnosis \cite{wardat2022deepdiagnosis}, DeepFD \cite{cao2022deepfd}) first detect and diagnose faults during program execution, identifying issues at a fine-grained level (such as bugs in specific layers or components like loss function). In contrast, traditional static methods use textual and structural information in bug reports to locate files, methods, or lines where bugs exist \cite{han2023bjxnet, DNNLOC, buglocator, bluir, qi2021dreamloc}. Our study focuses on file-level bug localization, which aligns more with developers' practical steps when addressing a bug, first locating the relevant file in large codebases and then examining its details.

\textit{Scope of Bugs Addressed:} Another important consideration is that DeepLocalize \cite{deeplocalize}, DeepDiagnosis \cite{wardat2022deepdiagnosis}, and DeepFD \cite{cao2022deepfd} are specifically designed to address training faults. For example, DeepLocalize \cite{deeplocalize} targets training-phase numerical faults by analyzing dynamic traces. DeepFD \cite{cao2022deepfd}  uses machine learning classifiers (e.g., KNN, Decision Tree, and Random Forest) to predict and localize specific training faults (e.g., issues with loss functions, epochs, optimizers, activation functions, or learning rates). DeepDiagnosis \cite{wardat2022deepdiagnosis} focuses on CNN architectures by mapping runtime error symptoms to root causes. Since our dataset, which reflects real-world scenarios, contains a diverse mix of deep learning bugs beyond just training faults, applying these techniques across the entire dataset is neither practical nor appropriate.

\textit{Framework and Architecture Limitation:} All these techniques are specifically designed to detect and address faults within the TensorFlow/Keras framework only \cite{deeplocalize, wardat2022deepdiagnosis, cao2022deepfd}. Moreover, neither DeepLocalize \cite{deeplocalize} nor DeepDiagnosis \cite{wardat2022deepdiagnosis} supports RNN architectures. \\

\textbf{Evaluation of Dynamic Approaches:}
To ensure a fair and comprehensive evaluation, we followed a structured approach:
\begin{enumerate}
    \item First, we selected 350 bug reports from our dataset (i.e., Denchmark) using stratified sampling, ensuring that they exclusively use TensorFlow/Keras framework. This subset was chosen to achieve a 95\% confidence level with a 5\% margin of error, ensuring it is statistically representative of our deep learning system (DLSW) dataset. 
    \item Second, to better understand the performance of dynamic approaches on the training faults they are designed to handle, we manually identified training bugs within our stratified sample. Among the 350 bug reports, we identified 123 unique DL models. This is because multiple bug reports often correspond to the same DL model in the Denchmark dataset for DLSW. We also excluded the RNN-based DL models since DeepDiagnosis and DeepLocalize do not support them, which left us with 101 DL models.
    \item Finally, two independent annotators (the authors) conducted a two-phase labeling process. In the first phase, each DL buggy model was categorized as either a training bug or another type of bug. In the second phase, the training bugs were further classified by their root causes, following the definitions provided by dynamic-based approaches \cite{humbatova2020taxonomy}. Before starting the complete annotation, the annotators jointly labeled a sample of 10 buggy DL models to ensure a shared understanding of the classification criteria. The inter-annotator reliability, measured by Cohen’s kappa coefficient, was 0.88, indicating a strong level of agreement. Any disagreements were resolved through a systematic consensus-building process that involved re-examining the cases using clear taxonomy definitions and additional resources such as literature. 
\end{enumerate}
This process resulted in 21 unique DL models containing training bugs with well-defined root causes. We then evaluated \cite{deeplocalize}, DeepDiagnosis \cite{wardat2022deepdiagnosis}, and DeepFD \cite{cao2022deepfd} on these models.\\

\textbf{Results and Discussion:}
\begin{table}[h]
    \centering
    \caption{Performance of Dynamic Approaches for Locating Training Bugs from DLSW}
    \begin{tabular}{lcccc}
        \hline
        \textbf{Technique} & \textbf{Accuracy} & \textbf{Precision} & \textbf{Recall} & \textbf{F1-Score} \\
        \hline
        DeepLocalize & 0.23 & 0.25 & 0.20 & 0.22 \\
        DeepFD & 0.28 & 0.30 & 0.25 & 0.27 \\
        DeepDiagnosis & 0.33 & 0.35 & 0.30 & 0.32 \\
        \hline
    \end{tabular}
    \label{tab:dynamic_result}
\end{table}

The performance of dynamic approaches across the 21 models reflects the challenges posed by the heterogeneous nature of DL bugs. While these dynamic approaches demonstrate strengths in localizing specific training bugs (e.g., numerical anomalies, learning rate issues), their limitations underscore the multifaceted nature of DL bugs and the inadequacy of existing state-of-the-art techniques for bug localization. Table \ref{tab:dynamic_result} shows the performance of these techniques against each of the 21 bugs. Based on our evaluation, we found several limitations of these techniques, as follows.

\textit{1. Limited Scope:} All three dynamic techniques demonstrate significant limitations, even when applied to a small subset of training bugs, due to their narrow scope (see Table \ref{tab:dynamic}). DeepLocalize, for instance, focuses on identifying numerical anomalies like NaN or Inf values during bug tracing. While it successfully located a faulty layer caused by an incorrect activation function by detecting NaN values in the loss (Bug 17), it failed to detect bugs that did not involve numerical abnormalities (e.g., Bug 1, 2, 3). Similarly, DeepDiagnosis relies on predefined heuristics to map symptoms to root causes, which restricts its ability to identify issues like an incorrect dropout rate (Bug 21). DeepFD, which uses classifiers trained on runtime features, also struggled with bugs outside its predefined fault categories, such as regularization issues (Bug 10). Our analysis above reveals a troubling trend: even for the specific category of bugs for which they were targeted  (e.g., training bugs), these state-of-the-art techniques often fail to detect issues. This suggests that they may not be sufficient for effective bug localization in real-world deep-learning projects, where the bugs tend to be more complex and varied, even within the same category.

\textit{2. Inability to Handle Extrinsic Bugs:}  A key finding from RQ3 is that 40\% of DL bugs are extrinsic, meaning they arise from external dependencies such as GPU issues, third-party libraries, or training data. We found that these dynamic bug localization techniques, which rely on analyzing runtime information, are inherently unable to detect such bugs. For example, in our evaluation (see Table \ref{tab:dynamic}), the three methods failed to identify bugs caused by extrinsic factors like incorrect or mislabeled training data (Bug 4, 16). These issues originate from issues in the data pipeline rather than the model’s execution, making them invisible to runtime analysis. This limitation highlights a critical gap — extrinsic bugs require approaches that go beyond code execution or runtime monitoring. 

\textit{3. Challenges with Complex System Interactions:} A major insight from our evaluation of training bugs (21 buggy models) is that both static and dynamic methods struggle to address the intricate, often indirect interactions inherent in deep learning systems. For example, state-of-the-art static techniques like bjXnet \cite{han2023bjxnet} focus on code and textual bug reports, often overlooking issues such as unchanged model weights that only surface during execution (e.g., Bug 9). Meanwhile, dynamic tools like DeepLocalize, DeepDiagnosis, and DeepFD detect runtime anomalies (e.g., NaN values in Loss) but often fail to link these symptoms to their root causes, such as improper dropout rates causing NaN values in the loss (Table \ref{tab:dynamic}). Though dynamic approaches identified some training faults, their overall accuracy remained below 33\% (Table \ref{tab:dynamic_result}), highlighting their inability to address deeper systemic interactions.

Our findings demonstrate that deep learning software presents unique challenges that cannot be fully resolved by static or dynamic analysis alone. The limitations observed in both approaches suggest that bugs often arise from complex interactions between code, runtime behaviour, and system-level dependencies, which current tools are not equipped to handle comprehensively. These results align with our broader evaluation, emphasizing the need for more integrated techniques to localize bugs in DLSW effectively.

\begin{table*}[t]
\centering
\footnotesize
\caption{Performance Comparison of Dynamic Approaches for Training Bug Localization}
\setlength{\tabcolsep}{4pt}
\begin{tabular}{@{} c  >{\raggedright\arraybackslash}p{3.8cm}  >{\raggedright\arraybackslash}p{3.7cm} c c c @{}}
\toprule
\textbf{No.} & \textbf{Symptom} & \textbf{Root Cause (Bug)} & \textbf{DL} & \textbf{DF} & \textbf{DD} \\ 
\midrule
1  & Loss Not Decreasing             & Incorrect Loss Function                   & $\times$  & $\checkmark$ & $\checkmark$ \\
2  & Loss Not Decreasing             & Improper Activation Function              & $\times$  & $\times$    & $\checkmark$ \\
3  & Loss Increasing After Epochs    & Incorrect Optimizer                       & $\times$  & $\checkmark$ & $\times$    \\
4  & Model Overfitting               & Incorrect Training/Validation Data        & $\times$  & $\times$    & $\times$    \\
5  & Static Validation Loss          & Incorrect Dropout Usage                     & $\times$  & $\times$    & $\times$    \\
6  & Vanishing Loss \& Overfitting   & Small Training Data                         & $\times$  & $\times$    & $\times$    \\
7  & Overfitting                     & Incorrect Regularizer \& Small Data         & $\times$  & $\times$    & $\times$    \\
8  & Loss Oscillating \& Overfitting & Incorrect Dropout Rate                      & $\times$  & $\times$    & $\times$    \\
9  & Unchanged Weights \& Static Loss& Incorrect Learning Rate                     & $\times$  & $\checkmark$ & $\checkmark$ \\
10 & Poor Generalization             & Incorrect Normalization \& No Regularizer   & $\times$  & $\times$    & $\times$    \\
11 & NaN Values in Loss              & Incorrect Weight Initialization             & $\checkmark$ & $\times$   & $\checkmark$ \\
12 & Poor Performance (Low Accuracy) & Incorrect Epoch Number                      & $\times$  & $\checkmark$ & $\times$    \\
13 & Train/Val Loss Divergence  & Incorrect Batch Normalization               & $\checkmark$ & $\times$   & $\times$    \\
14 & Neurons Outputting Zero         & Incorrect Weight Initialization             & $\checkmark$ & $\times$   & $\checkmark$ \\
15 & Training Slowdown               & Incorrect Dropout Rate                      & $\times$  & $\times$    & $\times$    \\
16 & Overfitting                     & Incorrect Regularizer/Augmentation          & $\times$  & $\times$    & $\times$    \\
17 & NaN Values in Loss                        & Incorrect Activation Function               & $\checkmark$ & $\checkmark$ & $\checkmark$ \\
18 & NaN Values in Loss                        & Zero Sample Weights                         & $\checkmark$ & $\times$   & $\times$    \\
19 & Loss Plateaus                   & Improper Bias Initialization                & $\times$  & $\times$    & $\times$    \\
20 & Performance Degradation         & Incorrect Learning Rate                     & $\times$  & $\checkmark$ & $\checkmark$ \\
21 & Poor Generalization             & Incorrect Dropout Rate                      & $\times$  & $\times$    & $\times$    \\
\midrule
\multicolumn{3}{r}{\textbf{Accuracy (\%)}} & \textbf{23.8} & \textbf{28.6} & \textbf{33.3} \\
\bottomrule
\end{tabular}
\label{tab:dynamic}
\begin{flushleft}
\scriptsize{DL: DeepLocalize, DF: DeepFD, DD: DeepDiagnosis. $\checkmark$: Successfully identified, $\times$: Failed to identify.}
\end{flushleft}
\end{table*}

\FrameSep.3em
\begin{frshaded}
	\noindent
	\textbf{Summary of RQ$\mathbf{_1}$:} We evaluate the performance of five existing techniques in localizing bugs from deep learning systems and non-deep learning systems using three evaluation metrics. Our findings show that all five approaches perform significantly lower (e.g., \textbf{34.14\%} less MAP for DNNLOC, \textbf{27.87\%} less MAP for bjXnet) when localizing bugs from deep learning systems. We also compare their performance when localizing bugs from deep learning frameworks, libraries, and tools. We found that localizing bugs from frameworks is most challenging due to the complex interaction of their components.
\end{frshaded} 

\subsection{\textit{Answering RQ$\mathbf{_2}$: How do different types of bugs in deep learning systems impact bug localization?}}

In this research question, we investigate the characteristics and localization challenges of different types of bugs in deep learning systems through manual analysis. To classify different types of bugs in deep learning systems, we adopted the following steps:

\begin{figure}[h]
 \centering
 \includegraphics[width= 3.5in]{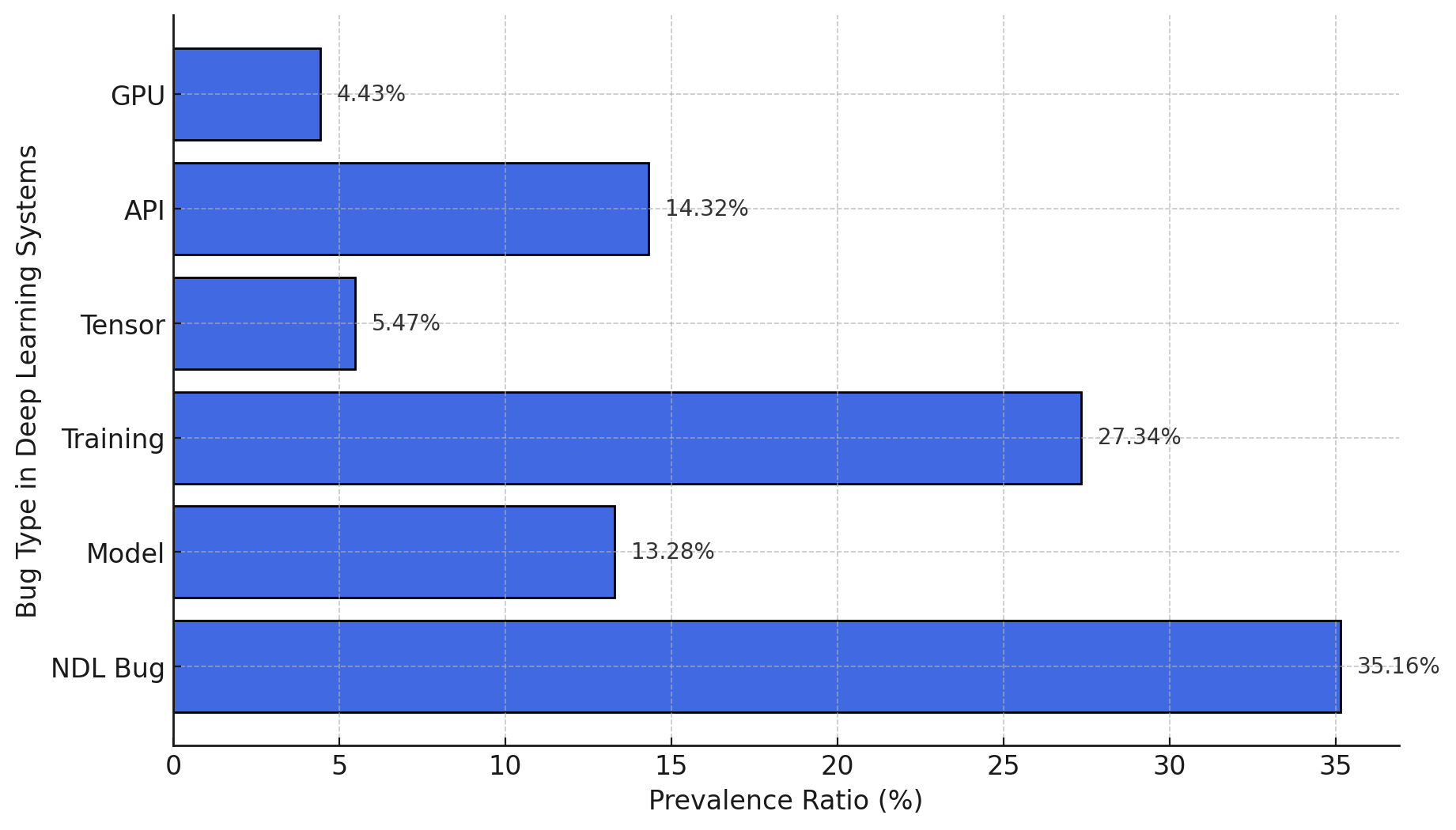}
  \vspace{-.4cm}
    \caption{Prevalence ratio of each category of bugs from deep learning systems} 
  \vspace{-.5cm}
 \label{fig:Prev_DLSW}
\end{figure}

\begin{itemize}
    \item We conducted our manual analysis on a sample of 385 bug reports, which have a 95\% confidence level and a 5\% margin of error. This allows our findings to be generalizable to the original population of deep learning system bugs.
    \item We classified deep learning-related bugs into five categories using the taxonomy of \citet{humbatova2020taxonomy}. This taxonomy provides explicit definitions for various DL bug types, including Model, Training, Tensor, API, and GPU issues. For instance, bugs involving neural network architecture or weight initialization were classified as 'Model' bugs, while issues with data preprocessing or augmentation during model training were categorized as 'Training' bugs. On the other hand, file handling errors or user interface glitches were categorized as Non-deep learning-related (NDL) bugs. 
    \item Two independent annotators (authors of this study) conducted a two-phase annotation. In the first phase, each bug was labeled as DL or NDL. In the second phase, DL bugs were further categorized into the five sub-types mentioned above. Prior to annotation, the annotators jointly labeled a sample of 20 bug reports to assess their understanding of the classification criteria. 
    \item The annotators cross-checked their annotations/classifications with project documentation and bug reports to prevent overlaps or misclassifications. Project documentation provided context and clarified the intended system functionality, helping to classify bugs accurately and avoid misclassification.
    \item We measured inter-annotator reliability using Cohen’s kappa coefficient \cite{cohenkappa}, resulting in a score of 0.80. This substantial agreement indicates a high level of consistency in our classification process. We resolved disagreements through a systematic consensus-building process involving joint re-examination of disputed cases, guided by taxonomy definitions and cross-referenced with additional documentation to ensure accuracy and consistency.
    \item We documented this manual analysis using an Excel sheet, with a total of $\approx$55 hours spent by each author, which is provided in our replication package \cite{Jahan2023replicationpackage}.\\
\end{itemize}

\textbf{Prevalence ratio of deep learning-related bugs:} We found that 64.80\% of the bugs from deep learning systems are related to deep learning algorithms (a.k.a DL bugs). That is, they are related to inputs, data, or training of deep learning models, underlying API endpoints, and computational resources. In particular, we found 27.30\% training bugs, 13.30\% model bugs, 5.50\% tensor bugs, 14.30\% API bugs, and 4.40\% GPU bugs (Fig. \ref{fig:Prev_DLSW}). Such a distribution informs us where the debugging efforts should be concentrated. Our findings also indicate that the majority of DL bugs are related to the training process. Training is a crucial step in deep learning that involves large amounts of data, complex learning algorithms, and optimization techniques, making it more susceptible to bugs and failures. \\
\indent

\textbf{Prevalence ratio of non-deep learning-related bugs:} We found that 35.20\% of the bugs from deep learning systems are not related to deep learning algorithms (a.k.a NDL bugs). These bugs do not directly affect the functionality of the deep learning model, but they still lead to unexpected, erroneous behaviors in a software system. Bug 1426 in Table \ref{tab:bug_extrinsic} is an NDL bug, which occurs when the tests from the CI pipeline are spread across multiple Windows machines. Although it is not directly connected to the deep learning module, it originated from the PyTorch-Ignite project, which is indeed a deep learning system.

\begin{figure*}[h]
  \centering
  \includegraphics[width=0.90\textwidth]{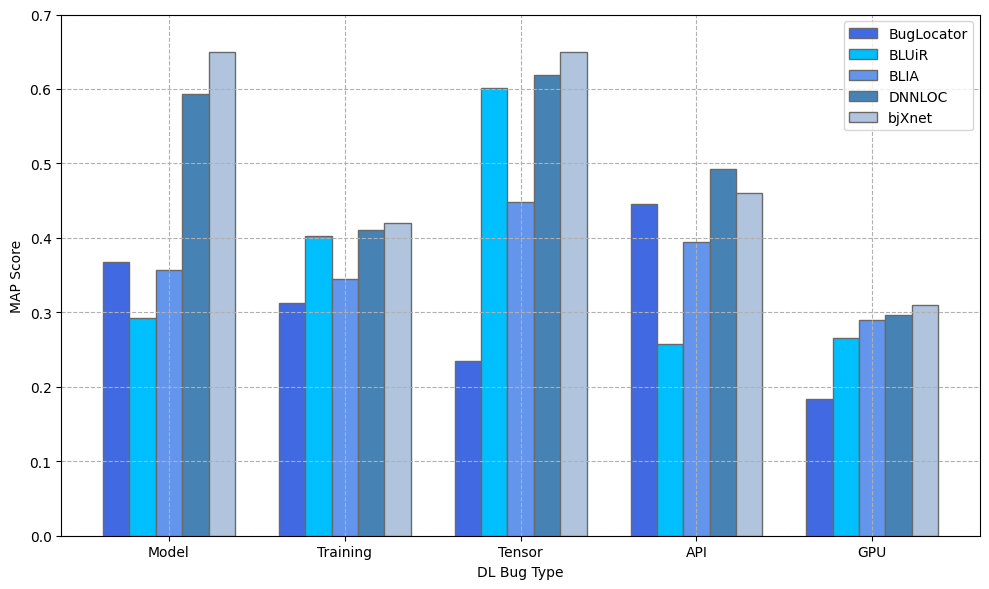}
  \caption{Performance of existing bug localization techniques (BugLocator, BLUiR, BLIA, DNNLOC, bjXnet) for each type of bug in deep learning systems}
  \label{fig:Result_DLSW}
  \vspace{-.5cm}
\end{figure*}

\textbf{Localization of bugs in deep learning systems:} 
To gain a deeper understanding of the challenges in localizing deep learning bugs, we analyzed the performance of our baseline techniques across different categories of bugs in DL systems. We employed stratified random sampling \cite{stratifiedrandomsampling} to select a balanced number of bug reports (100 bugs) from each category for performance comparison, ensuring fair representation and reducing selection bias. We repeated the process three times with different samples to ensure robustness. The results were averaged across these evaluations, providing reliable insights into the efficacy of the bug localization methods. In addition to the quantitative results, we conducted an in-depth analysis of each type of DL bug to understand their inherent challenges and how they impact the overall performance of the bug localization techniques. This detailed analysis, with examples, is provided below.

\begin{table}[h]
    \centering
    \caption{Experimental result of existing bug localization techniques (BugLocator, BLUiR, BLIA, DNNLOC, bjXnet) for each category of bugs in deep learning systems}
    \label{tab:result-dl-bug}
    \begin{adjustbox}{max width=\linewidth}
        \begin{threeparttable}
            \begin{tabular}{|c|c|c|c|c|c|}
                \hline
                \textbf{Method} & \textbf{Model} & \textbf{Training} & \textbf{Tensor} & \textbf{API} & \textbf{GPU} \\ 
                \hline 
                \multicolumn{6}{c}{\textbf{MAP}} \\ 
                \hline
                BugLocator & 0.368 & 0.312 & 0.235 & 0.446 & 0.183 \\ 
                \hline
                BLUiR & 0.293 & 0.403 & 0.601 & 0.258 & 0.266 \\
                \hline
                BLIA & 0.357 & 0.345 & 0.448 & 0.395 & 0.290 \\
                \hline
                DNNLOC & 0.593 & 0.411 & 0.619 & 0.493 & 0.296 \\
                \hline
                bjXnet & 0.650 & 0.420 & 0.650 & 0.460 & 0.310 \\
                \hline
                \multicolumn{6}{c}{\textbf{MRR}} \\ 
                \hline
                BugLocator & 0.532 & 0.386 & 0.358 & 0.378 & 0.223 \\
                \hline
                BLUiR & 0.387 & 0.427 & 0.621 & 0.424 & 0.311 \\
                \hline
                BLIA & 0.472 & 0.419 & 0.553 & 0.447 & 0.327 \\
                \hline
                DNNLOC & 0.585 & 0.467 & 0.682 & 0.548 & 0.336 \\
                \hline
                bjXnet & 0.720 & 0.480 & 0.680 & 0.510 & 0.360 \\
                \hline
            \end{tabular}   
            \centering
            \small
            \textbf{MAP}= Mean Average Precision \textbf{MRR}=Mean Reciprocal Rank
        \end{threeparttable}
    \end{adjustbox}
\end{table}

\textbf{Model bugs:} Model bugs present unique challenges in bug localization for deep learning systems. Our analysis in Table \ref{tab:result-dl-bug} reveals that bjXnet demonstrates superior performance in localizing model bugs. This can be attributed to its attention mechanisms, which enable focused examination of key subcomponents. Model bugs typically relate to a model's type, properties, and layers. We observed significant vocabulary overlap between bug reports describing model-related issues and the implementation of deep learning models. bjXnet's attention-based approach allows it to prioritize critical code sections, such as specific methods or functions directly implicated in the bug report. DNNLOC also performs well, followed closely by bjXnet, by employing deep neural networks to capture complex patterns and utilizing rVSM to assess textual relevance between bug reports and source code. In contrast, BLUiR shows the poorest performance in locating model bugs.

\setlength{\arrayrulewidth}{0.1mm}
\setlength{\tabcolsep}{11pt}
\renewcommand{\arraystretch}{1.0}
\begin{table}[h]
\centering
\caption{Example of a model bug \cite{texar_pytorch_issue_313}}
\label{tab:model}
\begin{tabular}{c}
\hline
\multicolumn{1}{|c|}{\textbf{Model Bug (Bug ID: 313)}} \\ \hline
\textbf{Title} \\ \hline
\multicolumn{1}{|l|}{A bug in GPT2Tokenizer} \\ \hline
\textbf{Description} \\ \hline
\multicolumn{1}{|l|}{\begin{tabular}[c]{@{}l@{}}GPT2Tokenizer fails to recover a sentence \\ \textbackslash{}"BART is a seq2seq model.\textbackslash{}"\\ with encoded ids of it. \\ The output sentence is \textbackslash{}"BART is a seqseq model.\textbackslash{}".\\ It should be related to numbers' processing. \\ A script to show the bug is here:\\ https://github.com/tanyuqian/texar-pytorch\\ /blob/master/examples/bart/gpt2\_tokenizer\_bug.py\\ 
Detailed BR:https://github.com/asyml/texar-pytorch/issues/313 \\
\end{tabular}}  \\ \hline
\end{tabular}
\end{table}

Table \ref{tab:model} \cite{texar_pytorch_issue_313} illustrates an example of a model bug from the \texttt{CASL.ai} project, with the corresponding code snippet shown in Fig. \ref{fig:model_bug_actual} \cite{texar_pytorch_issue_313} (Appendix \ref{appendix:A}). The bug occurs in the \texttt{bpe} method of the \texttt{GPT2Tokenizer}, causing faulty tokenization that impacts the GPT2 language model's functionality.

We noticed that IR-based techniques struggle with locating model bugs. For instance, BugLocator incorrectly identifies 'SentencePieceTokenizer.py' (Fig. \ref{fig:model_bug_false_class}, Appendix \ref{appendix:A}) as the Top@1 result, which is likely to stem from lexical overlap between the bug report and source code. Four key terms from the bug report (Table \ref{tab:model}) -- \textit{GPT2Tokenizer, recover, seq2seq, and model} -- appear in the incorrect file, potentially leading to this incorrect localization. Notably, BugLocator retrieves the actual ground truth file at the 8th position. 

On the other hand, BLIA ranks the ground truth relatively low at the 17th position, potentially because meta components, such as stack trace, version control history, or commit history, are less effective in localizing model bugs. Model bugs typically involve deep architectural issues within the neural network, which are not easily traced through surface-level meta-information, making BLIA less suitable for these types of bugs compared to traditional software bugs.

BLUiR performs even worse, ranking the ground truth code at the 131st position. This suggests that the similarity between bug reports and source code elements may be insufficient for accurately locating model bugs. In the example from Table \ref{tab:model}, the bug manifests when tokenizing the sentence 'BART is a seq2seq model.' BLUiR mistakenly retrieves the source code file containing the \texttt{Seq2Seq} class as the Top@1 result. This error likely occurs because several important keywords from the bug report align with similar code elements (e.g., 'seq2seq', 'encode', and 'model' matching the \texttt{seq2seq} class). Such misalignment results from BLUiR's strong emphasis on structural elements (e.g., class, method, variable) and insufficient consideration of the bug report's semantic aspects.

In contrast, both bjXnet and DNNLOC successfully retrieve the buggy file at Top@1. Unlike IR-based methods, DNNLOC's hybrid approach leveraging neural networks can capture deeper semantic links beyond lexical similarities. For instance, despite textual mismatches, it can better connect terms like 'seq2seq' and 'tokenizer' in the bug report to their corresponding code-level implementations in GPT2Tokenizer. DNNLOC's ability to leverage non-linear relationships likely contributes to its superior performance in localizing model bugs in deep learning systems. bjXnet's success can be primarily attributed to its attention mechanism, which enables the targeting of critical subcomponents within the code, such as the \texttt{bpe} method in the GPT2Tokenizer. By focusing on the most relevant methods and variables implicated in the bug report, bjXnet achieves higher precision, especially in modular deep-learning components like tokenization and layer definitions, where pinpoint accuracy is crucial for fault identification. \\
\indent

\textbf{Training bugs:} All five approaches perform poorly in localizing training bugs, as shown in Table \ref{tab:result-dl-bug}. bjXnet, DNNLOC, and BLUiR perform comparatively better than the other techniques, though their overall MAP values remain low (40.30\% -- 42.00\%). Table \ref{tab:training} \cite{fastai_issue_3048} illustrates a training bug from the \texttt{fast.ai} project, where the interaction between \texttt{Gradient Accumulation} and the \texttt{MixedPrecision Callback} leads to artificially high training loss values.

\setlength{\arrayrulewidth}{0.1mm}
\setlength{\tabcolsep}{11pt}
\renewcommand{\arraystretch}{1.0}
\begin{table}[h]
\centering
\caption{Example of a training bug \cite{fastai_issue_3048}}
\label{tab:training}
\begin{tabular}{c}
\hline
\multicolumn{1}{|c|}{\textbf{Training Bug (Bug ID: 3048)}} \\ \hline
\textbf{Title} \\ \hline
\multicolumn{1}{|l|}{\text{\begin{tabular}[c]{@{}l@{}}Gradient Accumulation + Mixed Precision\\ shows artificially high training loss\end{tabular}}} \\ \hline
\textbf{Description} \\ \hline
\multicolumn{1}{|l|}{\begin{tabular}[c]{@{}l@{}}OB: The bug occurs when Gradient Accumulation \\ and the MixedPrecision Callback are both used. \\ Gradient Accumulation runs before Mixed Precision \\ and causes the after\_backwards to not be run,\\ meaning that the loss is not unscaled before it is logged. \\ This means that very large losses, such as 6000000+, are to be logged.\\ S2R: seed=random.randint(0,2**32-1)\\ with no\_random(seed):   \\ db=synth\_dbunch(bs=8,n\_train=1,n\_valid=1,cuda=True)\\     learn = synth\_learner(data=db)\\      learn.fit(1, lr=0.01)\\  \#start without gradient overflow\\ max\_loss\_scale=2048.0\\  with no\_random(seed):\\      db=synth\_dbunch(bs=1,n\_train=8,n\_valid=8,cuda=True)     \\ learn = synth\_learner(data=db,cbs={[}GradientAccumulation(n\_acc=8){]})   \\ learn.to\_fp16(max\_loss\_scale=max\_loss\_scale)\\      learn.fit(1, lr=0.01)\\  The training loss will be very high, 5000+ for fp16. \\ fp32 will be reasonable\\ EB: Similar training loss \\ between the fp32 and fp16 versions. \textless{}2 difference in loss.\\
Detailed BR: https://github.com/fastai/fastai/issues/3048 \\
\end{tabular}} \\ \hline
\centering
\small
\vspace{10 pt}
\textbf{EB}=Expected Behaviour, \textbf{S2R}=Steps to Reproduce, \textbf{OB}=Observed Behaviour
\end{tabular}
\end{table}

For this example, DNNLOC retrieved the ground truth file at Top@4, while BLUiR ranked it at Top@7. bjXnet performs slightly better than DNNLOC in localizing training bugs and retrieving the ground truth file at Top@3. bjXnet's inability to rank the correct file at Top@1 likely stems from the distributed nature of training bugs. These bugs often span across multiple interconnected components like callbacks, loss functions, and training loops, making it harder to pinpoint a single root cause. While bjXnet's attention mechanism can highlight important sections like callbacks and loss scaling, training bugs tend to involve interactions between different parts of the code. This complexity means that the root cause may not be confined to a single location, requiring a broader contextual understanding of the training process, which limits bjXnet’s ability to achieve Top@1 localization

On the other hand, BLIA also performs poorly, ranking the correct file at Top@21. Meta information like stack traces and version history from BLIA might not be particularly helpful for training bugs, as they often involve complex, abstract issues (e.g., improper configurations of loss functions or callbacks), which do not leave clear traces in version control or stack traces. Moreover, BugLocator performs the worst (e.g., Top@23 position for the example bug in Table \ref{tab:training}) in localizing the training bugs (Fig. \ref{fig:training_bug_false}) (as shown in Appendix \ref{appendix:A}), which suggests that similarity analysis between bug reports and source code might not be sufficient for identifying these bugs. Training bugs, which occur during the model's training phase and might involve aspects such as an improperly defined loss function, are more conceptual. These bugs might not be easily located using these existing methods and might require techniques that can offer a deeper insight into the model's architecture and training process. \\
\indent

\textbf{Tensor bugs:} Tensors are central to deep learning and often involve intricate dimensional and mathematical issues. From Fig. \ref{fig:Result_DLSW}, we notice that bjXnet, DNNLOC, and BLUiR are more effective in localizing tensor bugs than other techniques (e.g., BugLocator, BLIA). In the example bug from Table \ref{tab:tensor} \cite{mxnet_issue_13760}, the bug report described the issue with the \texttt{`nd.slice'} function in MXNet, which should return an empty tensor when the `begin' and `end' parameters are equal. Instead, the function returns a tensor with the same shape as the data.
\setlength{\arrayrulewidth}{0.1mm}
\setlength{\tabcolsep}{11pt}
\renewcommand{\arraystretch}{1.0}
\begin{table}[h]
\centering
\caption{Example of a tensor bug \cite{mxnet_issue_13760}}
\label{tab:tensor}
\begin{tabular}{l}
\hline
\multicolumn{1}{|c|}{\textbf{Tensor Bug (Bug ID: 13760)}} \\ \hline
\multicolumn{1}{c}{\textbf{Title}} \\ \hline
\multicolumn{1}{|l|}{nd.slice does not return empty tensor when begin=end} \\ \hline
\multicolumn{1}{c}{\textbf{Description}} \\ \hline
\multicolumn{1}{|l|}{OB: For mxnet.ndarray.slice(data, begin, end),} \\
\multicolumn{1}{|l|}{if begin=end, it does not return an empty tensor.} \\
\multicolumn{1}{|l|}{Instead, it returns a tensor with the same shape as the data.} \\
\multicolumn{1}{|l|}{Environment info: } \\
\multicolumn{1}{|l|}{ ...... }  \\
\multicolumn{1}{|l|}{ ...... }  \\
\multicolumn{1}{|l|}{ ...... }  \\
\multicolumn{1}{|l|}{S2R: import mxnet.ndarray as nd} \\
\multicolumn{1}{|l|}{a = nd.normal(shape=(4, 3))} \\
\multicolumn{1}{|l|}{nd.slice(a, begin=0, end=0)} \\
\multicolumn{1}{|l|}{nd.slice(a, begin=2, end=2)} \\ 
\multicolumn{1}{|l|}{Detailed BR: https://github.com/apache/mxnet/issues/13760} \\
\hline
\textbf{BR}=Bug Report, \textbf{S2R}=Steps to Reproduce, \textbf{OB}=Observed Behaviour
\end{tabular}
\end{table}

For this specific tensor bug, bjXnet, DNNLOC, and BLUiR all successfully retrieved the correct buggy file at Top@1 due to their targeted handling of tensor-related operations. bjXnet’s attention mechanism focuses on key operations, such as the \texttt{slice} function, which mishandles tensor shapes when \texttt{begin} equals \texttt{end}. This targeted focus might allow bjXnet to effectively identify the specific code causing the bug. DNNLOC captures the complex interactions between the bug report and tensor manipulations, mapping the error in how the \texttt{nd.slice} function handles dimensions directly to the buggy code. Interestingly, by parsing the AST of the code, BLUiR identified the relevant code snippet in the test$\text{\_}$slice() function, which shares significant keyword overlap with the bug report. BLUiR determines the relevance of low-level code elements (e.g., class names, method names) against a bug report, which helps to reduce noise in the code segments where tensors could be handled or manipulated. 

On the other hand, BLIA retrieved the ground truth file at the top 12 positions. However, our analysis revealed that incorporating the stack trace information negatively impacted its bug localization performance. Tensor bugs are typically related to data manipulation and computations \cite{humbatova2020taxonomy} rather than code execution flow or call stack \cite{velez2022debugging} that can be found in stack traces. By excluding the stack trace from the BLIA approach, we were able to improve the ranking of the buggy file from the 12th position to the 9th position. 

Lastly, BugLocator's difficulty in accurately locating tensor bugs is demonstrated by its retrieval of the ground truth file at the 47th position (Fig. \ref{fig:tensor_bug_actual}) (as shown in Appendix \ref{appendix:A}) and the incorrect file at Top@1 (Fig. \ref{fig:tensor_bug_false}) (as shown in Appendix \ref{appendix:A}). It can be attributed to its reliance on textual similarity. We found that BugLocator's heavy reliance on textual similarity and the overlapping of trivial words (e.g., environment, system, and hardware) with the incorrect file (Fig. \ref{fig:tensor_bug_false}) (as shown in Appendix \ref{appendix:A}) led to the incorrect ranking. \\
\indent

\textbf{API bugs:} From Fig. \ref{fig:Result_DLSW}, we observe that DNNLOC performs well in localizing API bugs, whereas BLUiR is the least effective. This could be due to BLUiR's heavy reliance on structural information from the source code, which might not capture the specifics of API bugs (e.g., incorrect API calls). Another factor could be the rapid evolution of deep learning APIs, which affects versioning and compatibility \cite{zhang2019empirical}. Due to frequent structural changes in the code, the technique might be less effective in assessing the relevance between code and bug reports. 

\setlength{\arrayrulewidth}{0.1mm}
\setlength{\tabcolsep}{11pt}
\renewcommand{\arraystretch}{1.0}
\begin{table}[h]
\centering
\caption{Example of an API bug \cite{mxnet_issue_13862}}
\label{tab:API}
\begin{tabular}{l}
\hline
\multicolumn{1}{|c|}{\textbf{API Bug (Bug ID: 13862)}} \\ \hline
\multicolumn{1}{c}{\textbf{Title}} \\ \hline
\multicolumn{1}{|l|}{\begin{tabular}[c]{@{}l@{}}{[}1.4.0{]} unravel\_index no longer works with\\  magic '-1' in shape parameter as in 1.3.1\end{tabular}} \\ \hline
\multicolumn{1}{c}{\textbf{Description}} \\ \hline
\multicolumn{1}{|l|}{\begin{tabular}[c]{@{}l@{}}OB: The unravel\_index op seems to no longer correctly\\ work with 'magic' shape values, such as '-1's. \\ The following example still works with mxnet 1.3.1, \\ but does not on the latest master \\ (it returns all zeros in the result \\ without throwing an error) or 1.4.0.\end{tabular}} \\
\multicolumn{1}{|l|}{We have a use case for this in Sockeye.} \\
\multicolumn{1}{|l|}{Environment info (Required):} \\
\multicolumn{1}{|l|}{…} \\
\multicolumn{1}{|l|}{S2R: Input data taken from Sockeye unit tests.} \\
\multicolumn{1}{|l|}{x = mx.nd.array({[}335, 620, 593, 219, 36{]}, dtype='int32')} \\
\multicolumn{1}{|l|}{mx.nd.unravel\_index(x, shape=(-1, 200))} \\
\multicolumn{1}{|l|}{With mxnet==1.5.0b20190111, the result is incorrect:} \\
\multicolumn{1}{|l|}{With mxnet==1.3.1, the result is correct:} \\
\multicolumn{1}{|l|}{\begin{tabular}[c]{@{}l@{}}However, if the shape parameter is fully specified\\ (shape=(5,200)), mxnet==1.5.0b20190111\\  returns the correct values.\end{tabular}} \\
\multicolumn{1}{|l|}{Detailed BR:  https://github.com/apache/mxnet/issues/13862} \\ \hline
\centering
\small
\textbf{BR}=Bug Report, \textbf{S2R}=Steps to Reproduce, \textbf{OB}=Observed Behaviour
\end{tabular}
\end{table}

The bug from Table \ref{tab:API} \cite{mxnet_issue_13862} involves the \texttt{unravel-index} function in MXNet, where its behavior incorrectly varies with certain input parameters across different versions (Fig. \ref{fig:API_bug_actual}) (as shown in Appendix \ref{appendix:A}). DNNLOC located the correct buggy code for the example bug (Table \ref{tab:API}) at the 2nd position, while BugLocator ranked it in the top 5th. DNNLOC might be effective for this API bug because textual similarity (through rVSM) matches API-specific terms in bug reports with source code, while neural networks (through DNN) can capture complex, non-obvious relationships in the API's usage and functionality. Moreover, our findings indicate that relying on either element (rVSM, DNN) in isolation is less effective, resulting in a noticeable drop in performance.

On the other hand, BLUiR retrieved the buggy file at the 38th position. BLUiR incorrectly retrieved the file containing the \texttt{NDArray} class. In the case of API bugs, the class file may not always be relevant. This is because API bugs frequently arise at the interface between different layers of abstraction \cite{dig2006apis}. Such bugs could be attributed to the interaction of a higher-level function (e.g., \texttt{unravel-index} from Table \ref{tab:API}) with lower-level components rather than an issue within the code of the function itself. Meanwhile, BLIA initially placed the buggy file at the top 7th position without the stack traces but improved to 5th when stack trace information was included. It indicates that stack trace information, highlighting execution flow and function calls, is beneficial for locating API bugs. 

Interestingly, despite its advanced attention mechanisms, bjXnet ranked the correct buggy file at the 6th position, lower than DNNLOC and BugLocator. We found that bjXnet's attention mechanism might over-prioritize certain structural sections of the code, causing it to miss key interactions between components, which are critical for localizing API bugs. For example, in the unravel-index bug (Table \ref{tab:API}), bjXnet correctly identified the key \texttt{unravel-index} function but failed to capture how the interactions between the API and different framework versions caused the bug. The issue arose from the \texttt{shape} parameter's change in behavior across MXNet versions, but bjXnet’s focus on specific code segments may have led it to miss this broader context. \\
\indent

\textbf{GPU bugs:} Our investigation reveals that GPU bugs are the most difficult to localize for all five existing approaches. One possible explanation could be the complex nature of GPU bugs, as they can be triggered by a variety of factors, such as the compatibility between hardware (e.g., GPU device) and software (e.g., PyTorch). It might not even be located in the source code (a.k.a extrinsic GPU bug) \cite{humbatova2020taxonomy}. From Table \ref{tab:manual analysis}, we notice that only 17.65\% of the GPU bugs can be found in the source code (a.k.a intrinsic GPU bug). Some examples of intrinsic GPU bugs that we found -- the wrong reference to a GPU device, failed parallelism, incorrect state sharing between subprocesses, and faulty data transfer to a GPU device.

\setlength{\arrayrulewidth}{0.1mm}
\setlength{\tabcolsep}{11pt}
\renewcommand{\arraystretch}{1.0}
\begin{table}[t!]
\centering
\caption{Example of a GPU bug \cite{autokeras_issue_1238}}
\label{tab:GPU}
\begin{tabular}{l}
\hline
\multicolumn{1}{|c|}{\textbf{GPU Bug (Bug ID: 1238)}} \\ \hline
\multicolumn{1}{c}{\textbf{Title}} \\ \hline
\multicolumn{1}{|l|}{How to use multiple GPUs?} \\ \hline
\multicolumn{1}{c}{\textbf{Description}} \\ \hline
\multicolumn{1}{|l|}{\begin{tabular}[c]{@{}l@{}}I want to use a single machine with multiple\\ GPUs for training, but it has no actual effect.\end{tabular}} \\
\multicolumn{1}{|l|}{\begin{tabular}[c]{@{}l@{}}OB: Only one single GPU is doing all the computations, \\ the other three remain idle. When following @FontTian \\ and inserting distribution\_strategy=strat into the \\ initialization of the image classifier, the same \\ error RuntimeError: Too many failed attempts \\ to build the model occurs. The same happens \\ when adding tuner='random' to ak.ImageClassifier.\end{tabular}} \\
\multicolumn{1}{|l|}{\begin{tabular}[c]{@{}l@{}}As suggested by @haifeng-jin, I ran a basic \\ KerasTuner example on 4 GPUs which \\ worked just fine. Furthermore, in \#440 (comment), \\ I read that the clear\_session() before every \\ run might wipe out the GPU configuration. \\ Removing this line from the code did not change\\ anything with respect to the errors/problems stated above.\end{tabular}} \\
\multicolumn{1}{|l|}{\begin{tabular}[c]{@{}l@{}}I am specifying 4 GPUs (out of 8) to train \\ the current model in a distributed fashion, \\ using tf.distribute.MirroredStrategy( ) since \\ tf.keras.utils.multi\_gpu\_model( ) is \\ deprecated and removed since April 2020.\end{tabular}} \\
\multicolumn{1}{|l|}{S2R: def make\_model(ckpt\_path, max\_try = 1):} \\
\multicolumn{1}{|l|}{......} \\
\multicolumn{1}{|l|}{run\_search(checkpoint, max\_try = 3)} \\
\multicolumn{1}{|l|}{\begin{tabular}[c]{@{}l@{}}Detailed BR: https://github.com/keras-team/autokeras/issues/1238\\
\end{tabular}} \\ \hline
\centering
\small
\textbf{BR}=Bug Report, \textbf{S2R}=Steps to Reproduce, \textbf{OB}=Observed Behaviour
\end{tabular}
\end{table}

Table \ref{tab:GPU} \cite{autokeras_issue_1238} shows a GPU bug triggered by the codebase (a.k.a intrinsic). The bug is connected to the use of multiple GPUs during training. According to the report, the machine contains multiple GPU devices, but only one GPU is used during computation. All five bug localization techniques performed poorly in locating the actual buggy code for this GPU bug. BLUiR retrieved the buggy file at the 50th position, followed by BugLocator at the 65th, BLIA at the 47th, DNNLOC at the 43rd, and bjXnet at the 42nd position in their respective rankings.

We observed that there is almost no keyword overlapping and no structural similarity between the bug report (Table \ref{tab:GPU}) and the actual buggy code (Fig. \ref{fig:GPU_bug_actual}) (as shown in Appendix \ref{appendix:A}). This suggests that these techniques struggled due to the lack of both textual and code-wise similarity between the bug report and source code, making it challenging to identify the buggy code for the GPU bug.

We observed a minimal keyword overlap and structural similarity between the bug report (Table \ref{tab:GPU}) and the actual buggy code (Fig. \ref{fig:GPU_bug_actual}) (as shown in Appendix \ref{appendix:A}), presenting a significant challenge in locating such bugs using BugLocator. Additionally, BLUiR's analysis of smaller code segment similarity also proves ineffective for GPU bugs, as it solely concentrates on source code and overlooks the hardware-software interactions and runtime specifics crucial for comprehending such bugs. Similarly, BLIA's integration of stack trace information, commit history, or version control history fails to encapsulate the unique aspects of GPU bugs as well. Moreover, despite DNNLOC's and bjXnet's capability to capture non-linear complex relationships through deep neural networks, it did not help to locate GPU bugs. One reason might be that these bugs often involve complex hardware-software dynamics that are not typically addressed by standard source code analysis or within the codebase's non-linear mappings.\\
\indent
\begin{table}[h]
    \centering
    \caption{Experimental result of existing bug localization techniques (BugLocator, BLUiR, BLIA, DNNLOC, bjXnet) for non-deep learning-related (NDL) bugs Bugs in deep learning systems}
    \label{tab:result-ndl-bug}
    \begin{adjustbox}{max width=\linewidth}
        \begin{threeparttable}
            \begin{tabular}{|c|c|c|}
                \hline
                \textbf{Method} & \textbf{MAP} & \textbf{MRR} \\
                \hline
                BugLocator & 0.362 & 0.417 \\
                \hline
                BLUiR & 0.292 & 0.334 \\
                \hline
                BLIA & 0.437 & 0.381 \\
                \hline
                DNNLOC & 0.588 & 0.439  \\
                \hline
                bjXnet & 0.495 & 0.450 \\
                \hline
            \end{tabular}
        \end{threeparttable}
    \end{adjustbox}
\end{table}

\textbf{NDL bugs:} We found that 35.20\% of bugs in deep learning applications are not directly related to deep learning, but they impact system behavior (e.g., failed CI build due to GPU compatibility issues). These bugs are known as Non-Deep Learning-related (NDL) bugs in deep-learning systems. From Table \ref{tab:result-ndl-bug}, we find that the performance of existing techniques in localizing these bugs is also poor. DNNLOC outperforms other techniques, whereas BLUiR performs the lowest in locating NDL bugs. We observed that NDL bugs are less complex than their deep-learning counterparts. However, they are more prone to be extrinsic than the traditional bugs (48.15\% from Table \ref{tab:manual analysis}); we provided the details about extrinsic bugs in RQ${_3}$. Since existing baseline techniques focus on code-level artifacts only, they might fall short in detecting these bugs from deep learning systems.\\
\indent

\textbf{Bug report quality for bugs in deep learning systems:} Our analysis showed that bug reports from deep learning systems contain more code snippets (83.11\%) than traditional software systems (33.24\%). Unfortunately, that does not help much in bug localization, as code snippets alone might not be sufficient. Deep learning bugs often involve intricate dependencies that extend beyond specific code components (e.g., training data bugs and GPU bugs). Complex bugs (e.g., gradient instability during training) warrant a deeper understanding of the model architecture, its dynamic behavior, and training processes, which the code snippets may not always capture.

\FrameSep.3em
\begin{frshaded}
        \noindent
        \textbf{Summary of RQ$\mathbf{_2}$:} We found that \textbf{64.80\%} bugs in deep learning systems (DLSW) are related to deep learning algorithms, whereas the remaining bugs are not related to deep learning. Our analysis shows that Tensor bugs and API bugs are easier to localize than model and training bugs. However, GPU bugs are the most difficult to localize for each of the five approaches. Thus, our results not only inform the distribution of DL bugs but also highlight their localization challenges through extensive experiments. 
\end{frshaded}

\subsection{\textit{Answering RQ$\mathbf{_3}$: What are the implications of extrinsic bugs in deep learning systems for bug localization?}}

Most of the traditional bug localization techniques rely on the similarity between bug reports and source code \cite{buglocator, survey, bluir, DNNLOC, amalgam, blizzard, BLIA, BRTracer}. However, if a bug is of an extrinsic nature (e.g., originates from the operating system), simply relying on source code may not be effective for its localization.
\begin{figure*}[t!]
 \centering
 \includegraphics[width= 3.5in]{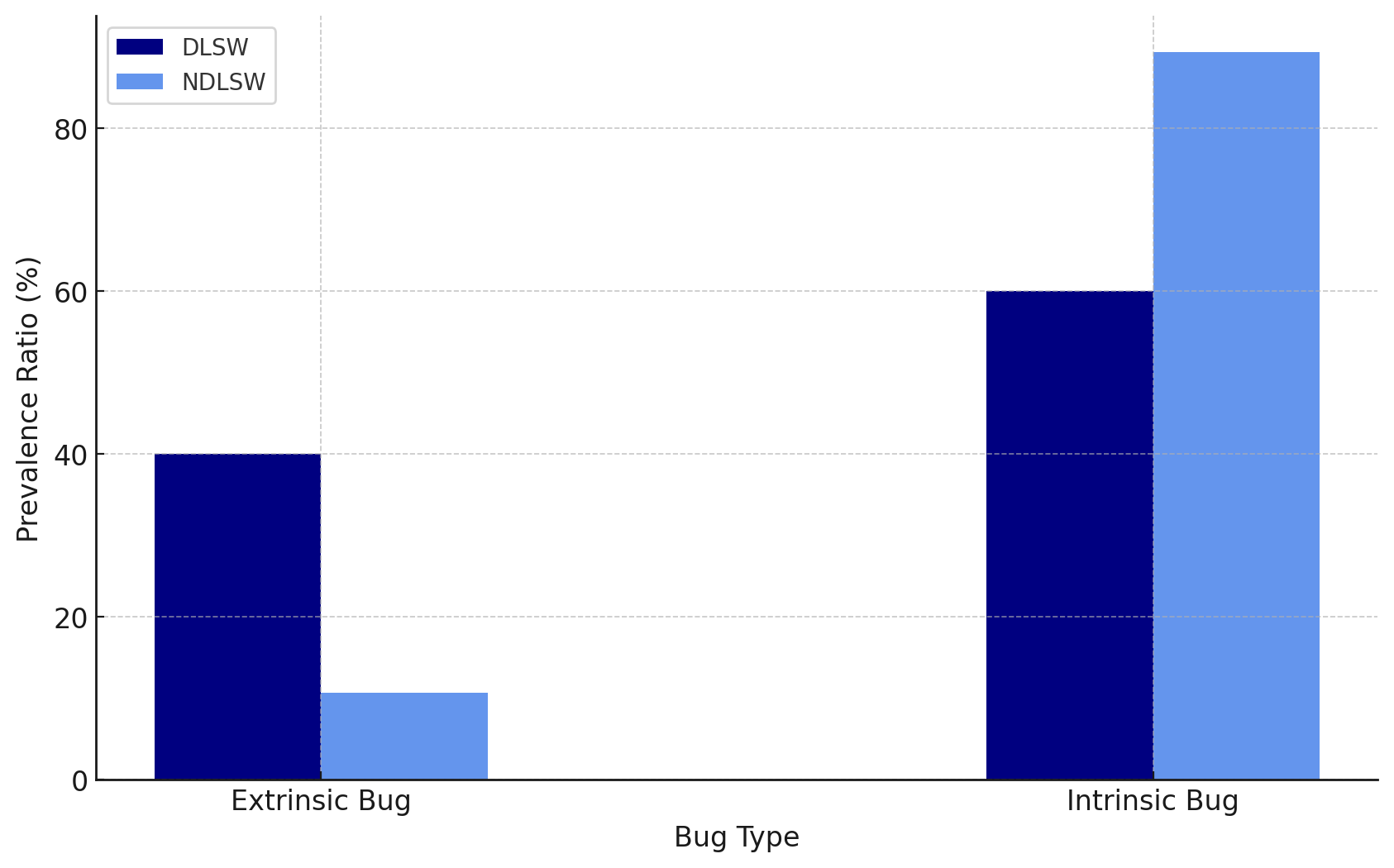}
  \vspace{-.4cm}
    \caption{Prevalence ratio of extrinsic and intrinsic bugs in deep learning systems (DLSW) and non-deep learning systems (NDLSW)} 
  \vspace{-.5cm}
 \label{fig:extrinsic_intrinsic}
\end{figure*} 

To investigate the impact of extrinsic bugs in deep learning systems, we performed another manual analysis following a set of steps as follows.

\begin{itemize}
    \item We conducted our manual analysis on a sample of 385 bug reports, which have a 95\% confidence level and a 5\% margin of error. This allows our findings to be generalizable to the original population of deep learning system bugs.
    \item We manually labeled 385 bug reports from deep learning software (DLSW) and 385 from non-deep learning software (NDLSW) as either extrinsic or intrinsic. We selected equal amounts of bug reports using randomization principles to ensure a fair comparison during the evaluation of bug localization techniques. We used stratified random sampling \cite{stratifiedrandomsampling} ensuring that our dataset was representative of both deep learning and traditional systems, reducing sampling bias that could affect the prevalence ratio of extrinsic and intrinsic bugs across categories.
    \item We strictly adhered to established heuristics \cite{extrinsic&intrinsic} to annotate extrinsic and intrinsic bugs. For example, a bug was classified as extrinsic only if there was clear evidence of external factors, such as system call failures or library version mismatches, documented in the bug report or related discussions.
    \item We achieved a Cohen’s kappa \cite{cohenkappa} score of 0.87, indicating substantial agreement and reinforcing the reliability of our manual labeling process. We resolved disagreements through a systematic consensus-building process involving joint re-examination of disputed cases, guided by taxonomy definitions and cross-referenced with additional documentation to ensure accuracy and consistency.
    \item We documented the manual analysis using an Excel sheet, which can be found in our replication package \cite{Jahan2023replicationpackage}. Each author spent a total of $\approx$20 hours on the analysis.
\end{itemize}

\textbf{Prevalence ratio of extrinsic \& intrinsic bugs:} We found 40.00\% extrinsic bugs within a total of 385 bug reports from deep learning systems (Denchmark dataset). The notion of extrinsic bugs is relatively new, especially in the case of bugs from deep learning systems. For a better comparison, we also manually inspected 385 bugs from non-deep learning systems (BuGL dataset) and determined the prevalence ratio of extrinsic and intrinsic bugs. We found only 10.65\% extrinsic bugs in non-deep learning systems. Thus, deep learning systems contain almost four times more extrinsic bugs (Fig. \ref{fig:extrinsic_intrinsic}) than non-deep learning systems.\\
\indent

\setlength{\arrayrulewidth}{0.1mm}
\setlength{\tabcolsep}{11 pt}  
\renewcommand{\arraystretch}{1.0}
\begin{table}[h]
    \centering
    \caption{Prevalence ratio of extrinsic and intrinsic bugs in deep learning systems}
    \label{tab:manual analysis}
    \resizebox{0.8 \textwidth}{!}{  
        \begin{threeparttable}
            \begin{tabular}{|c|c|c|c|}
                \hline
                \multicolumn{2}{|c|}{\textbf{Type}} & \textbf{Extrinsic (\%)} & \textbf{Intrinsic (\%)} \\ 
                \hline \hline
                \multicolumn{2}{|c|}{NDL} & 48.15 & 51.85 \\ 
                \hline
                \multirow{5}{*}{DL} & Model & 35.29 & 64.71 \\ \cline{2-4}
                                    & Training & 21.90 & 78.10 \\ \cline{2-4} 
                                    & Tensor & 38.10 & 61.90 \\ \cline{2-4} 
                                    & API & 38.19 & 61.81 \\ \cline{2-4} 
                                    & GPU & 82.35 & 17.65 \\ 
                \hline
            \end{tabular}
        \end{threeparttable}
    }
\end{table}

\textbf{Prevalence ratio of extrinsic \& intrinsic bugs from deep learning systems:} We randomly select 100 samples for each type of bug from deep learning systems (same subsets from RQ$_2$) and determined the prevalence ratio of extrinsic and intrinsic bugs for each type. Table \ref{tab:manual analysis} shows the results of our manual analysis for different bug categories in deep learning systems in terms of extrinsic and intrinsic bugs. We see that the prevalence ratios of deep learning-related extrinsic bugs range from 21.90\% to 82.35\%, whereas for non-deep learning-related bugs, the prevalence ratio is 48.15\%. This suggests that the deep learning components of a software system might be more likely to trigger extrinsic bugs than non-deep learning components.\\
\indent

\textbf{Localization of extrinsic \& intrinsic bugs from both systems:} To further analyze the impact of extrinsic bugs on bug localization, we experimented with our baseline techniques from RQ$_1$ on extrinsic and intrinsic bugs separately. We chose 100 random bugs to evaluate the performance of all five techniques in bug localization from each category of both benchmark datasets. We repeated the evaluation three times using three different random subsets and then calculated the average result for a fair comparison. 

\begin{figure*}[h]
  \centering
  \includegraphics[width=0.80\textwidth]{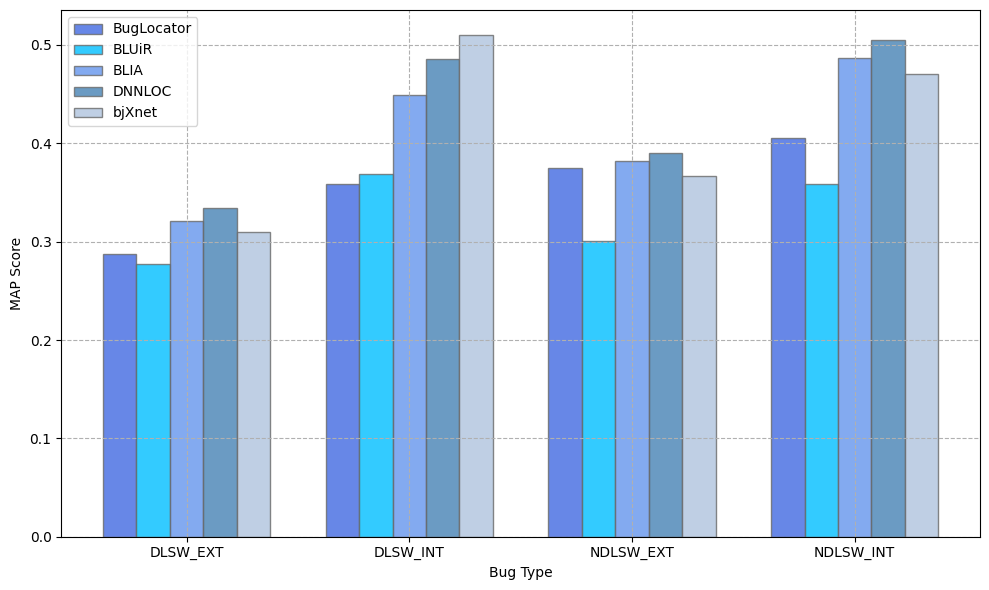}
  \caption{Performance of existing approaches (BugLocator, BLUiR, BLIA, DNNLOC, bjXnet) for localizing extrinsic and intrinsic bugs in both systems (DLSW \& NDLSW)}
  \label{fig:result_extrinsic_intrinsic}
  \vspace{-.5cm}
\end{figure*}

From Table \ref{tab:performance-extrinsic-intrinsic-dl}, we notice that DNNLOC performs slightly better than other techniques in localizing extrinsic bugs for both systems. However, it shows a clear performance gap in localizing extrinsic and intrinsic bugs. The technique is less effective with extrinsic bugs, particularly in DLSW. It suggests a shortcoming in handling external complexities despite being able to capture non-linear relationships between bug reports and source code. We notice that BugLocator's localization performance for extrinsic bugs in DLSW is lower than NDLSW. BugLocator might not be able to locate the extrinsic bugs in deep learning systems due to its naive approach, i.e., considering code as regular texts. We also found that BLUiR shows less performance gap between extrinsic and intrinsic bugs. It extracts different structured items (e.g., methods, classes) from the source code and bug reports. Thus, even if they reside outside the current codebase and are invoked from an external library, they could be matched with relevant keywords from a bug report. 

Interestingly, BLIA performs slightly better for extrinsic bugs in DLSW than NDLSW, which could be possible due to its diverse use of meta components (e.g., stack traces, version control history). Stack traces from deep learning systems (DLSW) often contain more intricate information and dependencies (e.g., complex neural network data flows, dependencies on specialized libraries, GPU-related synchronization issues) related to the bugs, unlike stack traces from non-deep learning systems, which might not provide the same level of detailed information \cite{o2020deep, karasov2022aggregation}.

\begin{table}[h]
    \centering
    \caption{Performance of Bug Localization Techniques in DL and TRAD Categories}
    \label{tab:performance-extrinsic-intrinsic-dl}
    \begin{adjustbox}{max width=\linewidth}
        \begin{threeparttable}
            \begin{tabular}{|c|c|c|c|c|}
                \hline
                \textbf{Method} & \textbf{DLSW+EXT} & \textbf{DLSW+INT} & \textbf{NDLSW+EXT} & \textbf{NDLSW+INT} \\ 
                \hline 
                \multicolumn{5}{c}{\textbf{MAP}} \\ 
                \hline
                BugLocator & 0.287 & 0.359 & 0.375 & 0.405 \\ 
                \hline
                BLUiR & 0.277 & 0.369 & 0.301 & 0.359 \\
                \hline
                BLIA & 0.321 & 0.449 & 0.382 & 0.487 \\
                \hline
                DNNLOC & 0.334 & 0.486 & 0.390 & 0.505 \\
                \hline
                bjXnet & 0.310 & 0.510 & 0.367 & 0.470 \\
                \hline
                \multicolumn{5}{c}{\textbf{MRR}} \\ 
                \hline
                BugLocator & 0.371 & 0.489 & 0.409 & 0.476 \\
                \hline
                BLUiR & 0.296 & 0.445 & 0.309 & 0.364 \\
                \hline
                BLIA & 0.385 & 0.518 & 0.404 & 0.505 \\
                \hline
                DNNLOC & 0.398 & 0.544 & 0.420 & 0.521 \\
                \hline
                bjXnet & 0.380 & 0.535 & 0.392 & 0.496 \\
                \hline
            \end{tabular}   
            \centering
            \small
            \textbf{DLSW+EXT} = Extrinsic Bug from Deep Learning Systems \\
            \textbf{DLSW+INT} = Intrinsic Bug from Deep Learning Systems \\
            \textbf{NDLSW+EXT} = Extrinsic Bug from Non-Deep Learning Systems \\
            \textbf{NDLSW+INT} = Intrinsic Bug from Non-Deep Learning Systems 
        \end{threeparttable}
    \end{adjustbox}
\end{table}

When considering bjXnet, its performance with extrinsic bugs is slightly lower than DNNLOC in DLSW but outperforms it in NDLSW. While bjXnet's attention mechanism effectively isolates critical components in the code, extrinsic bugs often involve dependencies on external libraries and system-wide interactions that are harder to capture through localized attention. This limitation could explain bjXnet’s slightly reduced performance for extrinsic bugs, particularly in deep learning systems where external factors, like third-party libraries or hardware dependencies, play a significant role. Despite this, bjXnet's ability to focus on key sections of the code still allows it to perform well with intrinsic bugs across both systems.

Overall, these results suggest that extrinsic bugs are hard to localize, whether related to deep learning or not. However, deep learning bugs with an extrinsic nature are the more difficult to localize. On the other hand, from Fig. \ref{fig:result_extrinsic_intrinsic}, we note that the performance of all five approaches for intrinsic bugs in NDLSW is higher compared to the intrinsic bugs in DLSW, which supports the fact that the bugs related to deep learning algorithms (a.k.a DL bugs) from deep learning systems are more challenging to localize. Furthermore, we conducted appropriate statistical significance tests and found that the performance differences between extrinsic and intrinsic bugs in both systems are statistically significant, with a p-value of 0.014 for the difference between DLSW+EXT and DLSW+INT, and a p-value of 0.021 for NDLSW+EXT and NDLSW+INT, both indicating medium effect sizes, which reinforce the observed trends.\\
\indent

\textbf{Correlation of extrinsic \& deep learning-related bugs:} To determine the potential correlation between the extrinsic bugs and the bugs in deep-learning systems, we performed a Chi-Square test to determine any significant association \cite{mchugh2013chi}. We conducted three iterations with different sample data to validate the Chi-Square test, averaging the results. We got a p-value of $\approx$1.79e-14. Such a low p-value (far below the conventional threshold of 0.05) indicates that the observed association is not a product of random chance. Instead, it implies a strong dependency between external factors contributing to bugs and the specific bug patterns within deep learning systems. Our manual analysis also supports the hypothesis, showing a higher prevalence of extrinsic bugs in deep learning systems (DLSW) compared to non-deep learning systems (NDLSW) (Fig. \ref{fig:extrinsic_intrinsic}). The prevalence ratio of extrinsic bugs varies from 21.90\% to 82.35\% across different types of deep-learning bugs, confirming a strong association between extrinsic factors and bugs from deep-learning systems. Our experiments also suggest that extrinsic bugs might have an underlying connection with deep-learning bugs (refer to RQ$\mathbf{_3}$: Localization of extrinsic \& intrinsic bugs from both systems), which contributes to the poor performance of the existing bug localization techniques. 
\FrameSep.3em
\begin{frshaded}
	\noindent
	\textbf{Summary of RQ$\mathbf{_3}$:} We found that deep learning systems (DLSW) contain almost \emph{four times} more extrinsic bugs than non-deep learning systems (NDLSW). The performance of our baseline bug localization techniques for extrinsic bugs is lower (e.g., \textbf{31.27\%} less MAP for DNNLOC) compared to that of intrinsic bugs. Our research also shows a strong connection between extrinsic bugs and bugs in deep learning systems.
\end{frshaded}

\subsection{Key findings}
Our empirical results reveal several key challenges that hinder bug localization in deep learning systems. All five baseline techniques (BugLocator \cite{buglocator}, BLUiR \cite{bluir}, BLIA \cite{BLIA}, DNNLOC \cite{DNNLOC}, bjXnet \cite{han2023bjxnet}) performed significantly worse with deep learning systems than traditional software systems (e.g., 
$\sim$30\% lower MAP on average), confirming a substantial performance gap. Through our three research questions, we identified the following factors contributing to this performance gap: \\

\textbf{1. Dominance of Extrinsic Bugs}

\textit{Challenge:} Deep learning systems contain a significantly higher number of extrinsic bugs.  Extrinsic bugs are issues triggered by external factors such as hardware-software compatibility, operating systems, third-party libraries, or training data \cite{extrinsic&intrinsic}. We found nearly four times more extrinsic bugs in deep learning systems than in traditional ones. Since these bugs stem from outside the source code (e.g., GPU driver incompatibilities), they often escape code-centric bug localization methods. All five baseline techniques from our study struggled to locate the extrinsic bugs. GPU bugs were among the hardest to localize, with all our techniques performing the worst in this category (e.g., MAP below 0.30 across all the baselines). The complex interplay of deep learning system layers (framework > CUDA library > GPU hardware) \cite{chakrabarti2012cuda} makes their localization challenging. Using the Chi-Square test, we also found a strong statistical correlation (e.g., a p-value of $\sim$1.79e-14) between deep learning projects and extrinsic bugs. 

\textit{Implication:} Existing bug localization approaches mainly focus on source code analysis and thus miss critical extrinsic bugs. Appropriate tools and processes capturing external factors and environmental context will be needed to address extrinsic bugs in deep learning systems. Integrating keyword search or stack trace pattern analysis \cite{sabor2020automatic} into bug triage might help identify externally triggered failures. Similarly,  categorizing extrinsic bugs into distinct classes – issues related to third-party libraries (e.g., version mismatches), hardware-related problems (e.g., GPU bugs), operating system incompatibilities, and dataset validity – can facilitate a more systematic localization process. For example, localizing hardware-specific extrinsic bugs may greatly benefit from targeted environmental checks, utilizing tools for GPU profiling \cite{yousefzadeh2023profiling} (e.g., NVIDIA Visual Profiler\footnote{https://developer.nvidia.com/nvidia-visual-profiler} and PyTorch Profiler\footnote{https://pytorch.org/tutorials/recipes/recipes/profiler\_recipe.html}).
\\

\textbf{2. Limitations of Textual Representation}

\textit{Challenge:} Many existing techniques for bug localization (e.g., IR-based) heavily depend on textual similarity between bug reports and source code, which might not be sufficient for localizing complex, deep-learning bugs. For example, bugs related to gradient vanishing require a deeper conceptual understanding of neural network behaviour beyond what textual descriptions in bug reports typically convey (e.g., ‘the model is not learning’). Similarly, the bug reports could contain highly abstract issues (e.g., incorrect model architectures, incorrect loss function), which warrant the knowledge of the model architecture, hyperparameters, loss function, training procedure, and dataset.   Our study revealed that existing techniques significantly struggled with such nuanced cases. Although the bug reports for deep learning systems might contain more code snippets compared to traditional software systems, these snippets did not substantially enhance bug localization, according to our findings. 

\textit{Implication:} Given the abstract concepts in bug reports and the complexities in many deep learning bugs, future localization methods or tools should carefully leverage domain-specific knowledge [6]. In particular,  a comprehensive understanding of neural network concepts (e.g., layers, gradients, loss functions), common framework usage patterns, and data-processing pipelines should be injected into these tools. For example, the bug report stating ‘the model isn’t learning’ was traced back to the vanishing gradient problem caused by multiple sigmoid layers. A domain-aware technique might recognize this underlying connection and point toward these sigmoid layers.  Another way to tackle the challenge is to incorporate rich contextual information into bug reports. Essential details (e.g., model configurations, hyperparameters, dataset characteristics, and runtime conditions) in the bug report might improve bug localization since they can provide critical information to map the errors with the root cause. \\

\textbf{3. Dynamic Complexity during Training}

\textit{Challenge:} Bugs emerging during model training represent another major challenge. We identified training bugs as the most prevalent bug type ($\sim$27\%) in deep learning systems. They involve dynamic behaviors such as model convergence and iterative parameter updates. Our baseline techniques often struggle to locate these bugs since they emerge from sequences of training steps and are not visible within the code. For example, a misconfigured batch size might only reveal itself after several epochs of training, rendering traditional, code-centric bug localization ineffective.

\textit{Implication:} To effectively localize bugs manifesting during the model training phase, future techniques should incorporate dynamic analysis along with static analysis [7]. By capturing runtime information and associating it with the relevant source code, training-phase bugs, especially silent bugs, might be localized and diagnosed. \\

\textbf{4. Lack of a Tailored Approach}

\textit{Challenge:} Our study indicates that no single technique performs consistently across all bug categories in deep learning systems. Techniques based on information retrieval (e.g., BLUiR) often perform well when textual cues in a bug report map directly to the source code (e.g., API bugs). On the other hand, deep learning–based approaches (e.g., DNNLOC, bjXnet) can more effectively address conceptual bugs (e.g., model bugs). However, neither approach fully addresses a major subset of bugs from deep learning systems, ranging from hardware compatibility problems to complex model-level inconsistencies, exposing the limitations of a ‘one-size-fits-all’ paradigm. 

\textit{Implication:} Unlike traditional software systems, different bug types in deep learning systems might benefit from different localization strategies. Future work should explore more holistic and adaptive techniques. Rather than relying on a single method, researchers could identify specific bug contexts (e.g., model, GPU, or API bug) and then apply appropriately tailored localization methods.

\section{Threats to Validity} \label{sec:loc_threats}
We identify a few threats to the validity of our findings. In this section, we discuss these threats and the necessary steps taken to mitigate them as follows. 

\textbf{Threats to internal validity} relate to experimental errors and human biases \cite{tian2014automated}. Traditional bug tracking systems (e.g., Bugzilla, GitHub, Jira) contain thousands of bug reports, and their quality cannot be guaranteed. This could be a source of threat as the bug reports are used as queries to locate the buggy files. Bug reports often contain poor, insufficient, missing, or even inaccurate information \cite{gupta2021systematic}. Hence, we used data from existing benchmarks (\cite{denchmark, BuGL}, where the authors took necessary steps to avoid low-quality or invalid bug reports. Thus, such threats might be mitigated.

Another potential source of threat could be the replication of existing work. The original replication package was unavailable; hence, we used the publicly available version of BugLocator, BLUiR \cite{lee2018bench4bl}, and DNNLOC \cite{DNNLOCreplicationpackage}. For BLIA and bjXnet, we reused the author's replication package \cite{BLIA, han2023bjxnet}. We validated our implementation of the existing methods using their original dataset and achieved comparable results (e.g., with differences $\approx$ 2.00\%--3.00\% using MAP).

\textbf{Threats to conclusion validity.} The observations from our study and the conclusions we drew from them could be a source of threat to conclusion validity \cite{garcia2012statistical}. In this research, we answer three research questions using two different datasets and re-implement five existing techniques. We use appropriate statistical tests (e.g., t-test) and report the test details (e.g., p-value, Cohen's D) to conclude. Thus, such threats might also be mitigated. \\
\indent
\textbf{Threats to construct validity} relate to the use of appropriate performance metrics. We evaluate all the methodologies using MRR, MAP, and Top@K, which have been used widely by the related work \cite{buglocator, bluir, amalgam, DNNLOC, BRTracer, deeplocator, wen2016locus}. Thus, such threats might also be mitigated.

\section{Related Work}\label{sec:loc_related_work}
\subsection{Software bug}
Understanding the nature and characteristics of bugs is essential for effective debugging and testing. They can differ across different programming languages and development frameworks \cite{bugsinpy}. Over the last 50 years, hundreds of studies have been conducted to tackle bugs in traditional software systems. Recently, bugs from deep learning systems have garnered much attention due to their great interest and significance. \citet{humbatova2020taxonomy}  proposed a taxonomy of bugs from deep learning systems with five main categories - model, training, tensor \& input, API, and GPU. \citet{chen2020comprehensive} focused on the unique obstacles for deep learning-based software deployment. According to \citet{islam2019comprehensive}, data bugs and logic bugs are the most severe in deep-learning software systems. Another study by \citet{islam2020repairing} showed that the bugs or repair patterns of deep learning models significantly differ from those of traditional systems. As a result, traditional software debugging approaches, such as bug localization techniques, might not be effective for deep learning systems. Therefore, empirical research like ours, which focuses on the challenges posed by deep learning bugs in the software debugging process (specifically in bug localization), is essential.

\subsection{Spectrum-based bug localization}
Spectrum-based bug localization techniques have been widely studied for bug localization in traditional software systems. These techniques analyze program traces from passing and failing tests to measure how frequently each program element is executed in both cases \cite{akbar2020large}. Based on these frequencies, spectrum-based techniques rank program elements by their suspiciousness, with the aim of prioritizing fault-prone locations. For example, Tarantula \cite{jones2005empirical} is a widely known spectrum-based localization technique that calculates suspiciousness by measuring the ratio of failed to total executions of a particular element, highlighting the ones more frequently associated with failures.

However, capturing execution traces for spectrum-based approaches is not as straightforward in deep learning systems. One of the main reasons for this is that not all bugs in DL systems result in crashes or stack traces \cite{islam2019comprehensive}. Instead, many DL system bugs manifest as silent bugs \cite{tambon2024silent}, leading to incorrect but seemingly valid behavior, making it difficult to generate program traces that are traditionally used in spectrum-based techniques. Moreover, the non-deterministic and data-driven nature of DL models further complicates the application of spectrum-based bug localization techniques. For instance, DeepFault \cite{eniser2019deepfault} attempts to apply spectrum-based techniques to Deep Neural Networks (DNNs) by focusing on neuron activation spectra to identify suspicious neurons. While this approach is promising, it fails to handle more complex bugs, such as silent bugs or those caused by external factors, including GPU-related issues.

Another major challenge of applying spectrum-based bug localization techniques to deep learning systems is the resource-intensive nature of these systems. DL models are computationally expensive to train and execute, requiring substantial hardware resources, such as GPUs and large memory capacities. The overhead of collecting and processing execution traces for spectrum-based approaches can, therefore, become prohibitive, particularly in large-scale DL projects. Given these limitations, spectrum-based bug localization techniques are not well-suited for our study, which focuses on localizing bugs in DL systems. 

\subsection{Information Retrieval-based bug localization}
One of the crucial steps toward fixing a software bug is to detect its location within the software code. Many existing approaches \cite{ buglocator, bluir, amalgam, blizzard, BLIA} use Information Retrieval (IR) to locate bugs by matching keywords between a query and the source code. 

\citet{buglocator} introduce BugLocator, which leverages textual similarity between bug reports and source code using rVSM for bug localization. \citet{bluir} propose BLUiR, which determines the textual similarity between source code and bug reports using the Okapi-BM25 algorithm \cite{bm25}. BLUiR also leverages structural items from both bug reports and source code, which boosts its localization performance. Later, \citet{amalgam} propose AmaLgam, which incorporates the textual similarity from BugLocator, structured items from BLUiR, and version control history into IR-based bug localization.

\citet{wang2015evaluating} analyzed IR-based fault localization techniques and found their effectiveness to be limited, mainly due to the frequent unavailability of high-quality bug reports. The quality of bug reports makes it challenging to localize bugs using traditional IR-based techniques. \citet{blizzard} propose BLIZZARD, which leverages the quality aspect of bug reports and introduces context-aware query reformulation into bug localization. \citet{BRTracer} proposed BRTracer, which improves upon BugLocator by combining source document segmentation and stack-trace analysis. 

\citet{le2014predicting} used an automated method to predict the effectiveness of IR-based bug localization by leveraging features extracted from bug reports and localization methods, with their findings focusing on the significance of metadata features (e.g., commit history, stack traces) in enhancing the performance of these techniques. Another technique, namely Locus \cite{wen2016locus}, uses the software change information from commit logs and change histories to improve bug localization. \citet{BLIA} proposed BLIA, which integrates bug reports, structured information of source files, and source code change history. It localizes bugs in two granularity levels - file level and method level -- and outperforms prior approaches.

All these IR-based approaches have been designed with a focus on traditional software bugs. Bugs in deep learning applications pose several unique challenges: (a) non-deterministic behavior due to factors like random initialization and stochastic optimization \cite{sajjadi2016regularization}, (b) complex relationships between high-dimensional data and model behavior and the influence of data-specific issues without direct code-level manifestations \cite{islam2019comprehensive}, (c) strong external dependencies on hardware (e.g., PyTorch leverages GPU) \cite{foutse}. Although IR-based bug localization techniques have shown promising results in traditional software systems, their performance might decline while localizing bugs in deep learning systems. Our experiments also show relevant evidence to support this observation. Please check Section \ref{sec:loc_result} for further details on our experiments.

Recently, \citet{irblrose} used basic IR-based techniques (e.g., VSM, rVSM, BM25) for locating bugs in deep learning systems but reported poor performance without any comprehensive analysis or explanation. Thus, the potential of existing IR-based solutions for bugs in deep-learning applications is not well understood yet. Our work in this article fills in that significant gap in the literature.

To the best of our knowledge, no IR-based techniques have been developed exclusively for localizing bugs in DL systems. Existing approaches for bug detection in DL systems, such as DeepLocalize \cite{wardat2021deeplocalize}, DeepDiagnosis \cite{wardat2022deepdiagnosis}, and DeepFD \cite{cao2022deepfd}, focus on specific types of DL bugs (e.g., model bugs, training bugs) and rely on dynamic analysis or heuristic-based methods. While these approaches are valuable, they are limited in terms of scope, scalability, and generalizability. They cannot be applied to non-DL bugs. For example, DeepLocalize is tightly coupled with the Keras library and exhibits low accuracy in bug detection. DeepDiagnosis depends on hard-coded rules, restricting its adaptability across different contexts. Moreover, these techniques are highly specific to DL models and cannot be adapted to traditional software bugs, making direct comparisons infeasible. On the other hand, IR-based techniques like BugLocator, BLUiR, and BLIA are more general-purpose and can be applied to both DL and non-DL systems.

By including traditional IR-based and recent deep learning-based baselines in our study, we attempt to identify the strengths and limitations of these techniques when applied to DL systems.

\subsection{Deep learning-based bug localization}
Unlike the above IR-based methods, deep learning can detect non-linear relationships between bug reports and source code for bug localization \cite{obulesu2018machine, almeida2002predictive, bitvai2015non}. \citet{polisetty2019usefulness} evaluated deep learning-based bug localization models against traditional machine learning (ML) models, finding that while deep neural network (DNN) models generally outperform conventional ML models in performance, they require substantial resources such as GPUs and memory. \citet{DNNLOC} propose DNNLOC combining with information retrieval (e.g., rVSM \cite{buglocator}) and deep learning for bug localization. \citet{deeplocator} propose DeepLocator, where they use CNN and AST to extract features from bug reports and source documents,  respectively. To learn unified features from natural language and source code during bug localization, \citet{NPCNN} propose NP-CNN, which integrates both lexical and program structure information. \citet{CAST} propose CAST, combining a tree-based CNN (TB-CNN) with customized AST to locate buggy files. However, these deep learning-based techniques are developed and evaluated using the source code from traditional software systems (e.g., JDT, SWT, Tomcat, AspectJ). These software systems do not represent deep learning applications, and thus, the designed techniques above might not be sufficient to tackle all the challenges of deep learning-related bugs. 

\citet{deeplocalize} propose an approach to locate Deep Neural Network (DNN) bugs through dynamic and statistical analysis. However, their method's sole focus on model and training bugs, low accuracy, and over-reliance on the Keras library pose challenges for practical adoption. Deep learning-based approaches also lack explainability and heavily rely on source code, which may not be sufficient for the bugs with external dependencies (a.k.a extrinsic bugs) in deep learning applications. 

Additionally, \citet{cao2022understanding} characterize the performance problems in deep learning (DL) systems and identify key stages where these problems are introduced and exposed. They highlight the unique challenges of detecting and localizing bugs in DL systems, particularly due to dependencies on GPUs and external libraries like TensorFlow and Keras. However, their focus is limited to performance problems related to hardware dependencies, overlooking broader system components such as DNN model architecture and training dynamics. For example, they do not address how issues like vanishing gradients during model training or incorrect hyperparameter configurations (e.g., learning rate or batch size) can also lead to performance degradation in deep learning systems. Therefore, empirical research like ours, which focuses on the challenges posed by deep learning bugs in the software debugging process (specifically in bug localization), is essential.

To address the above gap, in this empirical study, we replicated five existing techniques \cite{buglocator,bluir, BLIA, DNNLOC, han2023bjxnet} to locate bugs in deep learning systems. Unlike \citet{irblrose}, our study extends beyond bug localization from deep learning systems. Our study evaluates existing bug localization techniques, categorizes deep-learning bugs, analyzes their prevalence and challenges, and assesses each technique's effectiveness for different bug types. We also conduct extensive manual analysis and explain they are difficult to localize (e.g., extrinsic factors, multifaceted dependencies), which makes our work \emph{novel}.

\section{Conclusion}\label{sec:Conclusion}

Identifying the location of a bug within a software system (a.k.a. bug localization) is crucial to correct any bug. In recent years, bug localization techniques have received considerable attention in the context of traditional software systems. However, they might not be sufficient for deep learning systems as deep learning bugs pose a greater challenge due to their multifaceted dependencies. However, the potential of existing approaches for localizing bugs in deep learning systems is not well understood to date. In this work, we replicated five existing bug localization approaches and found that they show poor performance in localizing bugs from deep-learning systems. Secondly, through an in-depth analysis, we found that localizing certain categories of bugs (e.g., training bugs \& GPU bugs) is more difficult than other bugs in deep learning systems. Finally, we investigate and find that deep learning bugs are more likely to be extrinsic, i.e., connected to non-code artifacts (e.g., training data). Our research thus offers empirical evidence and actionable insights for deep learning software bugs, advancing automated software debugging research. Future work can focus on developing a new framework for automated software debugging based on the insights from this empirical study.

\section*{Statement and Declarations}

\textbf{Funding}\\
This work was supported by the Natural Sciences and Engineering Research Council of Canada (Discovery Grant RGPIN‑03236). \\
\noindent
\\
\textbf{Conflicts of Interests}\\
The authors declare they have no conflict of interest.\\
\noindent
\\
\textbf{Ethical Approval}\\
Not applicable.\\
\noindent
\\
\textbf{Informed Consent}\\
Not applicable.\\
\noindent
\\
\textbf{Author Contributions}\\
Conceptualization: Sigma Jahan; Methodology: Sigma Jahan; Formal analysis and investigation: Sigma Jahan; Writing—original draft preparation: Sigma Jahan; Writing—review \& editing: Sigma Jahan; Resources: Sigma Jahan, Mohammad Masudur Rahman; Supervision: Mohammad Masudur Rahman; Data annotation: Sigma Jahan, Mehil B Shah; Replication of baseline technique: Sigma Jahan, Mehil B Shah.\\
\noindent
\\
\textbf{Data Availability Statement} \\
The dataset and supplementary materials are publicly available at \url{https://doi.org/10.5281/zenodo.15091574}.

\newpage
\bibliography{3.1_reference_2025.bib}
\appendix
\newpage
\section{Appendix}\label{appendix:A}
\begin{figure*}[htbp]
  \centering
  \includegraphics[width=0.85\textwidth]{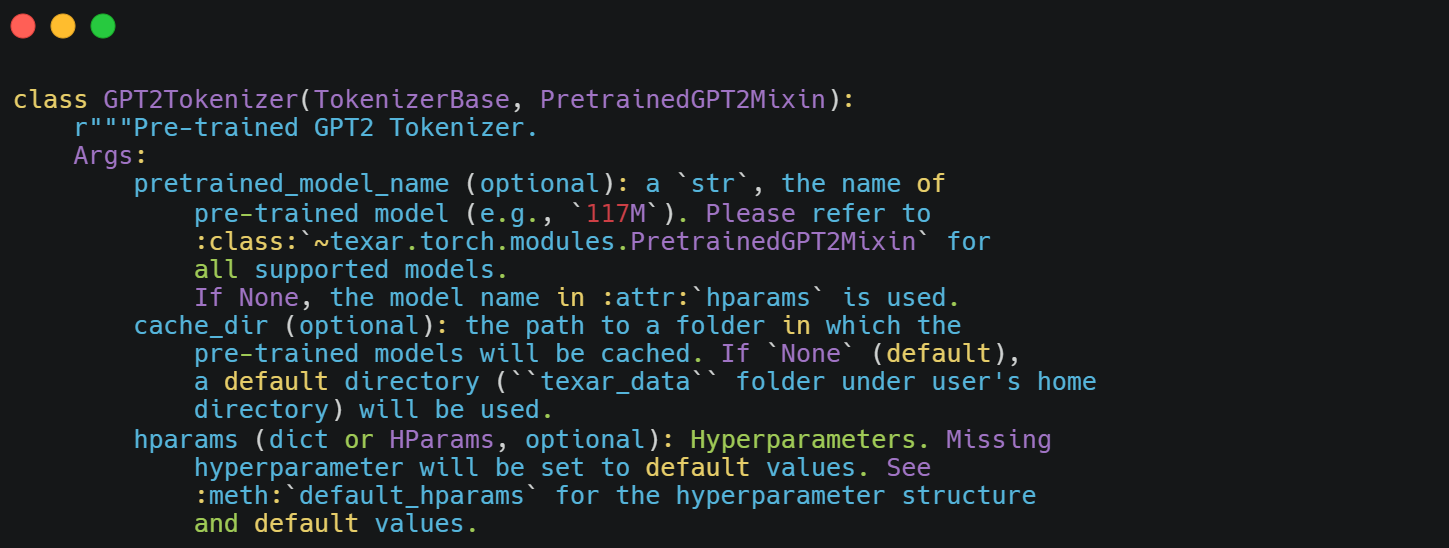}
  \caption{Code snippet of the ground truth for the model bug \cite{texar_pytorch_issue_313} \url{(https://bit.ly/3XJCdqq)}}
  \label{fig:my_label}
  \label{fig:model_bug_actual}
\end{figure*}
\begin{figure*}[htbp]
  \centering
  \includegraphics[width=0.85\textwidth]{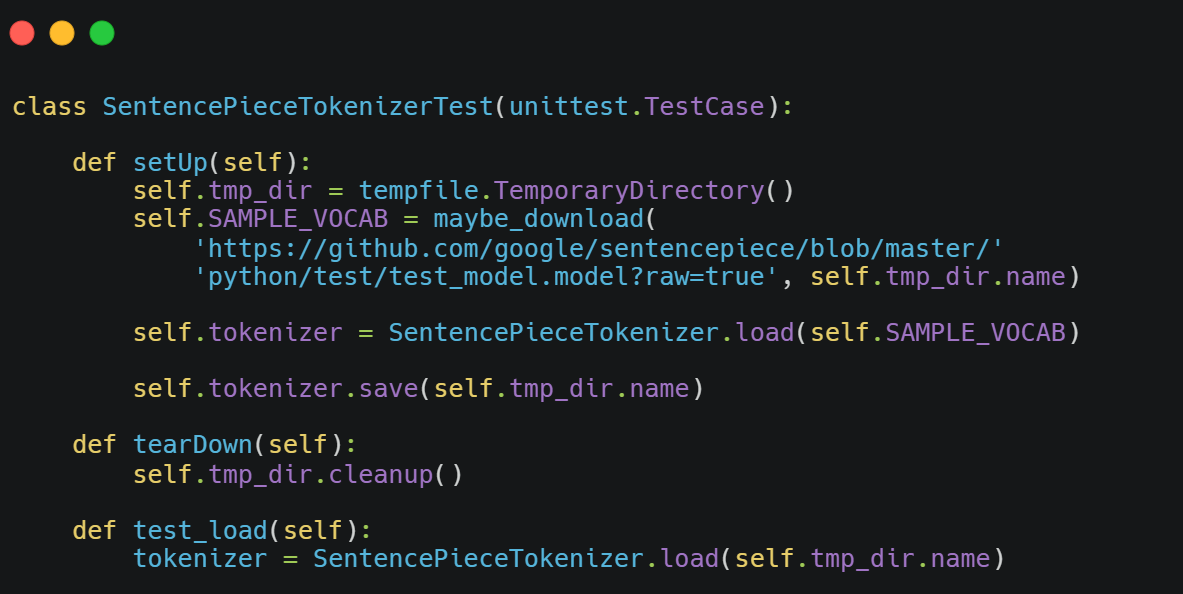}
  \caption{Code snippet of the incorrect source file for the model bug retrieved by BugLocator \url{(https://bit.ly/43klnj9)}}
  \label{fig:model_bug_false_class}
\end{figure*}
\begin{figure*}[htbp]
  \centering
  \includegraphics[width=0.85\textwidth]{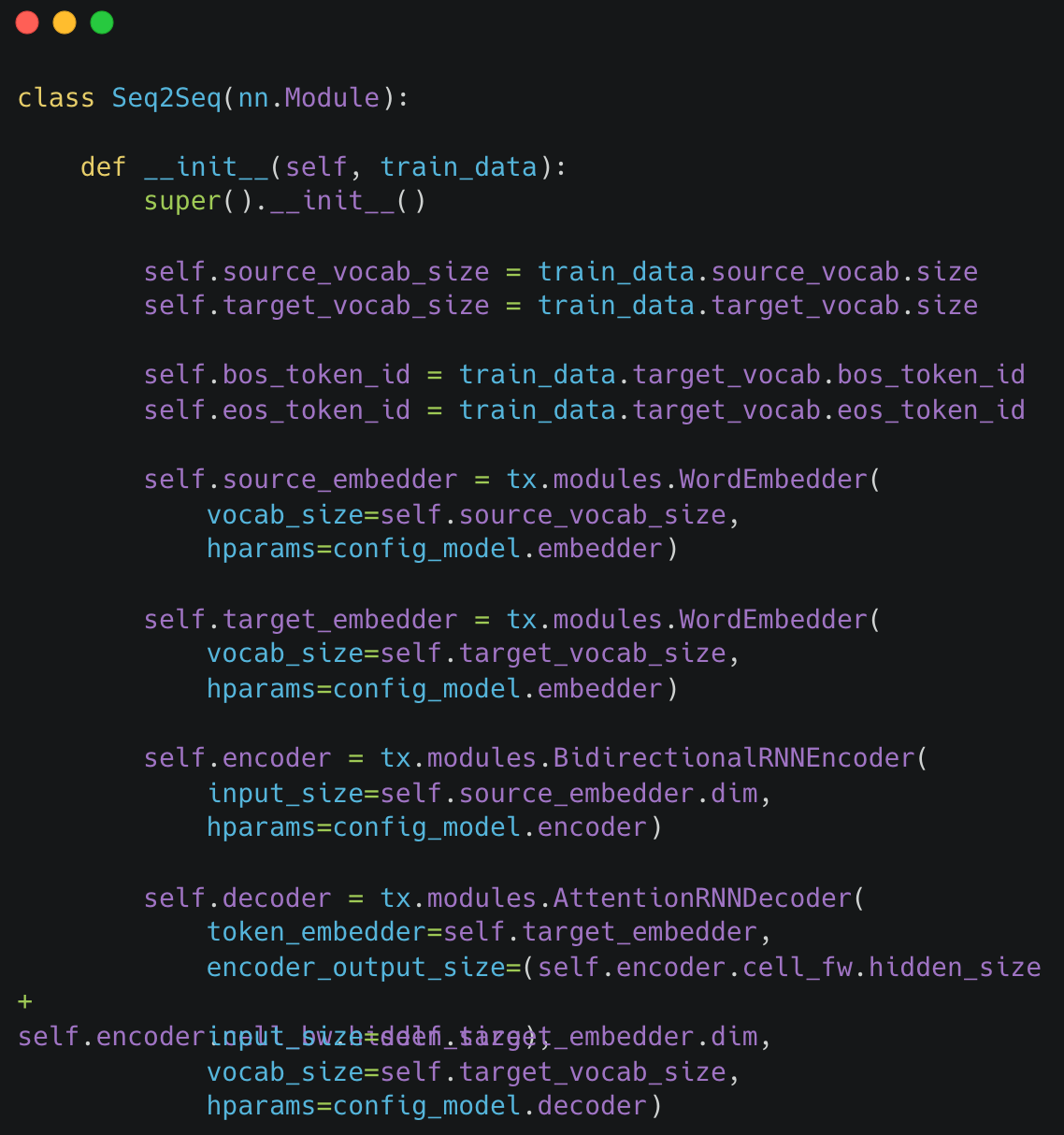}
  \caption{Code snippet of the incorrect source file for the model bug retrieved by BLUiR \url{(https://bit.ly/3PQo88U)}}
  \label{fig:model_bluir}
\end{figure*}
\begin{figure*}[htbp]
  \centering
  \includegraphics[width=0.85\textwidth]{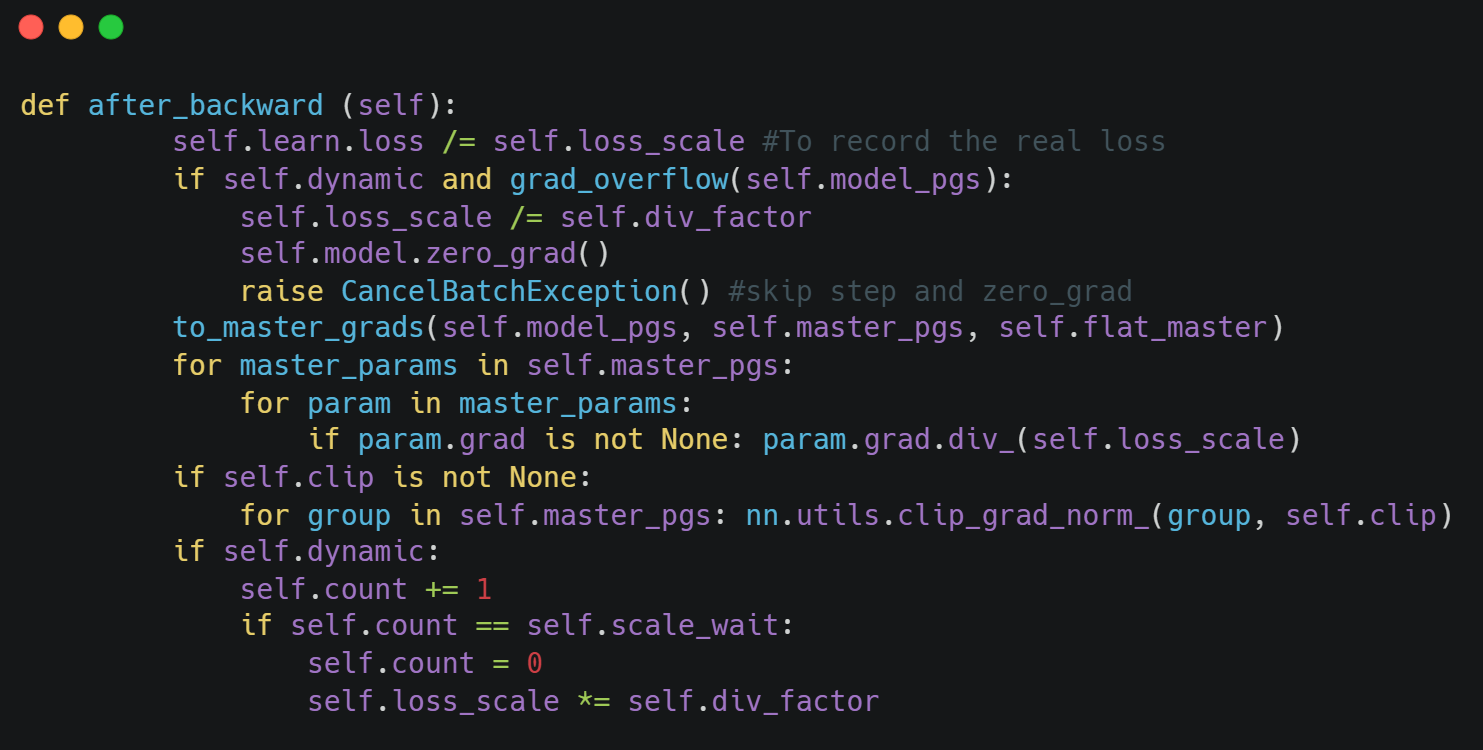}
  \caption{Code snippet of the ground truth for training bug \cite{fastai_issue_3048} \url{(https://bit.ly/3NKfRAH)}}
  \label{fig:training_bug_actual}
\end{figure*}
\begin{figure*}[htbp]
  \centering
  \includegraphics[width=0.85\textwidth]{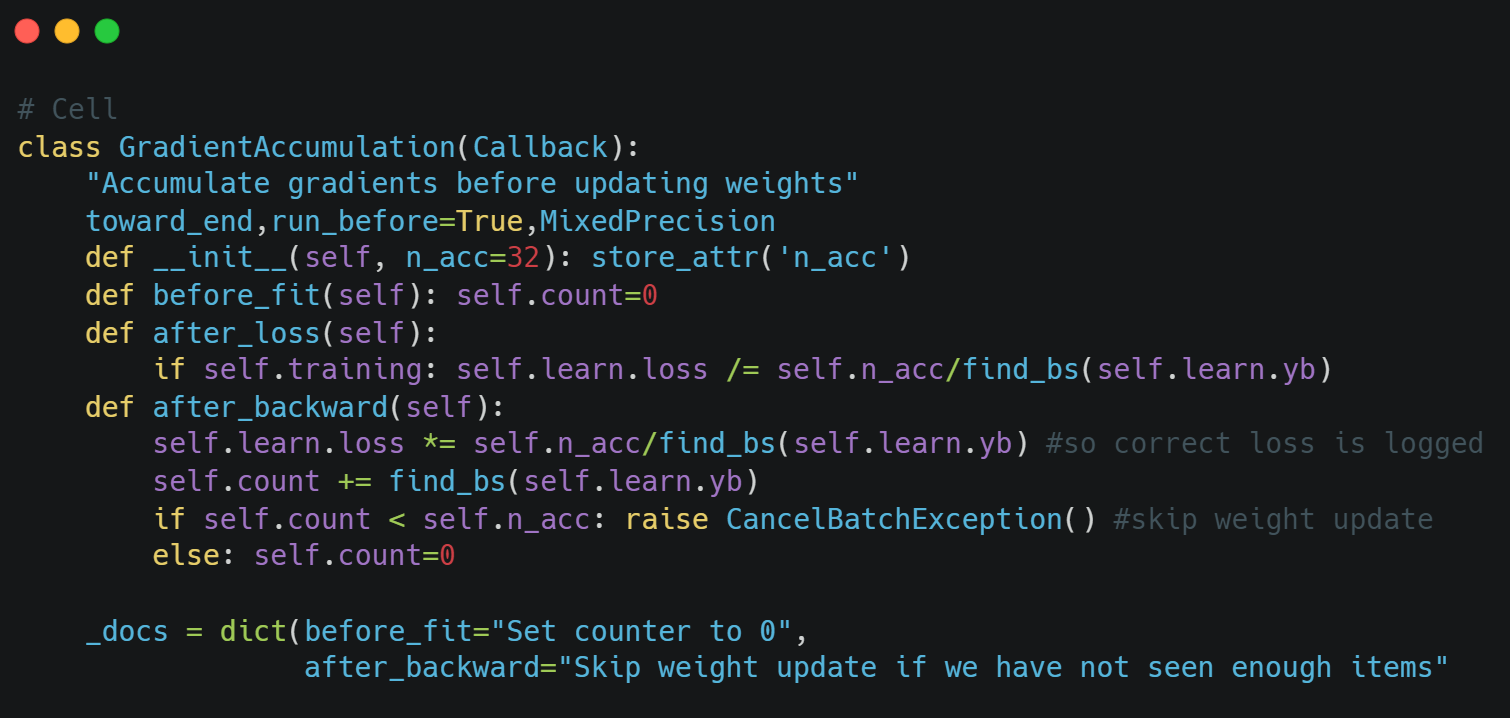}
  \centering
  \caption{Code snippet of the incorrect source file for the training bug retrieved by BugLocator \url{(https://bit.ly/3O5yEIa)}}
  \centering
  \label{fig:training_bug_false}
\end{figure*}
\begin{figure*}[htbp]
  \centering
  \includegraphics[width=0.85\textwidth]{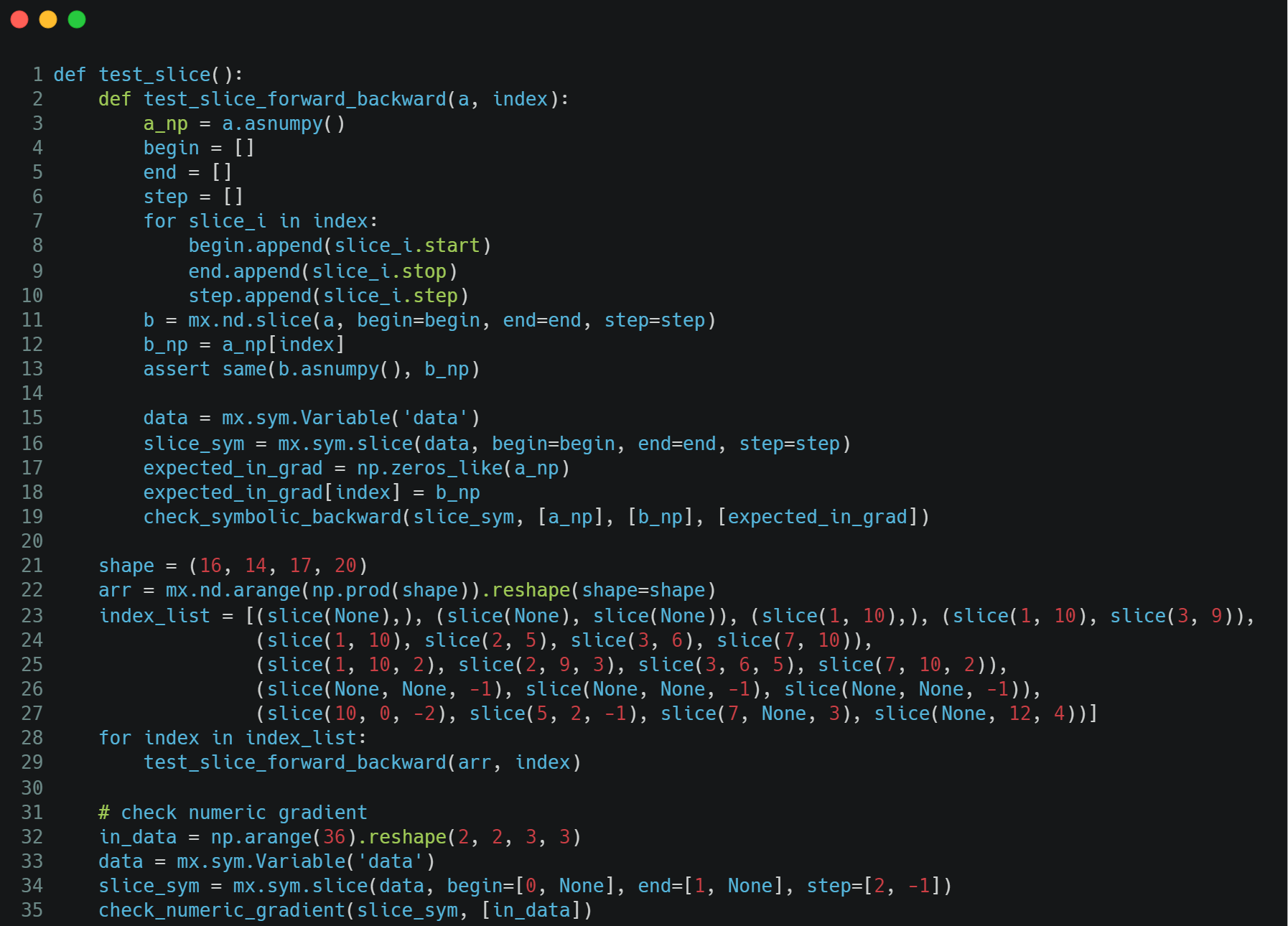}
  \caption{Code snippet of the ground truth for the tensor bug \cite{mxnet_issue_13760} \url{(https://bit.ly/3XLe82y)}}
  \label{fig:tensor_bug_actual}
\end{figure*}
\begin{figure*}[htbp]
  \centering
  \includegraphics[width=0.85\textwidth]{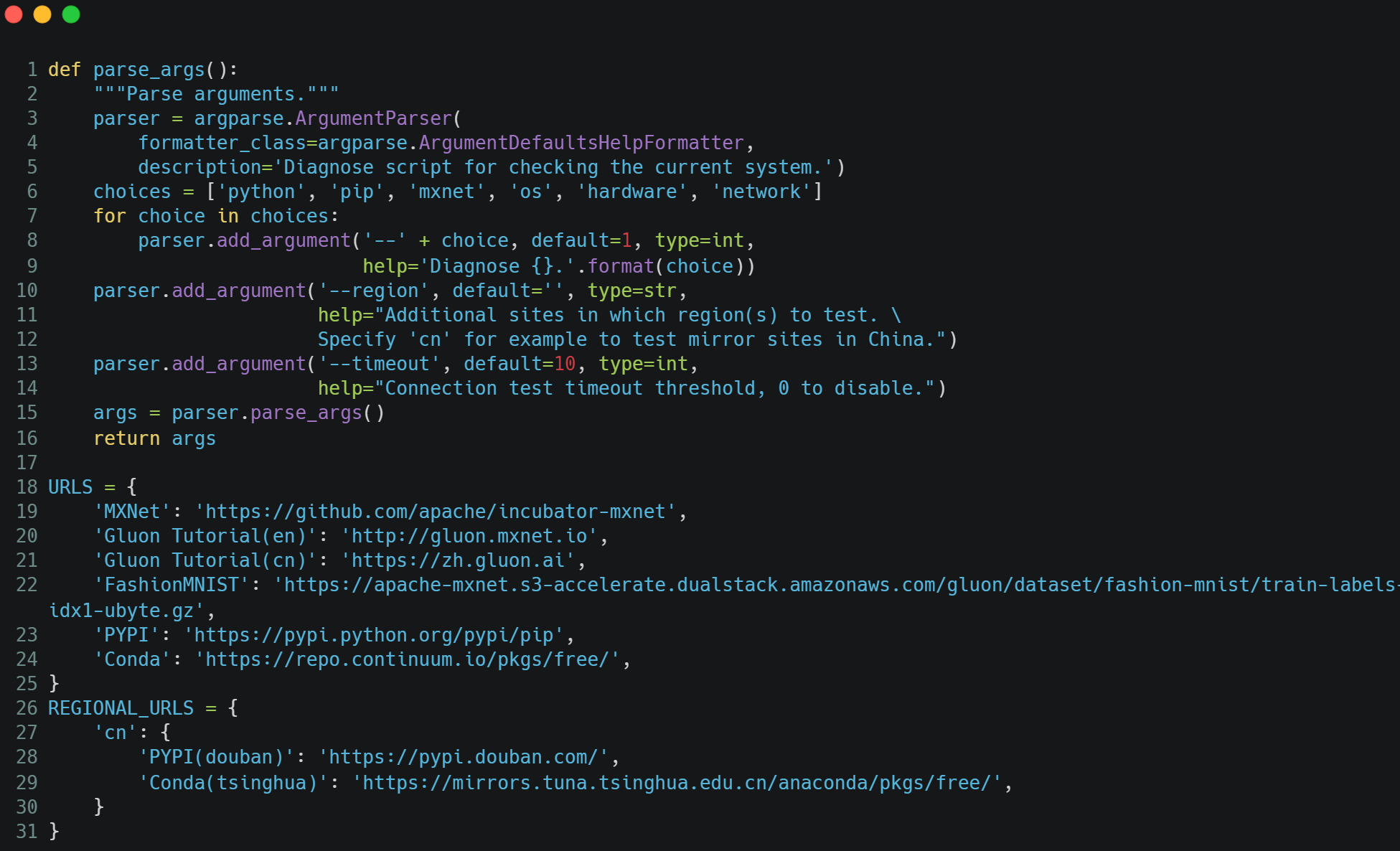}
  \centering
  \caption{Code snippet of the incorrect source file for the tensor bug retrieved by BugLocator \url{(https://bit.ly/3Dr5LA9)}}
  \centering
  \label{fig:tensor_bug_false}
\end{figure*}
\begin{figure*}[htbp]
  \centering
  \includegraphics[width=0.85\textwidth]{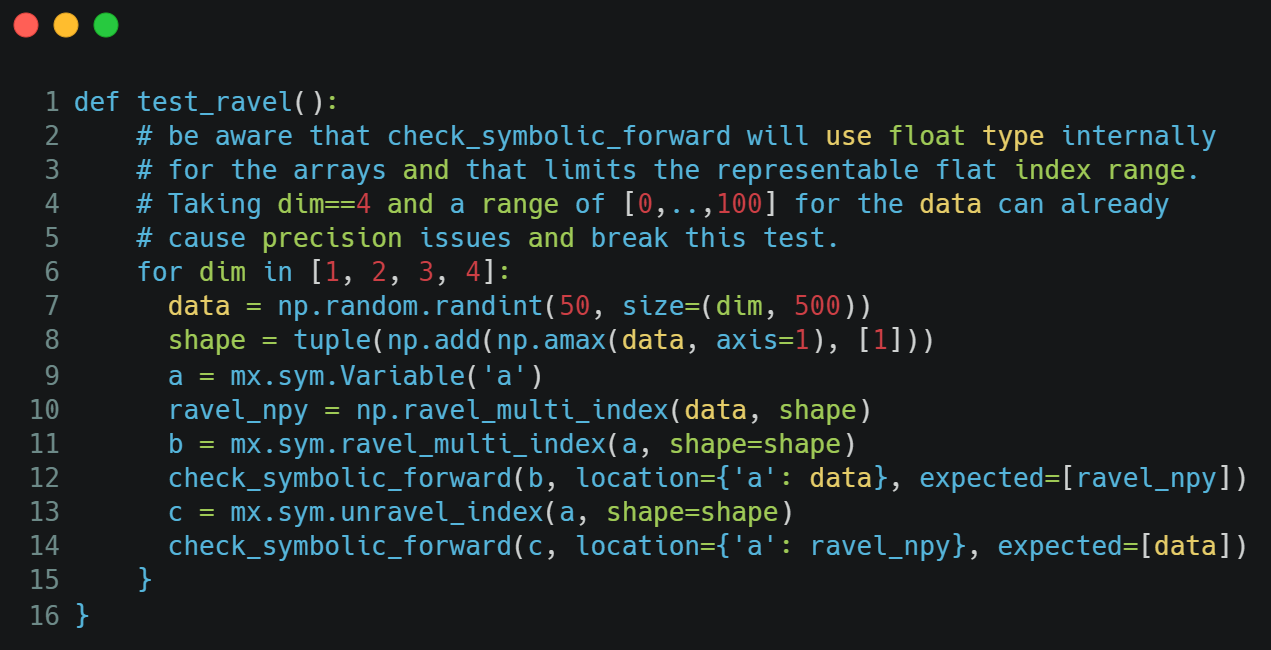}
  \caption{Code snippet of the ground truth for the API bug \cite{mxnet_issue_13862} \url{(https://bit.ly/3JT1JE8)}}
  \label{fig:API_bug_actual}
\end{figure*}
\begin{figure*}[htbp]
  \centering
  \includegraphics[width=0.85\textwidth]{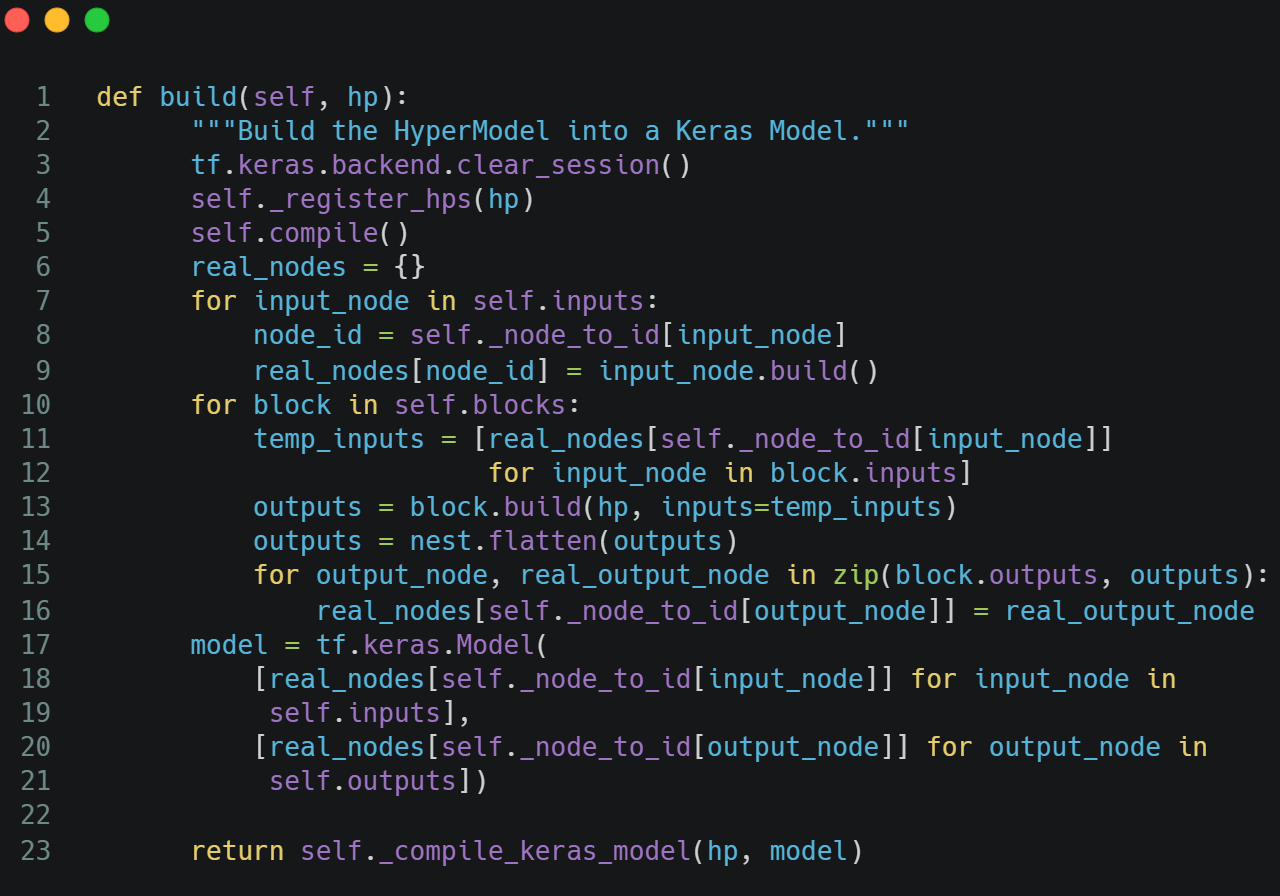}
  \caption{Code snippet of the ground truth for the GPU bug \cite{autokeras_issue_1238} \url{(https://bit.ly/44llZXc)}}
  \label{fig:GPU_bug_actual}
\end{figure*}
\end{document}